%% file: main.tex
\documentclass[conference]{IEEEtran}

\ifCLASSOPTIONonecolumn
\usepackage{setspace}
\fi

\IEEEoverridecommandlockouts
\usepackage{cite}
\usepackage{amsmath,amssymb,amsfonts}
\usepackage{adjustbox}
\usepackage{graphicx}
\usepackage{booktabs}
\usepackage{multirow}
\usepackage{color}
\usepackage{comment}
\usepackage{psfig}
\usepackage{multirow}
\usepackage{placeins}
\usepackage{environ}
\NewEnviron{nestedcomment}{\mycomment{\BODY}}
\usepackage{algorithm}
\usepackage{algpseudocode}
\ifCLASSOPTIONonecolumn
\usepackage[font=small,labelfont=bf]{caption}
\usepackage{etoolbox}
\makeatletter
\patchcmd{\@makecaption}
  {\scshape}
  {}
  {}
  {}
\makeatother

\fi
\def\urltilde{\kern -.15em\lower .7ex\hbox{\~{}}\kern .04em}

\begin{document}
\title{Model-Based Approaches to Channel Charting
\thanks{This work is partially supported by NSF grant 2030029.}
\thanks{This work was partially presented at the International Conference on Computing, Networking, and Communications (ICNC) 2023 and the IEEE International Conference on Communications (ICC) 2023.}
}

\ifCLASSOPTIONonecolumn
\author{\IEEEauthorblockN{ Amr Aly, {\em Student Member, IEEE}}\\
\IEEEauthorblockA{\textrm{CPCC, Department of EECS} \\
\textrm{University of California, Irvine, CA, USA}\\
alyas@uci.edu\\[7mm]}
\and
\IEEEauthorblockN{ Ender Ayanoglu, {\em Fellow, IEEE}}\\
\IEEEauthorblockA{\textrm{CPCC, Department of EECS} \\
\textrm{University of California, Irvine, CA, USA}
 \\ayanoglu@uci.edu
}}
\else
\author{\IEEEauthorblockN{ Amr Aly, {\em Student Member, IEEE}}
\IEEEauthorblockA{\textrm{CPCC, Department of EECS} \\
\textrm{University of California, Irvine, CA, USA}\\
alyas@uci.edu\\}

\and
\IEEEauthorblockN{Ender Ayanoglu, {\em Fellow, IEEE}}
\IEEEauthorblockA{\textrm{CPCC, Department of EECS} \\
\textrm{University of California, Irvine, CA, USA}\\
ayanoglu@uci.edu
}
}
\fi

\maketitle

\begin{abstract}
We present new ways of producing a channel chart \cite{b1} employing model-based approaches. We estimate the angle of arrival $\theta$ and the distance $\rho$ between the base station and the user equipment by employing our algorithms, inverse of the root sum squares of channel coefficients (ISQ) algorithm, linear regression (LR) algorithm, and 
MUSIC/MUSIC (MM) algorithm. We compare these methods with the channel charting algorithms principal component analysis (PCA), Sammon's method (SM), and autoencoder (AE) \cite{b1}. We show that ISQ, LR, and MM 
surpass
PCA, SM, and AE in performance.
We also compare our algorithm MM with an algorithm from the literature that uses the MUSIC algorithm jointly on $\theta$ and $\rho$. We call this algorithm the JM algorithm. JM performs very slightly better than MM but at a substantial increase in complexity. Finally, we introduce the rotate-and-sum (RS) algorithm which has about the same performance as the MM and JM algorithms. Unlike MUSIC, RS does not employ eigenvalue and eigenvector analysis. Thus, it is more suitable for direct register transfer logic (RTL) implementation.
\end{abstract}

\input{introduction}
\input{channelmodels}
\input{estimation}
\input{environment}

\input{performance}

\input{conclusion}

\ifCLASSOPTIONonecolumn
\clearpage\newpage
\fi
\bibliographystyle{IEEEtran}
\bibliography{ref}
\end{document}

%% file: introduction.tex
\section{Introduction}\label{ch:1}
A channel chart is a chart created from channel state information (CSI) that preserves the relative geometry of the radio environment consisting of a base station (BS) and user equipments (UEs) \cite{b1}. This chart helps the BS locate the UEs (relatively), which can help in many applications such as handover, cell search, user localization, and more.
Previous papers have proposed estimation of a channel chart using a number of dimensionality reduction techniques whereas in this paper we calculate the channel chart directly, employing model-based approaches.

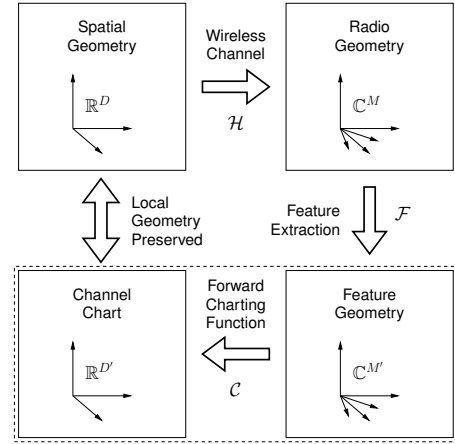
\begin{figure}[!ht]
\centering
\ifCLASSOPTIONonecolumn
\scalebox{0.5}{\input{cchart.pstex_t}}
\else
\scalebox{0.35}{\input{cchart.pstex_t}}
\fi
\caption{Summary of channel charting via dimensionality reduction \cite{b1}.}
\label{fig:cchart}
\end{figure}
Channel charting was introduced in \cite{b1}.
Consider Fig.~\ref{fig:cchart}, which is a redrawn and simplified version of Fig.~3 in \cite{b1}. As discussed in \cite{b1}, UE transmitters are located in spatial geometry $\mathbb{R}^D$, where $D=2$ or $3$. The BS receiver extracts CSI in radio geometry $\mathbb{C}^M$ where $M \gg D$.
In the next two steps, a channel chart is generated in $\mathbb{R}^{D'}$ where $D'\le D$ such that
the representation in $\mathbb{R}^{D'}$ preserves 
the relative positions of the UEs. Reference \cite{b1} discusses and compares three dimensionality reduction algorithms from the machine learning literature, namely principal component analysis (PCA), Sammon's mapping (SM), and autoencoder (AE).\footnote{Please see Sec.~\ref{ch:5} about the SM+ algorithm from \cite{b1}.} PCA is a linear method for dimensionality reduction. It maps a high-dimensional point set (e.g., CSI features) into a low-dimensional point set (e.g., the channel chart) in an unsupervised manner \cite{b1}. Although it is sometimes referred to as an unsupervised machine learning algorithm, it performs dimensionality reduction only for the data points used in calculations and does not form a function one can use to perform dimensionality reduction for future data points. SM is a nonlinear method for dimensionality reduction with the goal of retaining small pairwise distances between the two point sets \cite{b1}. Similarly to PCA, it does not form a function for dimensionality reduction of future data points. Whereas, an AE is a deep artificial neural network used for unsupervised dimensionality reduction \cite{b1}. Unlike PCA and SM, it performs learning and can be used for future data points.

In Fig.~\ref{fig:cchart}, four blocks to carry out channel charting are presented. In the upper left, the spatial geometry in $\mathbb{R}^D$ is shown. The upper right block creates the radio geometry in $\mathbb{C}^M$. The lower blocks perform feature extraction and forward charting to create channel charts. The approach proposed in this paper keeps the upper two blocks. It replaces the lower two blocks with model-based techniques to determine the angle of arrival and the distance from the BS of the UE directly. As a result, preservation of the relative positions of the UEs with respect to the BS, a goal of channel charting, is automatically satisfied.

Currently, channel charting is an active research area, see, e.g., \cite{DMMTS18,HCGMTGS19,LCTGS19,AUS20,GHL20,KAST20,FDOG20,PKTAV20,RLDJ20,FDOG21,ATL21,KPALT21,DTZJS21,LeMagoarou21,RGHASS21,ZW21,AUSMZH22,YMPC22,AUS22,ESGDT22,RLRAJ22,KAST22}. Potential ways to improve channel charting are discussed in \cite{HCGMTGS19,LCTGS19,AUS20,LeMagoarou21,AUSMZH22}. The use of channel charting for different applications is discussed in \cite{KAST20,RLDJ20,ATL21,KPALT21,ZW21,RLRAJ22,KAST22}. References \cite{DMMTS18,GHL20,PKTAV20,DTZJS21,AUS22} are on multipoint channel charting. References \cite{FDOG20,FDOG21,ESGDT22} discuss triplet-based channel charting, while \cite{RGHASS21,YMPC22} concentrate on triplet loss. Reference \cite{ZW21} tries to use the AE algorithm in a supervised fashion by allowing some of the UEs to have global positioning system (GPS) data for the exact location and use it to improve the AE learning of the geometry.
\subsection{Contributions of the Paper}
We estimate the angle of arrival (AOA) $\theta$ and the distance $\rho$ of the UE from the BS, based on the phase and magnitude of the CSI. Towards that end, we propose four techniques.
\begin{itemize}
\item
We estimate the AOA $\theta$ using the MUSIC algorithm based on the correlation matrix of the channel coefficients \cite{b5}. For the distance $\rho$ between the UE and the BS, we sum the magnitude of the channel gain on all antennas for each UE, then we take the inverse of the root of the sums and use it as $\rho$. We call this algorithm ISQ.
\item
We estimate the AOA $\theta$ using the MUSIC algorithm. For $\rho$, we consider linear regression of the known locations of a subset of UEs with the logarithm of the sum of the absolute value of the channel gain to estimate a slope and intercept. We call this algorithm LR.
\item
Our third algorithm requires multicarrier transmission from the UE to the BS. For $\theta$, we use MUSIC. For $\rho$, we also use MUSIC with multiple subcarriers to estimate the distance from the phase difference. We call this algorithm MM.
\item
We model the ideal CSI matrix in the absence of noise and fading. This model happens to have entries proportional to powers of $A_\theta=e^{j\pi \cos(\theta)}$ and $A_\rho=e^{-j2\pi\rho\Delta\! f/c}$ where $\Delta\! f$ is the difference in frequency between the subcarriers and $c$ is the speed of light. We then use a novel procedure to determine the best fits to $\theta$ and $\rho$ in the actual CSI matrix among all possible $\theta$ and $\rho$. We call this algorithm RS. RS does not employ eigenvalue and eigenvector analysis and may be simpler to implement in register transfer logic (RTL).
\end{itemize}
We then compare the results of these approaches and those of PCA, SM, and AE. We also compare their performance among themselves. This performance is measured by two quantities CT and TW introduced in Sec.~\ref{sec:CT+TW} as well as by using channel charts in terms of visual representation. We provide a comparison of the complexity and simulation times of PCA, SM, AE, ISQ, LR, and MM. Our algorithms perform better than the conventional algorithms in \cite{b1}. Note that in ISQ, LR, and MM, $\theta$ is calculated by employing the MUSIC algorithm. In each of these three cases, the novelty is in the calculation of $\rho$. This is the first use of the MUSIC algorithm in channel charting and this is the first paper that introduces model-based approaches to channel charting.
In addition,
\begin{itemize}
\item We compare the performance and complexity of an algorithm from the literature \cite{b6} we call JM with that of MM. JM performs about the same as MM, but has significantly more complexity.
\end{itemize}

We note that in our algorithms ISQ, LR, and MM, the main contribution is new methods for estimating the range (distance $\rho$ of the UE to the BS). In these algorithms, the AOA $\theta$ is calculated by using the conventional MUSIC algorithm. In our algorithms ISQ and LR, in addition to using the MUSIC algorithm for $\theta$, heuristic algorithms are used to estimate $\rho$. In our algorithm MM, in addition to using the MUSIC algorithm for $\theta$, the MUSIC algorithm is used for $\rho$ while employing multicarrier transmission. On the other hand, our algorithm RS does not employ the MUSIC algorithm either to estimate $\theta$ or $\rho$, instead it uses properties of the CSI matrix to estimate these quantities. Unlike the original paper on channel charting \cite{b1}, we do not explicitly attempt the preserve the local geometry of UEs. This property occurs as a consequence of estimating the UE positions accurately. As a result, the relative positions of UEs are mapped to the channel chart independently. This has many advantages such as reduced computational complexity, lower latency, incorporation of individual UE positions without waiting for the calculation of the channel chart for the whole ensemble, and easy accommodation of new services.

All of the four techniques ISQ, LR, MM, and RS are new. In ISQ, LR, and MM, the conventional MUSIC algorithm is employed to estimate $\theta$, but novel algorithms are introduced to calculate $\rho$. In algorithm RS, both $\theta$ and $\rho$ are calculated by using novel methods. 

%% file: cchart.pstex_t
\begin{picture}(0,0)%
\includegraphics{cchart.pstex}%
\end{picture}%
\setlength{\unitlength}{3947sp}%
\begingroup\makeatletter\ifx\SetFigFont\undefined%
\gdef\SetFigFont#1#2#3#4#5{%
  \reset@font\fontsize{#1}{#2pt}%
  \fontfamily{#3}\fontseries{#4}\fontshape{#5}%
  \selectfont}%
\fi\endgroup%
\begin{picture}(7974,7899)(814,-8248)
\put(2401,-886){\makebox(0,0)[b]{\smash{{\SetFigFont{17}{20.4}{\sfdefault}{\mddefault}{\updefault}{\color[rgb]{0,0,0}Spatial}%
}}}}
\put(2401,-1216){\makebox(0,0)[b]{\smash{{\SetFigFont{17}{20.4}{\sfdefault}{\mddefault}{\updefault}{\color[rgb]{0,0,0}Geometry}%
}}}}
\put(7201,-886){\makebox(0,0)[b]{\smash{{\SetFigFont{17}{20.4}{\sfdefault}{\mddefault}{\updefault}{\color[rgb]{0,0,0}Radio}%
}}}}
\put(7201,-1216){\makebox(0,0)[b]{\smash{{\SetFigFont{17}{20.4}{\sfdefault}{\mddefault}{\updefault}{\color[rgb]{0,0,0}Geometry}%
}}}}
\put(2401,-5686){\makebox(0,0)[b]{\smash{{\SetFigFont{17}{20.4}{\sfdefault}{\mddefault}{\updefault}{\color[rgb]{0,0,0}Channel}%
}}}}
\put(2401,-6016){\makebox(0,0)[b]{\smash{{\SetFigFont{17}{20.4}{\sfdefault}{\mddefault}{\updefault}{\color[rgb]{0,0,0}Chart}%
}}}}
\put(4801,-1036){\makebox(0,0)[b]{\smash{{\SetFigFont{17}{20.4}{\sfdefault}{\mddefault}{\updefault}{\color[rgb]{0,0,0}Wireless}%
}}}}
\put(4801,-1366){\makebox(0,0)[b]{\smash{{\SetFigFont{17}{20.4}{\sfdefault}{\mddefault}{\updefault}{\color[rgb]{0,0,0}Channel}%
}}}}
\put(4801,-5536){\makebox(0,0)[b]{\smash{{\SetFigFont{17}{20.4}{\sfdefault}{\mddefault}{\updefault}{\color[rgb]{0,0,0}Forward}%
}}}}
\put(4801,-5866){\makebox(0,0)[b]{\smash{{\SetFigFont{17}{20.4}{\sfdefault}{\mddefault}{\updefault}{\color[rgb]{0,0,0}Charting}%
}}}}
\put(4801,-6196){\makebox(0,0)[b]{\smash{{\SetFigFont{17}{20.4}{\sfdefault}{\mddefault}{\updefault}{\color[rgb]{0,0,0}Function}%
}}}}
\put(7201,-6016){\makebox(0,0)[b]{\smash{{\SetFigFont{17}{20.4}{\sfdefault}{\mddefault}{\updefault}{\color[rgb]{0,0,0}Geometry}%
}}}}
\put(7201,-5686){\makebox(0,0)[b]{\smash{{\SetFigFont{17}{20.4}{\sfdefault}{\mddefault}{\updefault}{\color[rgb]{0,0,0}Feature}%
}}}}
\put(2926,-4036){\makebox(0,0)[lb]{\smash{{\SetFigFont{17}{20.4}{\sfdefault}{\mddefault}{\updefault}{\color[rgb]{0,0,0}Local}%
}}}}
\put(2926,-4366){\makebox(0,0)[lb]{\smash{{\SetFigFont{17}{20.4}{\sfdefault}{\mddefault}{\updefault}{\color[rgb]{0,0,0}Geometry}%
}}}}
\put(2926,-4696){\makebox(0,0)[lb]{\smash{{\SetFigFont{17}{20.4}{\sfdefault}{\mddefault}{\updefault}{\color[rgb]{0,0,0}Preserved}%
}}}}
\put(6676,-4186){\makebox(0,0)[rb]{\smash{{\SetFigFont{17}{20.4}{\sfdefault}{\mddefault}{\updefault}{\color[rgb]{0,0,0}Feature}%
}}}}
\put(6676,-4516){\makebox(0,0)[rb]{\smash{{\SetFigFont{17}{20.4}{\sfdefault}{\mddefault}{\updefault}{\color[rgb]{0,0,0}Extraction}%
}}}}
\put(7651,-4261){\makebox(0,0)[lb]{\smash{{\SetFigFont{20}{24.0}{\familydefault}{\mddefault}{\updefault}{\color[rgb]{0,0,0}${\cal F}$}%
}}}}
\put(4801,-2611){\makebox(0,0)[b]{\smash{{\SetFigFont{20}{24.0}{\familydefault}{\mddefault}{\updefault}{\color[rgb]{0,0,0}${\cal H}$}%
}}}}
\put(4801,-7411){\makebox(0,0)[b]{\smash{{\SetFigFont{20}{24.0}{\familydefault}{\mddefault}{\updefault}{\color[rgb]{0,0,0}${\cal C}$}%
}}}}
\put(2101,-2311){\makebox(0,0)[lb]{\smash{{\SetFigFont{20}{24.0}{\familydefault}{\mddefault}{\updefault}{\color[rgb]{0,0,0}$\mathbb{R}^D$}%
}}}}
\put(2101,-7111){\makebox(0,0)[lb]{\smash{{\SetFigFont{20}{24.0}{\familydefault}{\mddefault}{\updefault}{\color[rgb]{0,0,0}$\mathbb{R}^{D'}$}%
}}}}
\put(6901,-2311){\makebox(0,0)[lb]{\smash{{\SetFigFont{20}{24.0}{\familydefault}{\mddefault}{\updefault}{\color[rgb]{0,0,0}$\mathbb{C}^M$}%
}}}}
\put(6901,-7111){\makebox(0,0)[lb]{\smash{{\SetFigFont{20}{24.0}{\familydefault}{\mddefault}{\updefault}{\color[rgb]{0,0,0}$\mathbb{C}^{M'}$}%
}}}}
\end{picture}%

%% file: channelmodels.tex
\section{Channel Models}
Throughout this paper, we employ three channel models, namely vanilla line-of-sight (LOS), Quadriga LOS (QLOS), and Quadriga  non-LOS (QNLOS) \cite{b4,b44,Jaeckel16}. These are the same models used in the paper that introduced the technique of channel charting \cite{b1} so that we can compare our results with that work. We start with the simplest, vanilla LOS. Vanilla LOS is one LOS ray
described as
\begin{equation}
h=\rho^{-r}\; e^{-j\left(\frac{2 \pi \rho}{\lambda}+\phi\right)} 
\label{eq20}
\end{equation}
where $\rho$ is the distance between the transmitter and the receiver and $r$ is known as the path loss exponent. 
In (\ref{eq20}), the first term in the channel phase is linearly proportional with the distance $\rho$. 
The second term $\phi$ is a uniformly distributed random variable in $[0, 2\pi)$. The channel amplitude is a random variable (Rician (QLOS) or Rayleigh (QNLOS)) which is inversely proportional to the distance square for free space, ${\sim}\rho^{-2}$, i.e., the path loss exponent $r = 2$. For more crowded environments, the path loss exponent $r$ can be 3 or 4. We will use $r=2$ for the vanilla LOS channel in this paper.
Note that the model in (\ref{eq20}) is an idealized one, we discuss a set of more realistic channel models next.

\ifCLASSOPTIONonecolumn
\begin{table}[!t]
\begin{center}
\begin{tabular}{||c | c||}
 \hline
 Parameter & Value \\ [0.5ex]
 \hline\hline
 Antenna array & Uniform Linear Array (ULA) with spacing  $\lambda/2 = 7.495$ cm \\
 \hline
 Number of array antennas & 32  \\
 \hline
 Number of transmitters (UEs)& 2048 \\
 \hline
 Carrier frequency & 2.0 GHz \\
 \hline
Bandwidth & 312.5 kHz \\
 \hline
 Number of clusters &0  \\ 
 \hline
  Number of subcarriers &1 (up to 32 in the case of the MM algorithm (Sec.~\ref{sec:MM}))  \\ 
 \hline
\end{tabular}
\end{center}
\caption{Simulation parameters.}
\label{tab1}
\end{table}
\else
\begin{table}[!t]
\begin{center}
\caption{Simulation parameters.}
\begin{tabular}{||c | c||}
 \hline
 Parameter & Value \\ [0.5ex]
 \hline\hline
 Antenna array & Uniform Linear Array (ULA)\\
               & with spacing  $\lambda/2 = 7.495$ cm \\
 \hline
 Number of array antennas & 32  \\
 \hline
 Number of transmitters (UEs)& 2048 \\
 \hline
 Carrier frequency & 2.0 GHz \\
 \hline
Bandwidth & 312.5 kHz \\
 \hline
Number of clusters &0  \\ 
 \hline
Number of subcarriers &1 (up to 32 in the case of the \\ 
&MM algorithm (Sec.~\ref{sec:MM}))  \\ 
 \hline
\end{tabular}
\end{center}
\label{tab1}
\end{table}
\fi
Next we discuss the Quadriga channel model \cite{b4,b44,Jaeckel16}. Quadriga stands for quasi deterministic radio channel generator. It is a statistical three-dimensional geometry-based stochastic channel model employing ray tracing. According to \cite{b4}, it has the following features: {\em i)\/} three-dimensional propagation (antenna modeling, geometric polarization, scattering clusters), {\em ii)\/} continuous-time evolution, {\em iii)\/} spatially correlated large- and small-scale fading, and {\em iv)\/} transition between varying propagation scenarios. The Quadriga model is very customizable. It has many features and details. 
We employ the following set of parameters: {\em i)\/} the coordinates of the transmitters and receivers, {\em ii)\/} the carrier frequency, the bandwidth, and the number of subcarriers, {\em iii)\/} the number of clusters, {\em iv)\/} the antenna shape, polarization, number of elements, and spacing between them, and {\em v)\/} QLOS or QNLOS scenario. The model was validated by measurements in downtown Dresden, Germany \cite[Ch. 4]{Jaeckel16} and in downtown Berlin, Germany \cite[Ch. 5]{Jaeckel16}, with accuracies better than the 3GPP-3D model \cite{Jaeckel16}. The channel noise is modeled as additive white Gaussian thermal noise and, in verifications of the model, the sensitivity of the measurement system is taken into account \cite{Jaeckel16}.
In this paper we used the parameters in Table~\ref{tab1} with the Urban Macro-Cell (UMa) version of the Quadriga mode in the simulations. Some details of the measurement setup are available in \cite[Sec.~III]{b4}, in specific detail in \cite[Table~II]{b4}.

The signal-to-noise ratio (SNR) in channel model is calculated by considering the power in the received signal ($P_r$) and the power in the noise measured at the receiver ($P_n$). We note that while the estimated channel would have some noise added to it, the most significant component of the noise at the receiver is additive white Gaussian thermal noise. Then, the SNR at the receiver is given as $\textrm{SNR} = P_r/P_n$ where $P_r$ takes into account the transmitted power and the channel model, see, e.g., Sec.~II-B in \cite{MYPC22}. In the code \cite{ChaChaCode} which we used as the basis for our simulations, the calculation of SNR is carried out by normalizing $P_r$ and then properly scaling the additive white Gaussian thermal noise power $P_n$ for all three channel models.

Many channel models have been developed within the last two decades. For example, see \cite{Jain07} for COST~231 and SUI models, \cite{PIIAK21} for 3GPP, 3GPP2, COST~259, COST~273, COST~2100, WINNER, and WINNER II models, and \cite{PZZHCL22} for Quadriga, NYUSIM, and MG5G models. A discussion of these and similar models is out of scope of this paper. In this paper, we employ the same models used in \cite{b1} so as to be able to compare our results with this original paper on channel charting. 

%% file: estimation.tex
\section{Estimating the Coordinates ${\theta}$ and ${\rho}$}\label{ch:2}
We will use the symbol $\theta$ for the AOA and $\rho$ for the distance between the BS and the UE. Estimating $\theta$ and $\rho$ can happen concurrently as they do not depend on each other. In this section, we will first discuss how to estimate $\theta$ by using the MUSIC algorithm and then we will discuss our first three algorithms to estimate $\rho$.
\subsection{Estimating $\theta$ Using MUSIC}
\begin{figure}[!tb]
\vspace{18mm}
\centering
\ifCLASSOPTIONonecolumn
\includegraphics[bb = 0 0 612 792, width=0.185\textwidth]{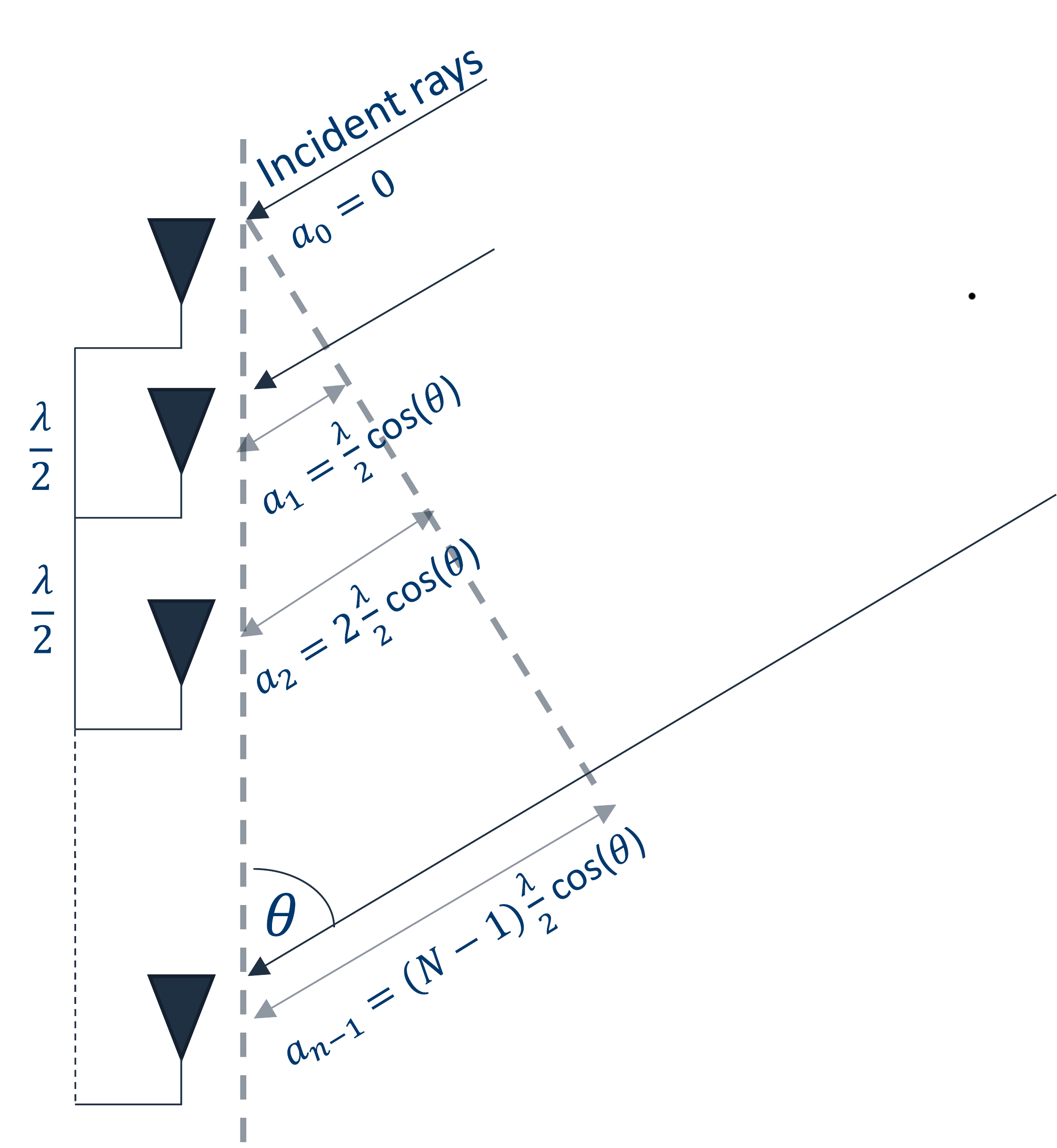}
\else
\hspace{-25mm}\includegraphics[bb = 0 0 612 792, width=0.185\textwidth]{fig1.eps}
\fi
\caption{Angle of arrival $(\theta)$ relation with phase.}
\label{fig1}
\end{figure}
From Fig.~\ref{fig1}, we can see that each antenna element will receive a ray that travels an additional distance $\frac{\lambda}{2} \cos(\theta)$ than the previous element. This means, for each antenna element, the incremental phase shift is $e^{j\pi \cos(\theta)}$. With this shift, we get what is called the steering vector
\begin{equation}
{\bf A}(\theta)= (1,e^{j\pi\cos(\theta)},e^{j\pi 2\cos(\theta)},\ldots,e^{j\pi (N_R-1)\cos(\theta)})^T,
\label{eq230}
\end{equation}
where $N_R$ is the number of receive antennas at the BS. This vector is essential in beamforming applications and in determining the AOA.

\begin{algorithm}[!t]
\caption{MUSIC Procedure for Estimating $\theta$}\label{alg:cap}
\begin{algorithmic}
    \State Calculate the CSI across antennas and subcarriers covariance matrix ${\bf R}$ = $\mathbb{E}[{\bf h}{\bf h}^{H}]$
    \State  Get the eigenvectors and eigenvalues of {\bf R}
    \State Separate system subspace $\cal S$ and noise subspace $\cal N$ by defining a threshold
    \State Calculate ${\bf N}$ by concatenating the eigenvectors of $\cal N$
         \For{\texttt{$\theta = 0: 180$} in increments of 1}
            \State Calculate the steering vector ${\bf A}(\theta)$
        \State Calculate the $\textrm{PMF}(\theta)=\frac{1}{\textrm{Norm}_2({\bf N}^H{\bf A}(\theta))}$
      \EndFor
    \State Search the PMF for a peak and find the corresponding $\theta$
\end{algorithmic}
\end{algorithm}
The steering vector ${\bf A}(\theta)$ is embedded within the CSI correlation matrix (${\bf R} =\mathbb{E} [{\bf h}{\bf h}^H]$), where ${\bf h}$ is the received channel vector at the BS. along with noise. The vector ${\bf h}$ is $N_R\times 1$. 
If we decompose $\bf R$ into its eigenvectors and examine the corresponding eigenvalues, we can separate the eigenvectors into a signal subspace $\cal S$ and a noise subspace $\cal N$, using the fact that the noise eigenvectors will correspond to very small eigenvalues compared to the signal space eigenvalues. The subspaces $\cal S$ and $\cal N$ are orthogonal to each other. Assume that the dimensionality of $\cal N$ is $p$. Form the $N_R \times p$ matrix ${\bf N}$ by concatenating the eigenvectors of $\cal N$ next to each other. The multiplication of the noise subspace eigenvectors matrix $\bf N$ and the steering vector will be almost zero. We can use this concept to find the correct angle by sweeping $\theta$ in the steering vector as illustrated in Algorithm~\ref{alg:cap} where PMF stands for probability mass function. Note that PMF($\theta$) is a PMF within a scale of constant.
%
\subsection{Estimating $\rho$}
We will now discuss how to estimate $\rho$. The simple channel ray model can be depicted as in (\ref{eq20}) with $r=2$.
\subsubsection{Estimating $\rho$ Using ISQ}
Our first proposal is a rather direct and simple approach. We calculate the square root inverse of the sum of CSI magnitudes for all antennas as
\begin{equation}
\rho=\frac{1}{\sqrt{\sum_{n=0}^{N_R-1} {\rm abs}(h_n)}} , \label{eq3}
\end{equation}
where $N_R$ is the number of antennas at the base station and $h_n$ is the channel between the UE and the $n$-th antenna at the BS.
We refer to this algorithm as ISQ (inverse square root sum). The motivation for this algorithm comes from (\ref{eq20}) with the path loss component $r=2$. Based on this formulation, $\rho = {1}/\sqrt{{\rm abs}(h_n)}$ and (\ref{eq3}) is a way of calculating this in an average sense.\footnote{Note that
\begin{equation}
\rho' =\frac{1}{\sqrt{\frac{1}{N_R}\sum_{n=0}^{N_R-1} {\rm abs}(h_n)}} = \sqrt{N_R} \rho.
\end{equation}
Thus the true average $\rho'$ is proportional to $\rho$. In other words, estimated $\rho$ is not to scale with the real $\rho$, but that will not affect the TW and CT.} 
\subsubsection{Estimating $\rho$ Using LR}
\begin{figure*}[!t]
\centering
\resizebox{\textwidth}{!}{
  \renewcommand{\arraystretch}{0}%
  \begin{tabular}{@{}c@{\hspace{0.25pt}}c@{}}
  \includegraphics[width=0.25in]{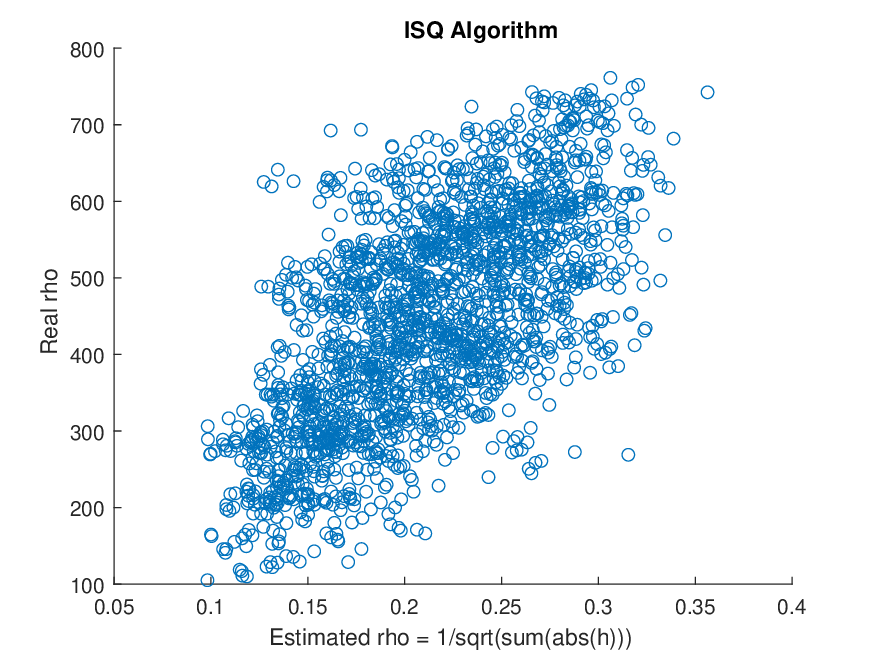} &
  \includegraphics[width=0.25in]{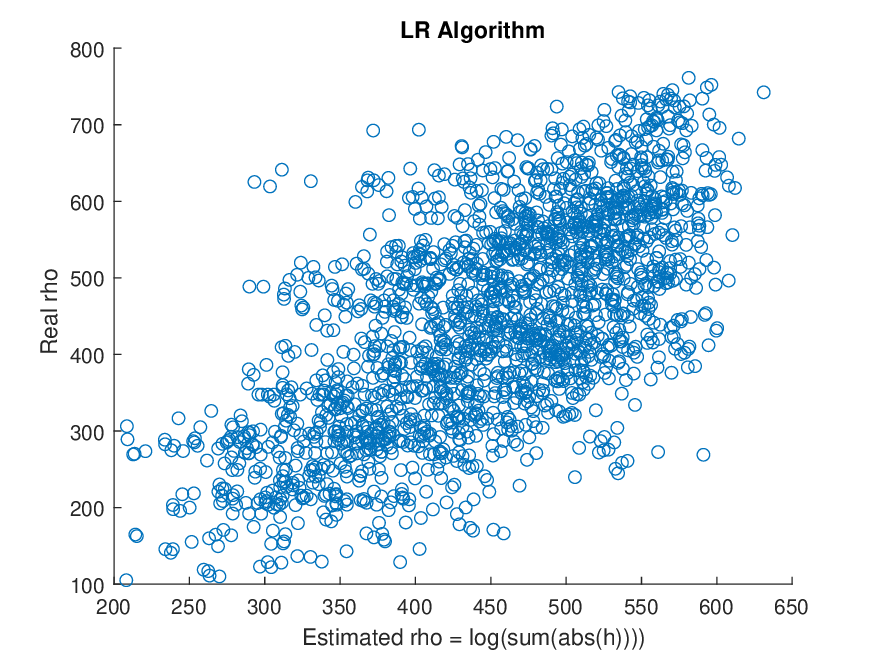} \\
  \end{tabular}
}
\caption{Correlation of real $\rho$ vs estimated $\rho$ under the channel model QNLOS.}
\label{fig20}
\end{figure*}
This is actually a learning-based, supervised approach where we assume we know the location of 256 (out of 2048) UEs and do a linear regression with the logarithm of the sum of CSI magnitudes for all antennas to find $a$ and $b$ in
\begin{equation}
\rho=aX+b,\ \ \text{where}\ X=\log\sum_{n=0}^{N_R-1} {\rm abs}(h_n) . \label{eq4}
\end{equation}
For the first 256 UEs, we use the known $\rho$ and $X$ values to generate $a$ and $b$ in a linear regression, and then for the rest of the UES, we use (\ref{eq4}) to estimate $\rho$ based on their $X$ values.
We call this algorithm the LR algorithm. As we will show later, the unsupervised performance of the ISQ algorithm is almost identical to the LR algorithm. Noting the $\log$ operation in (\ref{eq4}), and the fact that linear regression will generate $a<0$, this is a different way of expressing (\ref{eq3}).\footnote{Note that
\begin{equation}
X'=\log\bigg(\frac{1}{N_R}\sum_{n=0}^{N_R-1} {\rm abs}(h_n)\bigg) = X - \log(N_R).
\end{equation}
Therefore, the true average $X'$ differs from $X$ by a constant term, which can be absorbed by $b$ in (\ref{eq4}).} We would like to point out to the subtlety that while the LR algorithm employs the model in (\ref{eq4}), because of the use of linear regression being based on the first 256 (in our case) UE locations, it may be considered training based. We also note that the regression in (\ref{eq4}) is called a linear-log regression. Another approach could be to use log-log regression where $\rho$ in the left hand side of (\ref{eq4}) is replaced by $\log\rho$ \cite{Benoit11}. Although this is closer to the model in (\ref{eq3}), we did not observe a significant difference in numerical simulations considering linear-log and log-log regression techniques in terms of CT and TW. 
In addition, as can be observed from Fig.~\ref{fig20}, both ISQ and LR generate estimates that correlate linearly with real $\rho$.
\subsubsection{Estimating $\rho$ Using MUSIC}\label{sec:MM}
\begin{figure}[!t]
  \centering
  \includegraphics[bb = 0 0 651 572, width=0.4\textwidth]{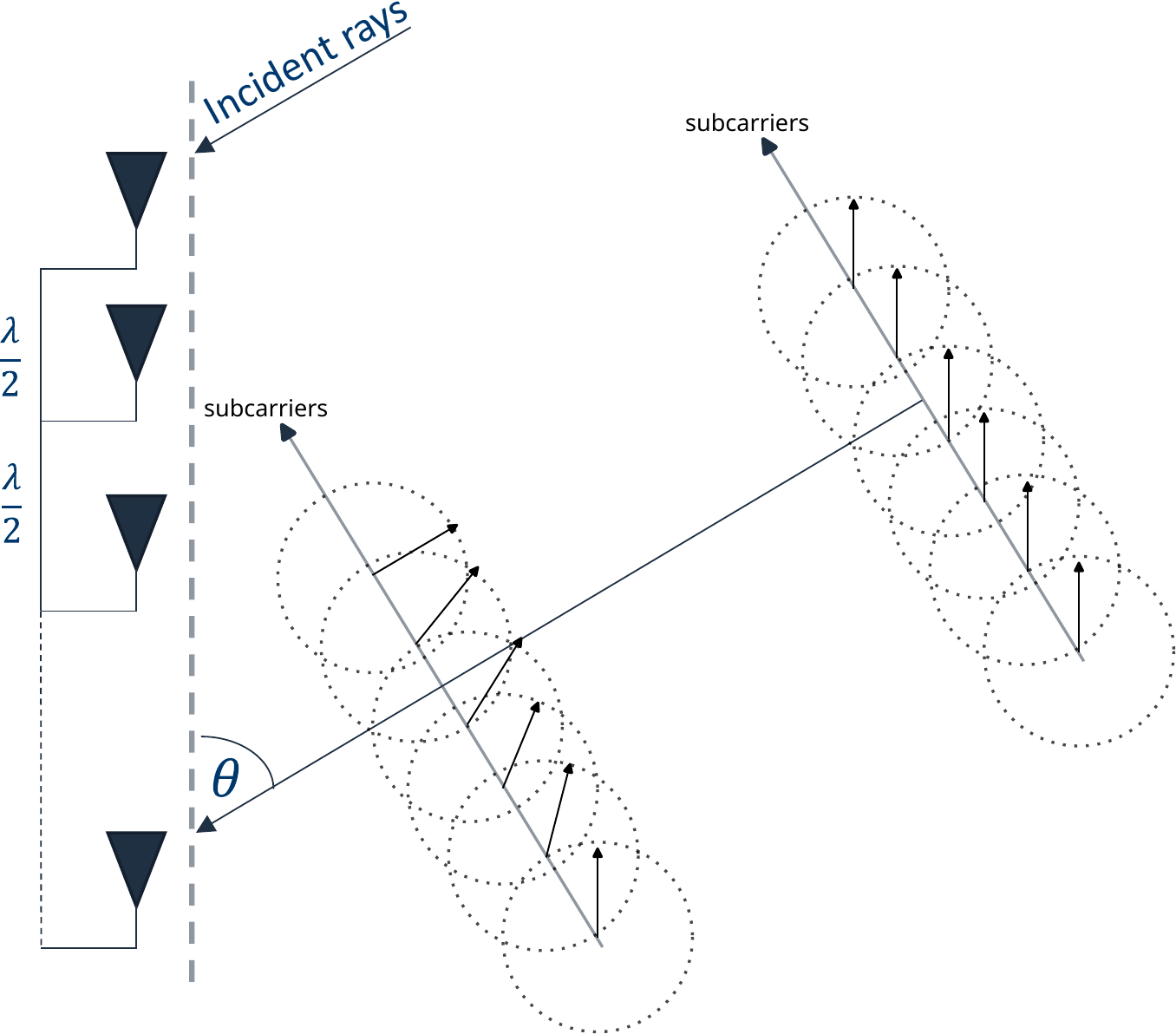}
  \caption{Phase change across subcarriers with distance.}
  \label{fig24}
\end{figure}
\begin{algorithm}[!t]
\caption{MUSIC Procedure for Estimating $\rho$}\label{alg:cap2}
\begin{algorithmic}
    \State Calculate the CSI across antennas and subcarriers covariance matrix ${\bf R}$ = $\mathbb{E}[{\bf h}{\bf h}^{H}]$
    \State  Get the eigenvectors and eigenvalues of {\bf R}
    \State Separate system subspace $\cal S$ and noise subspace $\cal N$ by defining a threshold
    \State Calculate ${\bf N}$ by concatenating the eigenvectors of $\cal N$
         \For{\texttt{$\rho = 0: 1000$} in increments of 1}
            \State Calculate vector ${\bf B}(\rho)$
        \State Calculate the $\textrm{PMF}(\rho)=\frac{1}{\textrm{Norm}_2({\bf N}^H {\bf B}(\rho))}$
      \EndFor
    \State Search the PMF for a peak and find the corresponding $\rho$
\end{algorithmic}
\end{algorithm}
Here we use the same concept for estimating $\rho$ as in estimating $\theta$. The only difference is that we use MUSIC to leverage the phase difference between subsequent subcarriers. As one can see from Fig.~\ref{fig24}, as the ray travels, the phases of the subcarriers keep changing each with rate according to their frequencies. If the subcarriers have a spacing of $\Delta\! f$ and we have $N_S$ subcarriers, their phase relation with distance is given as
\begin{equation}
{\bf B}(\rho)= (1,e^{-j2\pi\rho\Delta\!{f}/c},e^{-j2\pi\rho 2\Delta\!{f}/c},\ldots,e^{-j2\pi\rho (N_s-1) \Delta\!{f}/c})^T
\label{eq23}
\end{equation}
where 
$\rho$ is the distance 
and $c$ is the speed of light.
The vector ${\bf B}(\rho)$ will be used exactly as we used the steering vector ${\bf A}(\theta)$ in estimating $\theta$. The procedure is explained in Algorithm~\ref{alg:cap2}. Note that in Algorithm~\ref{alg:cap2}, PMF($\rho$) is a PMF within a scale of constant. We call the combination of using MUSIC to estimate $\theta$ and using MUSIC to estimate $\rho$ the MUSIC/MUSIC (MM) algorithm. A flowchart for this algorithm is given in Fig.~\ref{fig25}. In this flowchart, $N_{NS}$ is the number of noise subspace eigenvectors, i.e., $N_{NS}$ eigenvectors are selected for noise subspace.


In Sec.~\ref{ch:4} we will present the performance of the MM algorithm separately than the others, as it requires multiple subcarriers. For this reason, we cannot compare fairly with the other algorithms, but we will show the performance with different numbers of antennas and subcarriers. We came up with this algorithm within the context of channel charting. After calculating our results with it, we became aware that a related algorithm was proposed in a different context \cite{b6}. 
\ifCLASSOPTIONonecolumn
\begin{figure}[!t]
\centering
\scalebox{0.80}{\input{MUSICtheta.pstex_t}\hspace{40mm}\input{MUSICrho.pstex_t}}
\caption{MUSIC algorithm to find $\theta$ and $\rho$. The combination is called the MM algorithm..}
\label{fig25}
\end{figure}
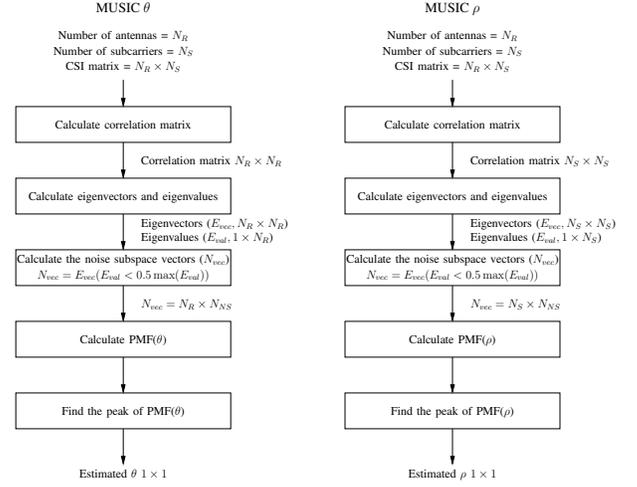
\else
\begin{figure}[!t]
\centering
\scalebox{0.375}{\input{MUSICtheta.pstex_t}\hspace{40mm}\input{MUSICrho.pstex_t}}
\caption{MUSIC algorithm to find $\theta$ and $\rho$. The combination is called the MM algorithm..}
\label{fig25}
\end{figure}
\fi
\subsection{JM Algorithm}\label{sec:JM}
We will now discuss the algorithm in \cite{b6}. This is a MUSIC algorithm that calculates $\theta$ and $\rho$ jointly, as opposed to our technique of calculating them separately in our MM algorithm in Sec.~\ref{sec:MM}. For this reason, we will call it the joint MUSIC (JM) algorithm in this paper.

The JM algorithm considers $N_R$ receive antennas at the BS, transmitted over $N_S$ subcarriers in a multicarrier transmission system. It forms an $N_S \times N_R$ matrix to use a joint MUSIC algorithm on, in order to jointly determine $\theta$ and $\rho$. To that end, it employs a subarray of dimension $N_{Sa} \times M_{Sa}$ to prepare the data in the $N_S \times N_R$ matrix for smoothing and generates an $N_{Sa}M_{Sa}\times (N_S-N_{Sa}+1)(N_R-M_{Sa}+1)$ matrix. The purpose of this two-dimensional smoothing is to remove noise at the expense of increasing the complexity of calculating the covarince matrix. After the smoothing operation, the resulting covariance matrix is $N_{Sa}M_{Sa} \times N_{Sa}M_{Sa}$. The MUSIC algorithm is then applied to the covariance matrix. With this algorithm, the search vector ${\bf A}$ is a function of both $\theta$ and $\rho$.
Then, the eigenvectors and eigenvalues of the covariance matrix are calculated. The resulting PMF is also a function of both $\theta$ and $\rho$. Furthermore, the search is carried out in two dimensions.
\subsection{RS Algorithm}\label{sec:rs}
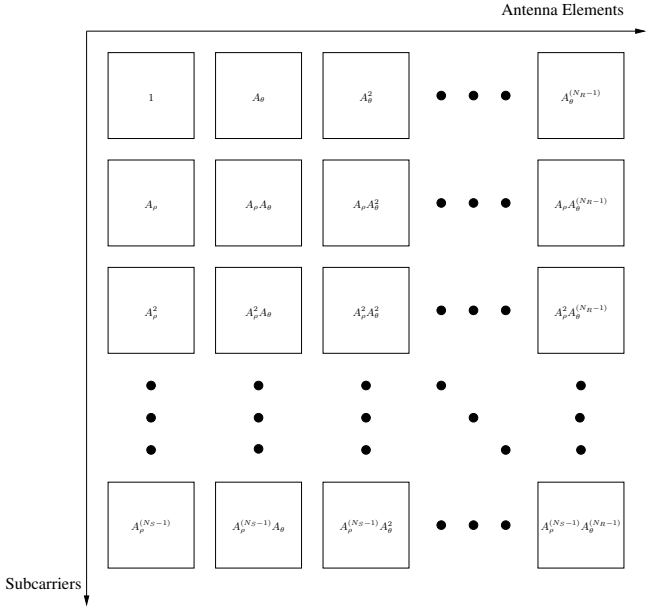
\begin{figure}[!t]
\centering
\ifCLASSOPTIONonecolumn
\scalebox{0.60}{\input{CSI_MAT_new.pstex_t}}
\else
\scalebox{0.45}{\input{CSI_MAT_new.pstex_t}}
\fi
\caption{Ideal CSI matrix $(N_S\times N_R)$ in the absence of fading and noise.}
\label{fig1001}
\end{figure}

In order to reduce complexity substantially, we propose a new model-based algorithm we call rotate and sum (RS). As shown in Fig.~\ref{fig1001}, the CSI matrix has $N_R$ column vectors where $N_R$ corresponds to the number of receive antennas at the BS, and $N_S$ row vectors where $N_S$ corresponds to the number of subcarriers. In the absence of fading and noise, for each column, we have a rotation factor $A_\theta$. This factor is the phase shift between two vertical neighboring elements or antenna elements. It is equal to $A_\theta = e^{j \pi \cos(\theta)}$, see (\ref{eq230}). For each row, we have a rotation factor $A_\rho$. This factor is the phase shift between two horizontal neighboring elements or subcarriers. It is equal to $A_\rho = e^{-j2\pi\rho\Delta\! {f}/c}$ where $\Delta\! f$ is the difference in frequency between the subcarriers and $c$ is the speed of light, see (\ref{eq23}).

\begin{algorithm}[!t]
\caption{Rotate-and-Sum Procedure for Estimating $\theta$}\label{alg:cap4}
\begin{algorithmic}
    \State Calculate the CSI matrix across antennas and subcarriers with size $N_S\times N_R$
    \State Set all entries of $N_R\times N_R$ matrix ${\bf S}$ equal to 0
    \For{${\tt subcarrier\_num}=1:N_S$}
    \State Let $\bf c$ be the row vector with index {\tt subcarrier\_num} of the CSI matrix
    \State Calculate ${\bf S} = {\bf S} + {\bf c}^H{\bf c}$
    \EndFor
    \State Divide all entries of ${\bf S}$ by $N_S$
    \For{\texttt{$\phi = 0:180$} in increments of 1}
    \State Let $A_\phi = e^{j\pi \cos\phi}$ where $\phi$ is in degrees
    \State Form matrix ${\bf B}$ as follows
    \For{$\ i = 1:N_R$ in increments of 1}
    \For{$\ j = i:N_R$ in increments of 1}
    \State ${\bf B}_{i,j} = A_\phi^{j-i}$
    \EndFor
    \For{$\ j = 1:i-1$ in increments of 1}
    \State ${\bf B}_{i,j} = [A_\phi^*]^{i-j}$
    \EndFor
    \EndFor
    \State Let $C(\phi) = \sum_{i=1}^{N_R} {\rm row}_i({\bf S}){\rm row}_i^H({\bf B})$
    \EndFor
    \State Set $\theta = {\rm arg\ max}\displaystyle_{\phi} C(\phi)$
\end{algorithmic}
\end{algorithm}
We will discuss two rotate-and-sum procedures in Algorithm~\ref{alg:cap4} and Algorithm~\ref{alg:cap5}, corresponding to the estimation of $\theta$ and $\rho$ respectively. The algorithms can be understood with the help of Fig.~\ref{fig1001}. In the case of Algorithm~\ref{alg:cap4},
consider the ideal CSI matrix shown in Fig.~\ref{fig1001}. Recall that this matrix does not include the effects of fading and noise. Let ${\bf c}$ be any row vector of this matrix and consider
\begin{equation}
{\bf c}^H{\bf c} \! = \! \left[\! \begin{array}{lllcl}
1 & A_\theta & A_\theta^2 & \cdots & A_\theta^{N_R-1} \\
A_\theta^* & 1 & A_\theta & \cdots & A_\theta^{N_R-2} \\
A_\theta^{*\, 2}&A_\theta^* & 1 & \ddots & A_\theta^{N_R-1} \\
\vdots & \vdots & \vdots & \ddots & \vdots \\[0.8mm]
A_\theta^{*\, N_R -1} & A_\theta^{*\, N_R -2}&A_\theta^{*\, N_R -3}&\cdots&1
\end{array}\! \right]\rule[-4.5em]{0pt}{0pt}
\label{eqn:ccH}
\end{equation}
which is independent of the row index. In the case of a real CSI matrix, the elements will deviate from the values of this matrix due to fading and noise. We form an average of ${\bf c}^H{\bf c}$ for all row vectors {\bf c} in the real CSI matrix expecting that the effects of fading and noise will be reduced. We call this average ${\bf c}^H{\bf c}$ matrix ${\bf S}$. Due to this averaging, we expect ${\bf S}$ to be close to the ideal ${\bf c}^H{\bf c}$ matrix in (\ref{eqn:ccH}) even when fading and noise are present.

\begin{algorithm}[!t]
\caption{Rotate-and-Sum Procedure for Estimating $\rho$}\label{alg:cap5}
\begin{algorithmic}
    \State Calculate the CSI matrix across antennas and subcarriers with size $N_S\times N_R$
    \State Set all entries of $N_S\times N_S$ matrix ${\bf S}$ equal to 0
    \For{${\tt antenna\_num}=1:N_R$}
    \State Let $\bf c$ be the column vector with index {\tt antenna\_num} of the CSI matrix
    \State Calculate ${\bf S} = {\bf S} + {\bf c}{\bf c}^H$
    \EndFor
    \State Divide all entries of ${\bf S}$ by $N_R$
    \For{\texttt{$\sigma = 0:1000$} in increments of 1}
    \State Let $A_\sigma = e^{-j2\pi\sigma \Delta\! f /c}$ where $\sigma$, $\Delta\! f/c$ are in m.
    \State Form matrix ${\bf B}$ as follows
    \For{$\ i = 1:N_S$ in increments of 1}
    \For{$\ j = i:N_S$ in increments of 1}
    \State ${\bf B}_{i,j} = [A_\sigma^*]^{j-i}$
    \EndFor
    \For{$\ j = 1:i-1$ in increments of 1}
    \State ${\bf B}_{i,j} = [A_\sigma]^{i-j}$
    \EndFor
    \EndFor
    \State Let $C(\sigma) = \sum_{i=1}^{N_S} {\rm row}_i({\bf S}){\rm row}_i^H({\bf B})$
    \EndFor
    \State Set $\rho = {\rm arg\ max}\displaystyle_{\sigma} C(\sigma)$
\end{algorithmic}
\end{algorithm}

Our next task is to figure out what $\theta$ is. To that effect, we generate matrices in the same form as (\ref{eqn:ccH}) but $\theta$ is replaced by $\phi$. We generate a number of such matrices for values of $\phi$ in $[0, \pi]$ radians. Suppressing $\phi$, each such matrix is called ${\bf B}$ in Algorithm~\ref{alg:cap4}. For each $\phi$ we compare ${\bf S}$ and ${\bf B}$, assigning a number representing how close they are. That value of $\phi$ for which the ${\bf B}$ matrix associated with it is closest to ${\bf S}$ is deemed to be the best estimate of $\theta$. 

%% file: MUSICtheta.pstex_t
\begin{picture}(0,0)%
\includegraphics{MUSICtheta.pstex}%
\end{picture}%
\setlength{\unitlength}{3947sp}%
\begingroup\makeatletter\ifx\SetFigFont\undefined%
\gdef\SetFigFont#1#2#3#4#5{%
  \reset@font\fontsize{#1}{#2pt}%
  \fontfamily{#3}\fontseries{#4}\fontshape{#5}%
  \selectfont}%
\fi\endgroup%
\begin{picture}(3624,8040)(3289,-7555)
\put(5101,-1636){\makebox(0,0)[b]{\smash{{\SetFigFont{12}{14.4}{\familydefault}{\mddefault}{\updefault}{\color[rgb]{0,0,0}Calculate correlation matrix}%
}}}}
\put(5101,-136){\makebox(0,0)[b]{\smash{{\SetFigFont{12}{14.4}{\familydefault}{\mddefault}{\updefault}{\color[rgb]{0,0,0}Number of antennas = $N_R$}%
}}}}
\put(5101,-391){\makebox(0,0)[b]{\smash{{\SetFigFont{12}{14.4}{\familydefault}{\mddefault}{\updefault}{\color[rgb]{0,0,0}Number of subcarriers = $N_S$}%
}}}}
\put(5101,-646){\makebox(0,0)[b]{\smash{{\SetFigFont{12}{14.4}{\familydefault}{\mddefault}{\updefault}{\color[rgb]{0,0,0}CSI matrix = $N_R \times N_S$}%
}}}}
\put(5101,314){\makebox(0,0)[b]{\smash{{\SetFigFont{14}{16.8}{\familydefault}{\mddefault}{\updefault}{\color[rgb]{0,0,0}MUSIC $\theta$}%
}}}}
\put(5401,-2236){\makebox(0,0)[lb]{\smash{{\SetFigFont{12}{14.4}{\familydefault}{\mddefault}{\updefault}{\color[rgb]{0,0,0}Correlation matrix $N_R \times N_R$}%
}}}}
\put(5101,-2836){\makebox(0,0)[b]{\smash{{\SetFigFont{12}{14.4}{\familydefault}{\mddefault}{\updefault}{\color[rgb]{0,0,0}Calculate eigenvectors and eigenvalues}%
}}}}
\put(5401,-3286){\makebox(0,0)[lb]{\smash{{\SetFigFont{12}{14.4}{\familydefault}{\mddefault}{\updefault}{\color[rgb]{0,0,0}Eigenvectors ($E_{vec}, N_R \times N_R$)}%
}}}}
\put(5401,-3511){\makebox(0,0)[lb]{\smash{{\SetFigFont{12}{14.4}{\familydefault}{\mddefault}{\updefault}{\color[rgb]{0,0,0}Eigenvalues ($E_{val}, 1 \times N_R$)}%
}}}}
\put(5101,-3886){\makebox(0,0)[b]{\smash{{\SetFigFont{12}{14.4}{\familydefault}{\mddefault}{\updefault}{\color[rgb]{0,0,0}Calculate the noise subspace vectors ($N_{vec}$)}%
}}}}
\put(5101,-4141){\makebox(0,0)[b]{\smash{{\SetFigFont{12}{14.4}{\familydefault}{\mddefault}{\updefault}{\color[rgb]{0,0,0}$N_{vec}=E_{vec}(E_{val}<0.5\max(E_{val}))$}%
}}}}
\put(5401,-4636){\makebox(0,0)[lb]{\smash{{\SetFigFont{12}{14.4}{\familydefault}{\mddefault}{\updefault}{\color[rgb]{0,0,0}$N_{vec} = N_R \times N_{NS}$}%
}}}}
\put(5101,-5236){\makebox(0,0)[b]{\smash{{\SetFigFont{12}{14.4}{\familydefault}{\mddefault}{\updefault}{\color[rgb]{0,0,0}Calculate PMF($\theta$)}%
}}}}
\put(5101,-6436){\makebox(0,0)[b]{\smash{{\SetFigFont{12}{14.4}{\familydefault}{\mddefault}{\updefault}{\color[rgb]{0,0,0}Find the peak of PMF($\theta$)}%
}}}}
\put(5101,-7486){\makebox(0,0)[b]{\smash{{\SetFigFont{12}{14.4}{\familydefault}{\mddefault}{\updefault}{\color[rgb]{0,0,0}Estimated $\theta\ 1\times 1$}%
}}}}
\end{picture}%

%% file: MUSICrho.pstex_t
\begin{picture}(0,0)%
\includegraphics{MUSICrho.pstex}%
\end{picture}%
\setlength{\unitlength}{3947sp}%
\begingroup\makeatletter\ifx\SetFigFont\undefined%
\gdef\SetFigFont#1#2#3#4#5{%
  \reset@font\fontsize{#1}{#2pt}%
  \fontfamily{#3}\fontseries{#4}\fontshape{#5}%
  \selectfont}%
\fi\endgroup%
\begin{picture}(3624,8040)(3289,-7555)
\put(5101,-1636){\makebox(0,0)[b]{\smash{{\SetFigFont{12}{14.4}{\familydefault}{\mddefault}{\updefault}{\color[rgb]{0,0,0}Calculate correlation matrix}%
}}}}
\put(5101,-136){\makebox(0,0)[b]{\smash{{\SetFigFont{12}{14.4}{\familydefault}{\mddefault}{\updefault}{\color[rgb]{0,0,0}Number of antennas = $N_R$}%
}}}}
\put(5101,-391){\makebox(0,0)[b]{\smash{{\SetFigFont{12}{14.4}{\familydefault}{\mddefault}{\updefault}{\color[rgb]{0,0,0}Number of subcarriers = $N_S$}%
}}}}
\put(5101,-646){\makebox(0,0)[b]{\smash{{\SetFigFont{12}{14.4}{\familydefault}{\mddefault}{\updefault}{\color[rgb]{0,0,0}CSI matrix = $N_R \times N_S$}%
}}}}
\put(5101,-2836){\makebox(0,0)[b]{\smash{{\SetFigFont{12}{14.4}{\familydefault}{\mddefault}{\updefault}{\color[rgb]{0,0,0}Calculate eigenvectors and eigenvalues}%
}}}}
\put(5101,-3886){\makebox(0,0)[b]{\smash{{\SetFigFont{12}{14.4}{\familydefault}{\mddefault}{\updefault}{\color[rgb]{0,0,0}Calculate the noise subspace vectors ($N_{vec}$)}%
}}}}
\put(5101,-4141){\makebox(0,0)[b]{\smash{{\SetFigFont{12}{14.4}{\familydefault}{\mddefault}{\updefault}{\color[rgb]{0,0,0}$N_{vec}=E_{vec}(E_{val}<0.5\max(E_{val}))$}%
}}}}
\put(5401,-4636){\makebox(0,0)[lb]{\smash{{\SetFigFont{12}{14.4}{\familydefault}{\mddefault}{\updefault}{\color[rgb]{0,0,0}$N_{vec} = N_S \times N_{NS}$}%
}}}}
\put(5101,314){\makebox(0,0)[b]{\smash{{\SetFigFont{14}{16.8}{\familydefault}{\mddefault}{\updefault}{\color[rgb]{0,0,0}MUSIC $\rho$}%
}}}}
\put(5101,-5236){\makebox(0,0)[b]{\smash{{\SetFigFont{12}{14.4}{\familydefault}{\mddefault}{\updefault}{\color[rgb]{0,0,0}Calculate PMF($\rho$)}%
}}}}
\put(5101,-6436){\makebox(0,0)[b]{\smash{{\SetFigFont{12}{14.4}{\familydefault}{\mddefault}{\updefault}{\color[rgb]{0,0,0}Find the peak of PMF($\rho$)}%
}}}}
\put(5401,-2236){\makebox(0,0)[lb]{\smash{{\SetFigFont{12}{14.4}{\familydefault}{\mddefault}{\updefault}{\color[rgb]{0,0,0}Correlation matrix $N_S \times N_S$}%
}}}}
\put(5401,-3286){\makebox(0,0)[lb]{\smash{{\SetFigFont{12}{14.4}{\familydefault}{\mddefault}{\updefault}{\color[rgb]{0,0,0}Eigenvectors ($E_{vec}, N_S \times N_S$)}%
}}}}
\put(5401,-3511){\makebox(0,0)[lb]{\smash{{\SetFigFont{12}{14.4}{\familydefault}{\mddefault}{\updefault}{\color[rgb]{0,0,0}Eigenvalues ($E_{val}, 1 \times N_S$)}%
}}}}
\put(5101,-7486){\makebox(0,0)[b]{\smash{{\SetFigFont{12}{14.4}{\familydefault}{\mddefault}{\updefault}{\color[rgb]{0,0,0}Estimated $\rho\ 1\times 1$}%
}}}}
\end{picture}%

%% file: CSI_MAT_new.pstex_t
\begin{picture}(0,0)%
\includegraphics{CSI_MAT_new.pstex}%
\end{picture}%
\setlength{\unitlength}{3947sp}%
\begingroup\makeatletter\ifx\SetFigFont\undefined%
\gdef\SetFigFont#1#2#3#4#5{%
  \reset@font\fontsize{#1}{#2pt}%
  \fontfamily{#3}\fontseries{#4}\fontshape{#5}%
  \selectfont}%
\fi\endgroup%
\begin{picture}(9056,8442)(257,-8098)
\put(9001,164){\makebox(0,0)[rb]{\smash{{\SetFigFont{14}{16.8}{\familydefault}{\mddefault}{\updefault}{\color[rgb]{0,0,0}Antenna Elements}%
}}}}
\put(1426,-7861){\makebox(0,0)[rb]{\smash{{\SetFigFont{14}{16.8}{\familydefault}{\mddefault}{\updefault}{\color[rgb]{0,0,0}Subcarriers}%
}}}}
\put(2401,-1036){\makebox(0,0)[b]{\smash{{\SetFigFont{9}{10.8}{\familydefault}{\mddefault}{\updefault}{\color[rgb]{0,0,0}$1$}%
}}}}
\put(3901,-1036){\makebox(0,0)[b]{\smash{{\SetFigFont{9}{10.8}{\familydefault}{\mddefault}{\updefault}{\color[rgb]{0,0,0}$A_\theta$}%
}}}}
\put(5401,-1036){\makebox(0,0)[b]{\smash{{\SetFigFont{9}{10.8}{\familydefault}{\mddefault}{\updefault}{\color[rgb]{0,0,0}$A_\theta^2$}%
}}}}
\put(8401,-1036){\makebox(0,0)[b]{\smash{{\SetFigFont{9}{10.8}{\familydefault}{\mddefault}{\updefault}{\color[rgb]{0,0,0}$A_\theta^{(N_R-1)}$}%
}}}}
\put(2401,-2536){\makebox(0,0)[b]{\smash{{\SetFigFont{9}{10.8}{\familydefault}{\mddefault}{\updefault}{\color[rgb]{0,0,0}$A_\rho$}%
}}}}
\put(3901,-2536){\makebox(0,0)[b]{\smash{{\SetFigFont{9}{10.8}{\familydefault}{\mddefault}{\updefault}{\color[rgb]{0,0,0}$A_\rho A_\theta$}%
}}}}
\put(5401,-2536){\makebox(0,0)[b]{\smash{{\SetFigFont{9}{10.8}{\familydefault}{\mddefault}{\updefault}{\color[rgb]{0,0,0}$A_\rho A_\theta^2$}%
}}}}
\put(8401,-2536){\makebox(0,0)[b]{\smash{{\SetFigFont{9}{10.8}{\familydefault}{\mddefault}{\updefault}{\color[rgb]{0,0,0}$A_\rho A_\theta^{(N_R-1)}$}%
}}}}
\put(2401,-4036){\makebox(0,0)[b]{\smash{{\SetFigFont{9}{10.8}{\familydefault}{\mddefault}{\updefault}{\color[rgb]{0,0,0}$A_\rho^2$}%
}}}}
\put(3901,-4036){\makebox(0,0)[b]{\smash{{\SetFigFont{9}{10.8}{\familydefault}{\mddefault}{\updefault}{\color[rgb]{0,0,0}$A_\rho^2 A_\theta$}%
}}}}
\put(5401,-4036){\makebox(0,0)[b]{\smash{{\SetFigFont{9}{10.8}{\familydefault}{\mddefault}{\updefault}{\color[rgb]{0,0,0}$A_\rho^2 A_\theta^2$}%
}}}}
\put(8401,-4036){\makebox(0,0)[b]{\smash{{\SetFigFont{9}{10.8}{\familydefault}{\mddefault}{\updefault}{\color[rgb]{0,0,0}$A_\rho^2 A_\theta^{(N_R-1)}$}%
}}}}
\put(2401,-7036){\makebox(0,0)[b]{\smash{{\SetFigFont{9}{10.8}{\familydefault}{\mddefault}{\updefault}{\color[rgb]{0,0,0}$A_\rho^{(N_S-1)}$}%
}}}}
\put(3901,-7036){\makebox(0,0)[b]{\smash{{\SetFigFont{9}{10.8}{\familydefault}{\mddefault}{\updefault}{\color[rgb]{0,0,0}$A_\rho^{(N_S-1)} A_\theta$}%
}}}}
\put(8401,-7036){\makebox(0,0)[b]{\smash{{\SetFigFont{9}{10.8}{\familydefault}{\mddefault}{\updefault}{\color[rgb]{0,0,0}$A_\rho^{(N_S-1)} A_\theta^{(N_R-1)}$}%
}}}}
\put(5401,-7036){\makebox(0,0)[b]{\smash{{\SetFigFont{9}{10.8}{\familydefault}{\mddefault}{\updefault}{\color[rgb]{0,0,0}$A_\rho^{(N_S-1)} A_\theta^2$}%
}}}}
\end{picture}%

%% file: environment.tex
\section{Simulation Environment}\label{ch:3}
In this paper we reused and integrated our algorithms into the simulation environment in \cite{b1} so that we can compare the performance improvement in a fair fashion. We adopted the simulation parameters in Table~\ref{tab1} at SNR = 0 dB. SNR is defined as the ratio of the signal power to the noise power and 0 dB means that the two are equal.
We used a three-dimensional environment exactly as in paper \cite{b1} as shown in Fig.~\ref{fig31}, where the antenna is 8.5 meters above the plane of the UEs. 
We call this three-dimensional scenario as 3D. A similar two-dimensional (2D) scenario is investigated in \cite{Aly22}.
The simulation environment is 1000m $\times$ 500m. As in \cite{b1}, the 2048 UEs are placed randomly except 234 of the UEs are selected to make the word ``VIP," so we can see if in the channel chart we preserve the shape.

\ifCLASSOPTIONonecolumn
\begin{figure*}[!t]
\centering
  \renewcommand{\arraystretch}{0}%
  \begin{tabular}{@{}c@{\hspace{1pt}}c@{\hspace{1pt}}c@{}}
  \includegraphics[width=3.0in]{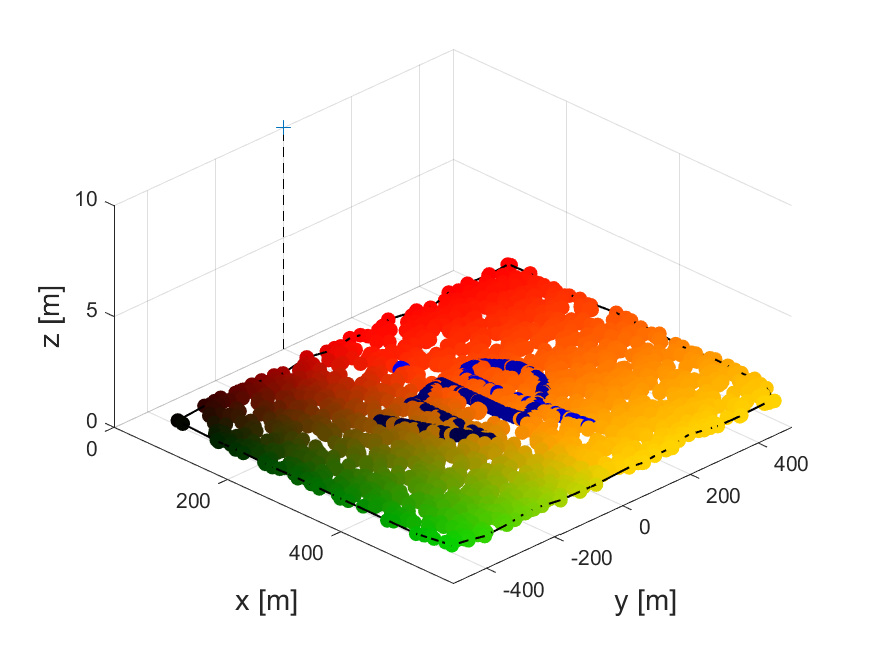} &
  \end{tabular}
  \caption{3D environment.}
  \label{fig31}
\end{figure*}
\else
\begin{figure}[!t]
\vspace{-7.5mm}
\centering
  \renewcommand{\arraystretch}{0}%
  \begin{tabular}{@{}c@{\hspace{1pt}}c@{\hspace{1pt}}c@{}}
  \includegraphics[width=3.0in]{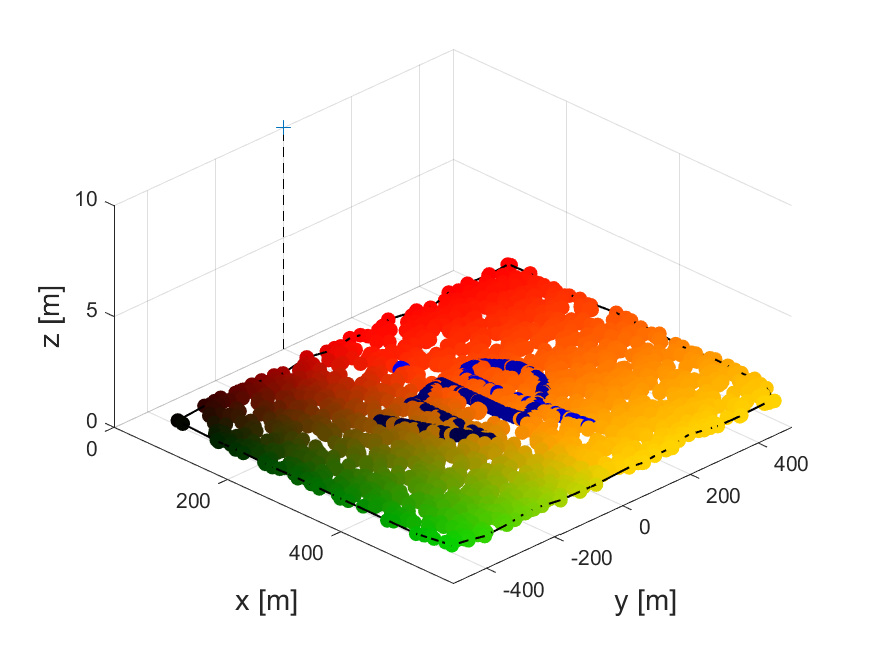} &
  \end{tabular}
  \caption{3D environment.}
  \label{fig31}
\end{figure}
\fi
The basis of the simulation program is the Matlab code \cite{ChaChaCode} released by the authors of \cite{b1}. In this code, PCA and SM are available in Matlab in pre-compiled and optimized form. We used the AE code kindly supplied to us by the first author of the original paper on channel charting \cite{b1}. The AE code was written in Python so we integrated it as an executable Python environment, which we can call from Matlab as a system call. We wrote the LR, ISQ, and MM codes. We integrated the latest Quadriga channel into the model. This is important because the Quadriga channel model employed in \cite{b1} has been changed online, with the channel employed in \cite{b1} being no longer available \cite{Struder21}.

%% file: performance.tex
\section{Performance Comparison and Complexity Analysis}\label{ch:4}
\subsection{Measures for Channel Charting: Continuity and Trustworthiness}\label{sec:CT+TW}
As in  \cite{b1}, we use continuity (CT) and trustworthiness (TW) as performance measures. CT measures if neighbors in the original space are close in the representation space.  TW measures how well the feature mapping avoids introducing new neighbor relations that were absent in the original space. Let ${\cal V}_K({\bf u}_i)$ be the $K$-neighborhood of point ${\bf u}_i$ in the original space. Also, let $\hat r(i,j)$ be the ranking of point ${\bf v}_j$ among the neighbors of point ${\bf v}_i$, ranked according to their similarity to ${\bf v}_i$. Then the point-wise continuity of the representation ${\bf v}_i$ of the point ${\bf u}_i$ is defined as
\begin{equation}
{\rm CT}_i(K) = 1 - \frac{2}{K(2N-3K -1)} \sum_{j\in {\cal V}_K({\bf u}_i)}({\hat r}(i,j) -K).
\end{equation}
The (global) continuity of a point set $\{{\bf u}_n\}_{n=1}^N$ and its representation $\{{\bf v}_n\}_{n=1}^N$ is
\begin{equation}
{\rm CT}(K)=\frac{1}{N}\sum_{i=1}^N{\rm CT}_i(K).
\end{equation}
Now, let ${\cal U}_K({\bf v}_i)$ be the set of ``false neighbors'' that are in the $K$-neighborhood of ${\bf v}_i$, but not of ${\bf u}_i$ in the original space. Also, let $r(i,j)$ be the ranking of point ${\bf u}_i$ in the neighborhood of point ${\bf u}_i$, ranked according to their similarity to ${\bf u}_i$. The point-wise trustworthiness of the representation of point ${\bf u}_i$ is then
\begin{equation}
{\rm TW}_i(K)=1-\frac{2}{K(2N-3K-1)}\sum_{j\in {\cal U}_K({\bf v}_i)}(r(i,j)-K).
\end{equation}
The (global) trustworthiness between a point set $\{{\bf u}_n\}_{n=1}^N$ and its representation $\{{\bf v}_n\}_{n=1}^N$ is
\begin{equation}
{\rm TW}(K)=\frac{1}{N}\sum_{i=1}^N{\rm TW}_i(K).
\end{equation}
Both point-wise and global CT and TW are between 0 and 1, with larger values being better \cite{b1}.
\subsection{LR and ISQ Performance Comparison}
We will now compare the performance of our algorithms LR and ISQ to the three algorithms PCA, SM, and AE from \cite{b1} in terms of TW and CT. These are given in Table~\ref{tab42} for the 3D channel. The value of $k$-nearest neighbors is 102 in both cases. 
We first note that LR and ISQ outperform the techniques in \cite{b1}, namely PCA, SM, and AE. Comparing our two techniques LR and ISQ, we find that LR performance is very close to that of ISQ. The difference is less than 1\%. 

In close examination, we note that TW values with our algorithms LR and ISQ are close to 1 for the LOS channel, while the other three algorithms have performance in the range 0.8272--0.8603, with PCA performing better. Yet, LR and ISQ have almost perfect TW performance for this channel. For the QLOS and QNLOS channels, LR and ISQ perform in the range of 0.9029--0.9092 for TW, whereas the other three algorithms have TW performance in the range of 0.8456--0.8574. Thus, LR and ISQ have significantly better TW performance than the other three algorithms for the LOS, QLOS, and QNLOS channels. Moving on to the CT performance, for the LOS channel, LR and ISQ have performance in the range of 0.9940--0.9968, which is significantly better than the other three algorithms whose performance is in the range of 0.8932--0.9288. For the QLOS and QNLOS channels, the performance of LR and ISQ is in the range of 0.9220--0.9416, while the other three algorithms have performance in the range of 0.9055--0.9278. 
\ifCLASSOPTIONonecolumn
\else
\clearpage
\begin{table*}
\begin{center}
\caption{Performance comparison for TW and CT at $k$-nearest = 102 for LR and ISQ algorithms in 3D channel.}
\begin{tabular}{|c|l|c|c|c|c|c|}
\hline
Measure & Channel & PCA & SM & AE & LR & ISQ \\ \hline
   & LOS   &     0.8603 &   0.8272  &  0.8286  &  0.9930 &   0.9885 \\ \cline{2-7}
TW & QLOS  &     0.8474 &   0.8512  &  0.8574  &  0.9089 &   0.9092 \\ \cline{2-7}
   & QNLOS &     0.8502 &   0.8456  &  0.8496  &  0.9029 &   0.9041 \\
\hline
   & LOS   &     0.9288 &   0.9051  &  0.8932  &  0.9968 &   0.9940 \\ \cline{2-7}
CT & QLOS  &     0.9223 &   0.9278  &  0.9055  &  0.9416 &   0.9304 \\ \cline{2-7}
   & QNLOS &     0.9237 &   0.9217  &  0.9057  &  0.9246 &   0.9220 \\
\hline
\end{tabular}
\end{center}
\label{tab42}
\end{table*}

\begin{figure*}
\centering 
\resizebox{\textwidth}{!}{%
  \renewcommand{\arraystretch}{0}%
\begin{tabular}{@{}c@{\hspace{1pt}}c@{\hspace{1pt}}c@{\hspace{1pt}}c@{\hspace{1pt}}c@{}}
  \includegraphics[width=0.5in]{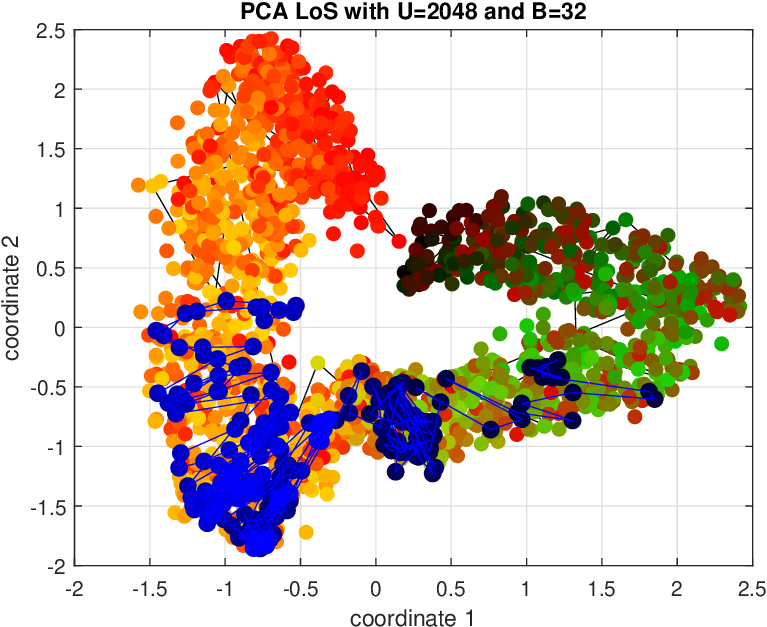} &
  \includegraphics[width=0.5in]{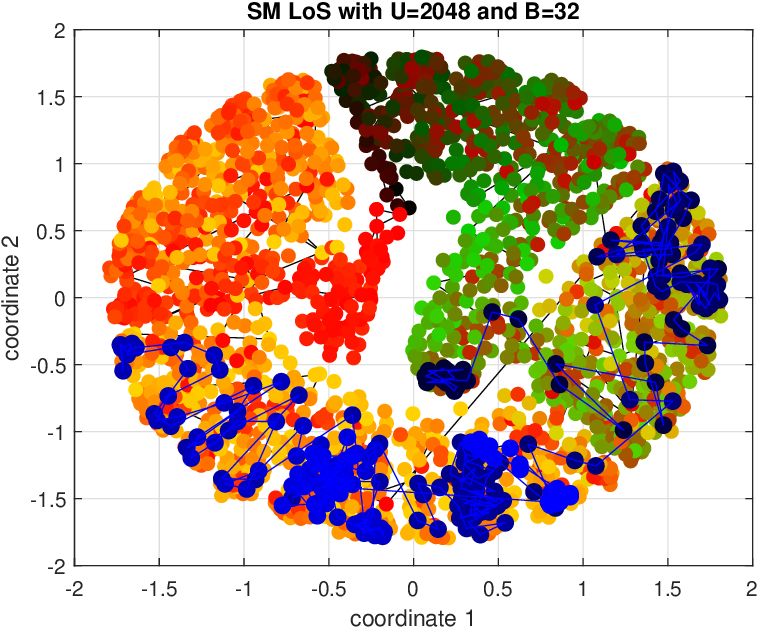} &
  \includegraphics[width=0.5in]{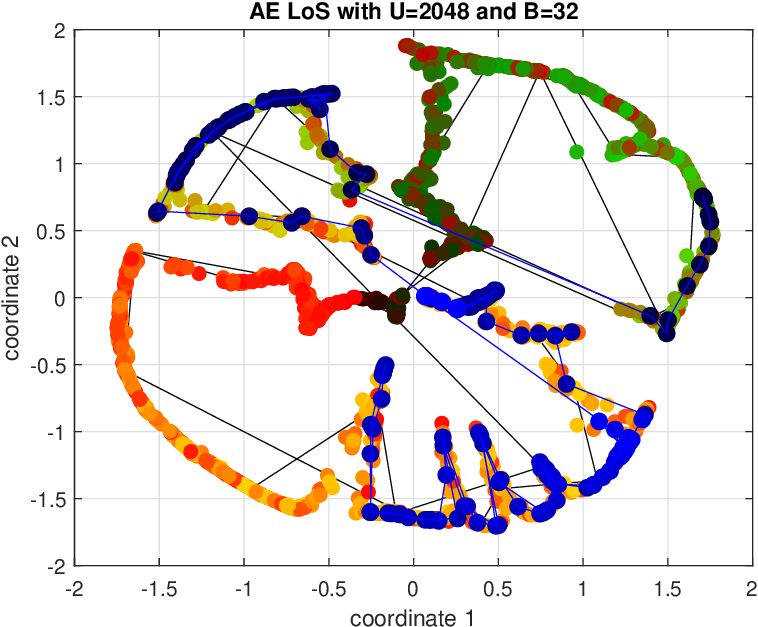} &
  \includegraphics[width=0.5in]{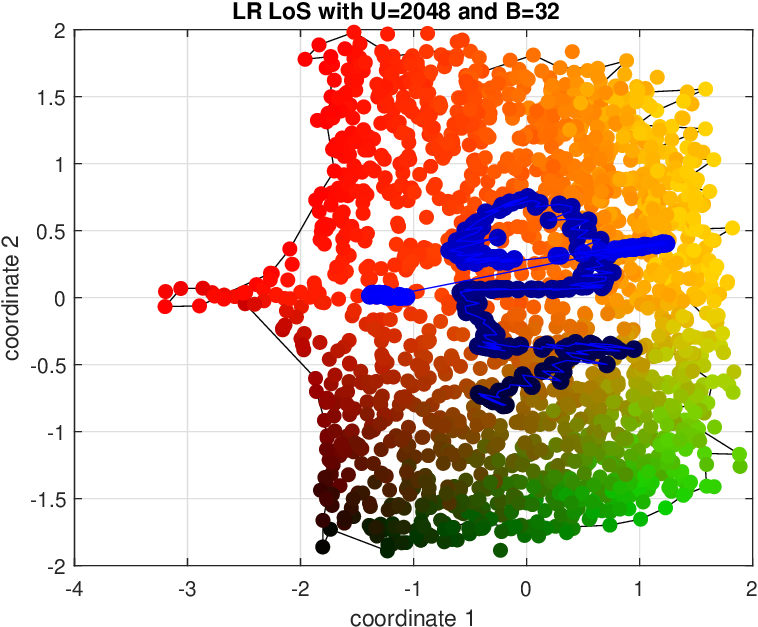} &
  \includegraphics[width=0.5in]{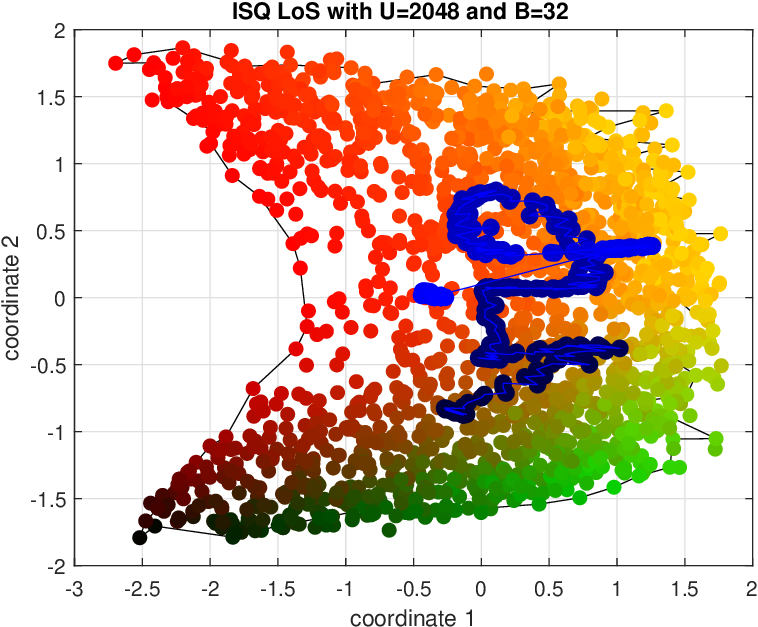} \\
  \addlinespace[2pt]
  \includegraphics[width=0.5in]{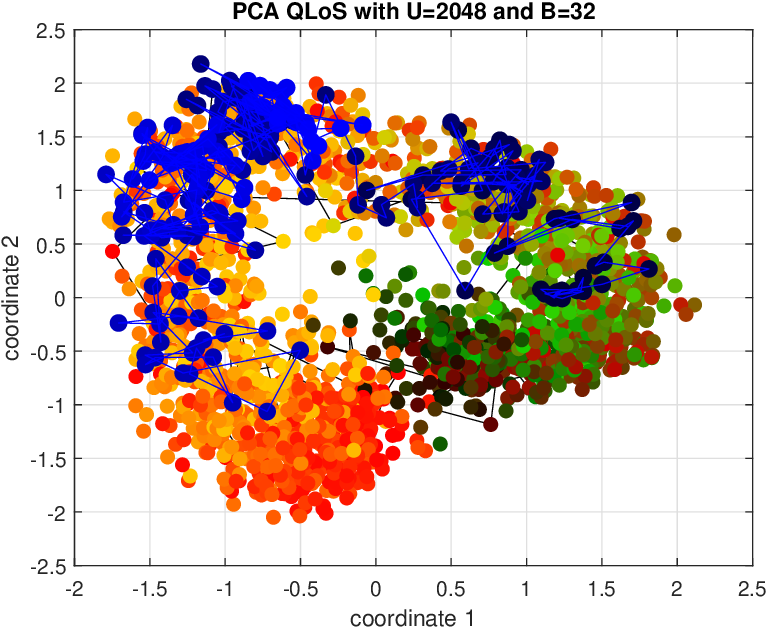} &
  \includegraphics[width=0.5in]{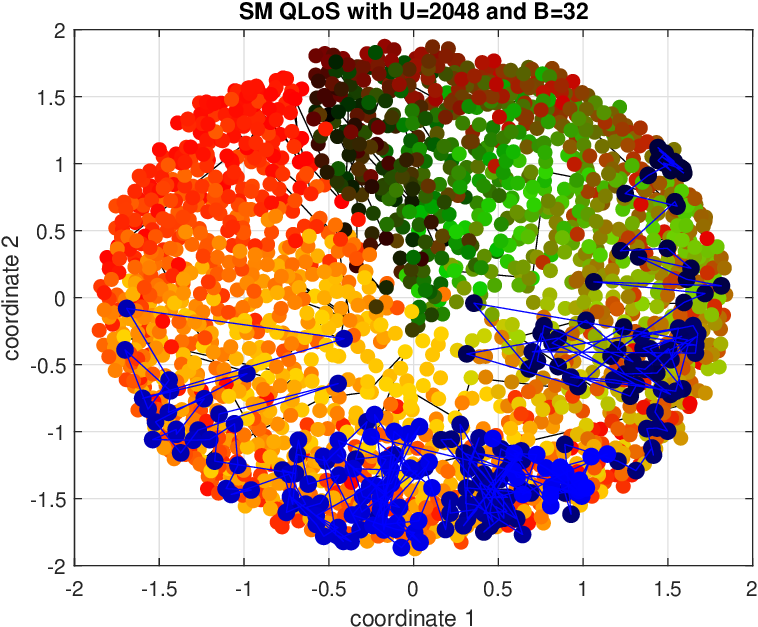} &
  \includegraphics[width=0.5in]{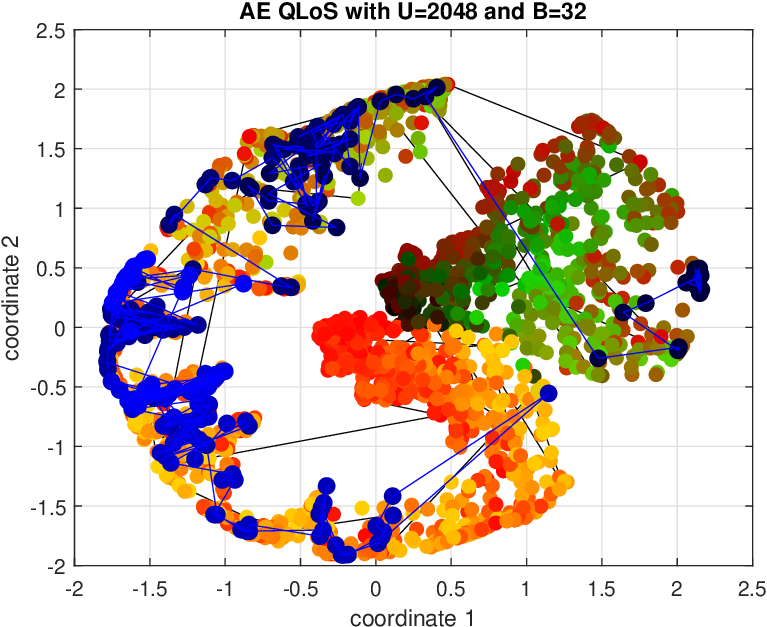} &
  \includegraphics[width=0.5in]{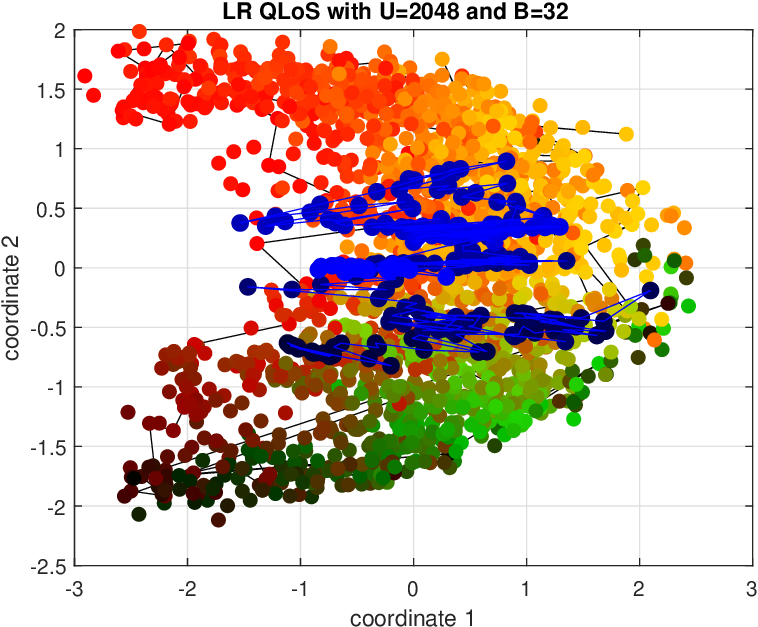} &
  \includegraphics[width=0.5in]{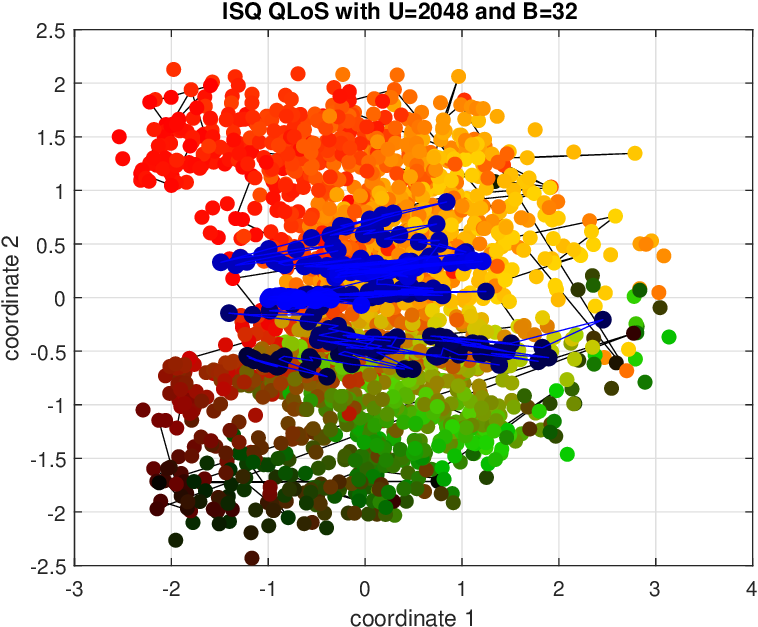} \\
  \addlinespace[2pt]
  \includegraphics[width=0.5in]{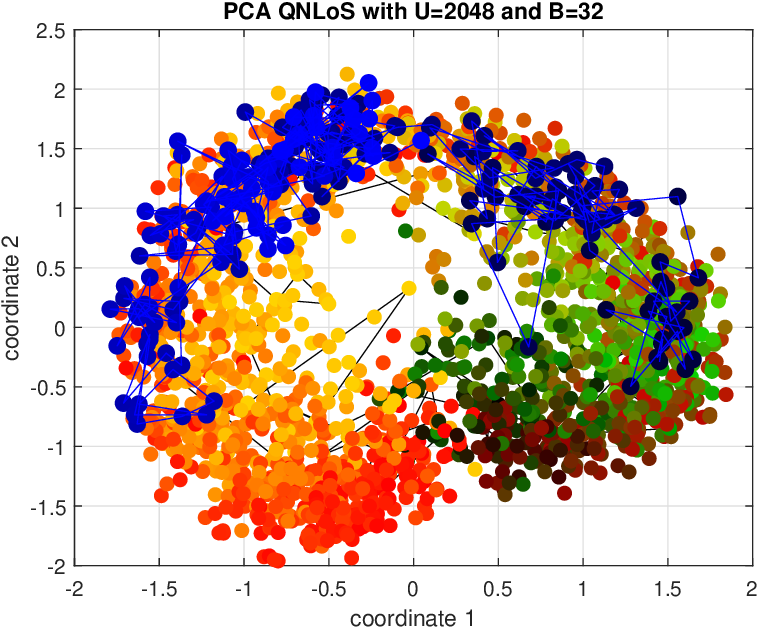} &
  \includegraphics[width=0.5in]{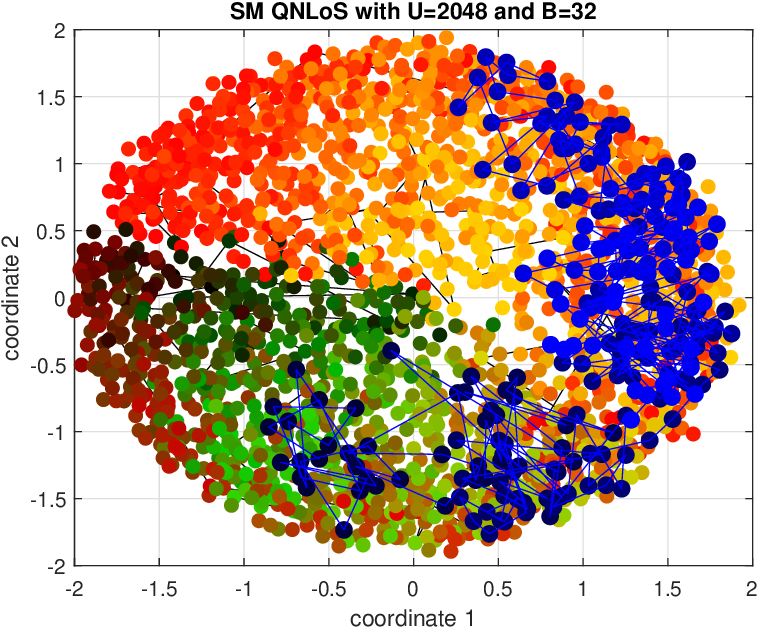} &
  \includegraphics[width=0.5in]{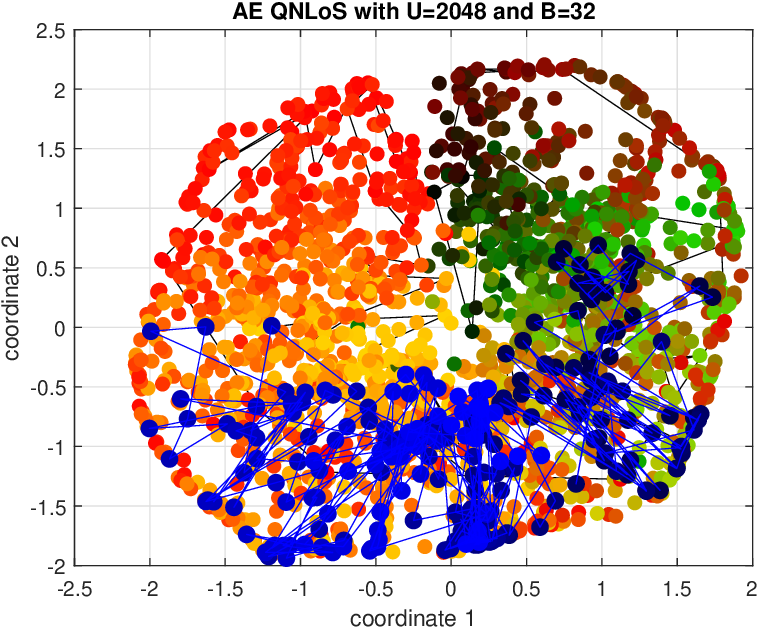} &
  \includegraphics[width=0.5in]{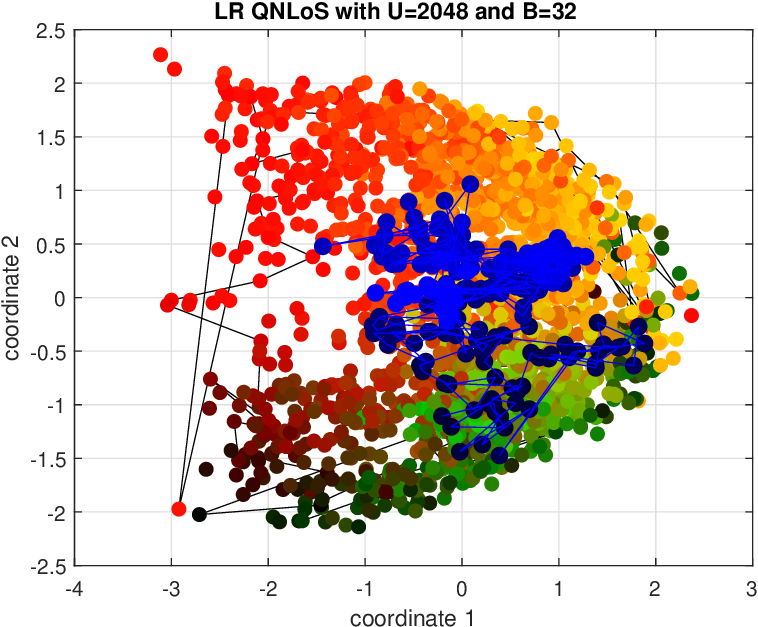} &
  \includegraphics[width=0.5in]{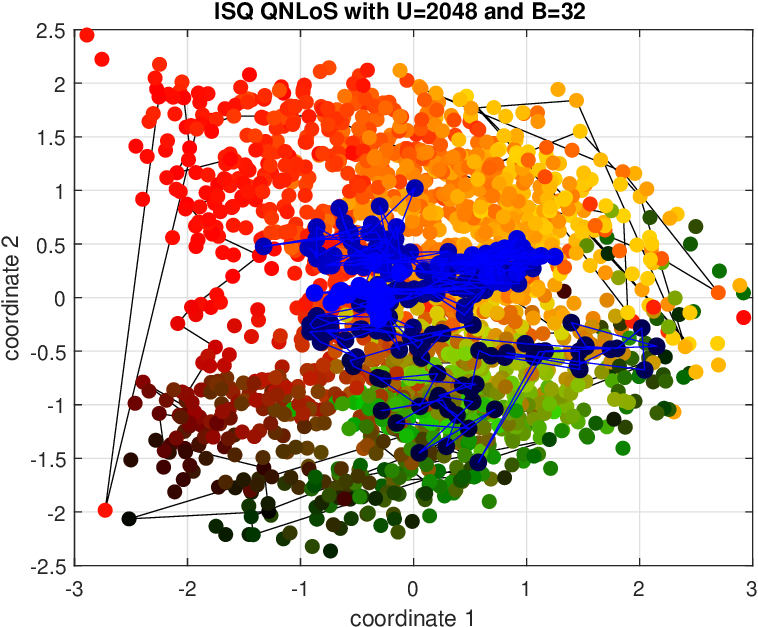} \\
  \end{tabular}
}
 \caption{Channel charts with PCA, SM, AE, LR, and ISQ algorithms for the 3D LOS, QLOS, and QNLOS channels.}
  \label{fig12}
\end{figure*}

\begin{figure*}[htbp]
\centering 
\begin{adjustbox}{width=1.0\textwidth}
  \renewcommand{\arraystretch}{0}%
  \begin{tabular}{c}
     \addlinespace[1pt]
  \includegraphics[width=0.5in]{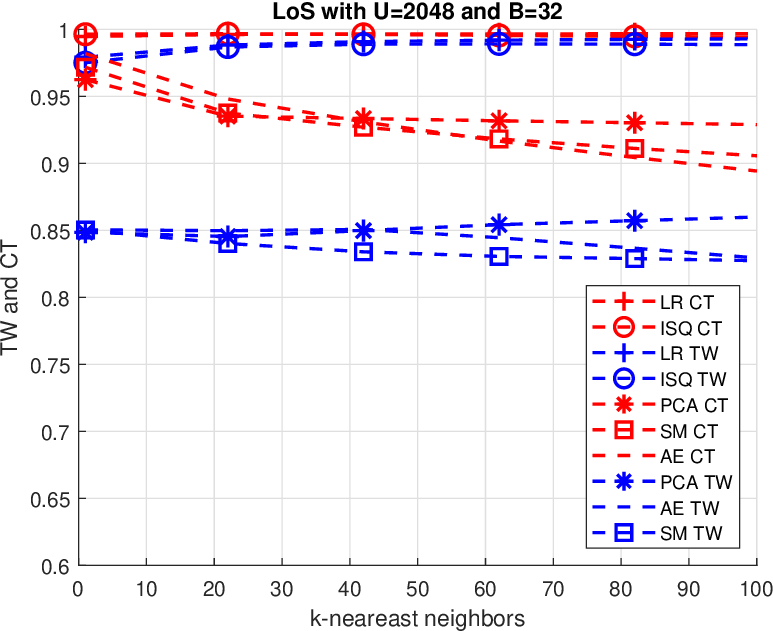}
  \includegraphics[width=0.5in]{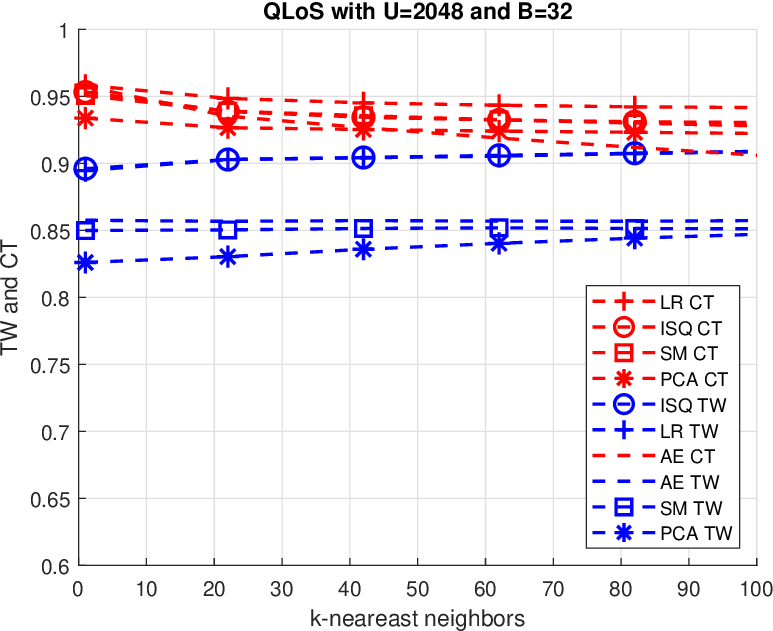}
   \includegraphics[width=0.5in]{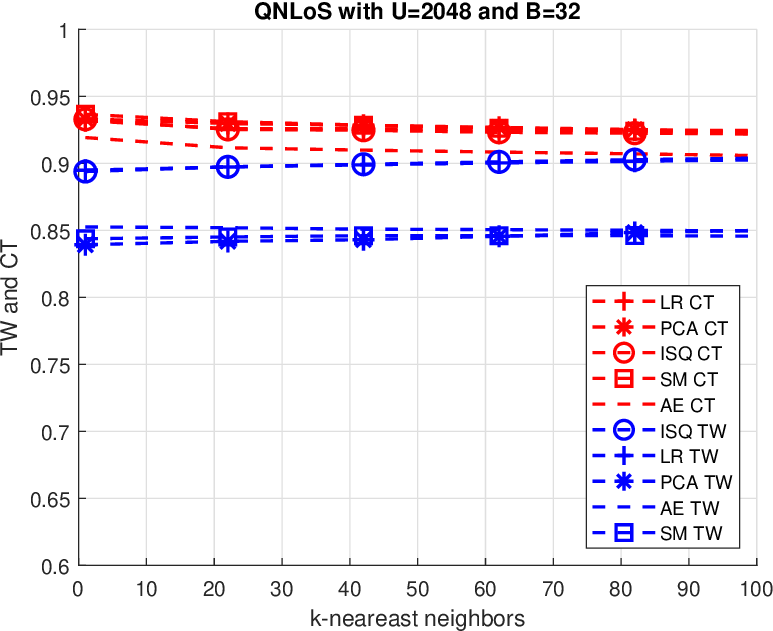} \\
  \end{tabular}
\end{adjustbox}
\caption{TW and CT performance against $k$-nearest neighbors for LR and ISQ algorithms in 3D. Left: LOS, middle: QLOS, right: QNLOS.}
 \label{fig9}
 \end{figure*}
\clearpage
\fi

Fig.~\ref{fig12} presents the channel charts for the 3D channel. In Fig.~\ref{fig12}, columns 1 through 5 correspond to PCA, SM, AE, LR, and ISQ algorithms, respectively. Therefore, one should compare the fourth and the fifth columns with the first three columns on a row-by-row basis. The considered system geometry is given in Fig.~\ref{fig31} or \cite[Fig. 1(a)]{b1}. The goal of the channel chart is to employ CSI and then derive a chart which preserves the distances in the system geometry. It can be seen from Fig.~\ref{fig12} that our algorithms LR and ISQ do a significantly better job than PCA, SM, and AE in that regard. In particular, the letters VIP present in Fig.~\ref{fig31} (\cite[Fig. 1(a)]{b1}) can be seen to be much more preserved with LR and ISQ. LR and ISQ channel charts are similar with a very slight preference towards LR. In close examination, we first look at the first row of Fig.~\ref{fig12}, corresponding to the LOS channel. Our algorithms LR and ISQ have visually very appealing performance, reproducing the word ``VIP'' in very close representation. Whereas, the other three algorithms do not have performance that is close. Among the other three algorithms, AE is a clear worst, PCA being better, and perhaps SM the best. Yet, their performance is nowhere close to those of ISQ or LR. The visual appearance of the channel charts for the LR and ISQ algorithms on the QLOS and QNLOS channels deteriorate but the word ``VIP'' is still somewhat visible. Interestingly, the PCA, SM, and AE algorithms appear to result in  better channel charts on the QLOS and QNLOS channels. But, as in the case with the numerical results in Table~\ref{tab42}, LR and ISQ yield clearly better channel charts than PCA, SM, and AE.

TW and CT performance curves for the 3D channel are given in Fig.~\ref{fig9} against $k$-nearest neighbors. The blue curves are for TW, while the red curves are for CT. ISQ results are given by dashed curves with a circular symbol, while LR curves are given by dashed curves with a plus symbol. It can be observed that while the performance of LR and ISQ are close, they consistently beat PCA, SM, and AE results. In particular, considering the LOS results in Fig.~\ref{fig9}, it can be observed that LR and ISQ results in terms of both TW and CT are very close to each other, while PCA, SM, and AE results are much worse. Furthermore, LR and ISQ results improve with $k$-nearest neighbors in terms of both TW and CT, while PCA, SM, and AE results get worse with increasing $k$-nearest neighbors in the case of CT, and remain about the same in the case of TW. Considering NLOS and QNLOS cases, in the case of TW, LR and ISQ perform significantly better than the other three algorithms. In the case of CT, LR and ISQ still have better performance than 
AE, and essentially the same as PCA and SM.
%
\subsection{MM Performance Comparison}
\ifCLASSOPTIONonecolumn
\else
\begin{table*}
\begin{center}
\caption{Performance comparison for TW and CT at $k$-nearest = 102 for MM algorithm in 3D channel at 2, 8, 14, 20, 26, and 32 subcarriers.}
\begin{tabular}{|c|l|c|c|c|c|c|c|}
\hline
Measure & Channel & 2sc & 8sc & 14sc & 20sc & 26sc & 32sc \\ \hline
   & LOS   & 0.9975 &   0.9986 &   0.9992 &   0.9997  &  0.9996  &  0.9998 \\ \cline{2-8}
TW & QLOS  & 0.9817 &   0.9958 &   0.9970 &   0.9973  &  0.9976  &  0.9976 \\ \cline{2-8}
   & QNLOS & 0.9622 &   0.9802 &   0.9818 &   0.9825  &  0.9837  &  0.9856 \\
\hline
   & LOS   & 0.9992 &   0.9999 &   1.0000 &   0.9999  &  0.9999  &  1.0000 \\ \cline{2-8}
CT & QLOS  & 0.9816 &   0.9972 &   0.9985 &   0.9990  &  0.9993  &  0.9992 \\ \cline{2-8}
   & QNLOS & 0.9626 &   0.9803 &   0.9821 &   0.9833  &  0.9845  &  0.9865 \\
\hline
\end{tabular}
\end{center}
\label{tab44}
\end{table*}

\begin{figure*}
\centering 
\resizebox{\textwidth}{!}{%
  \renewcommand{\arraystretch}{0}%
  \begin{tabular}{@{}c@{\hspace{1pt}}c@{\hspace{1pt}}c@{\hspace{1pt}}c@{}}
  \includegraphics[width=0.5in]{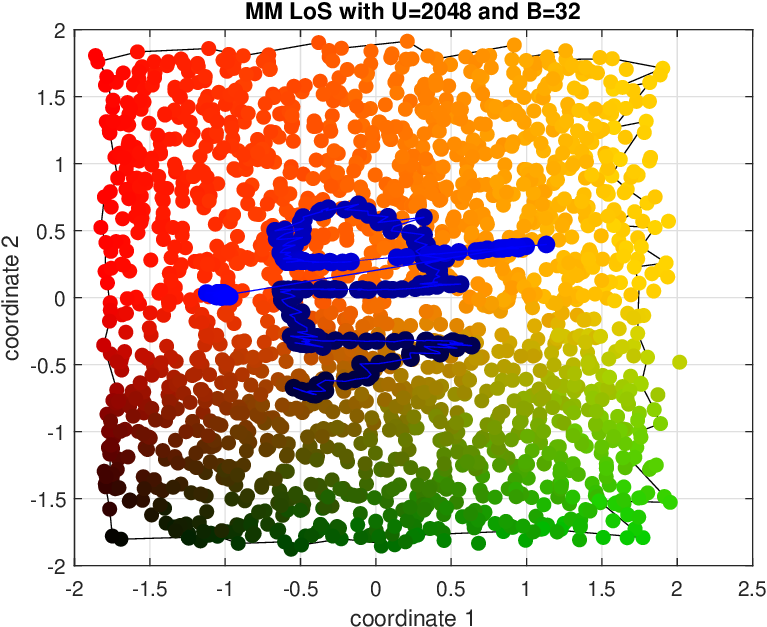} &
  \includegraphics[width=0.5in]{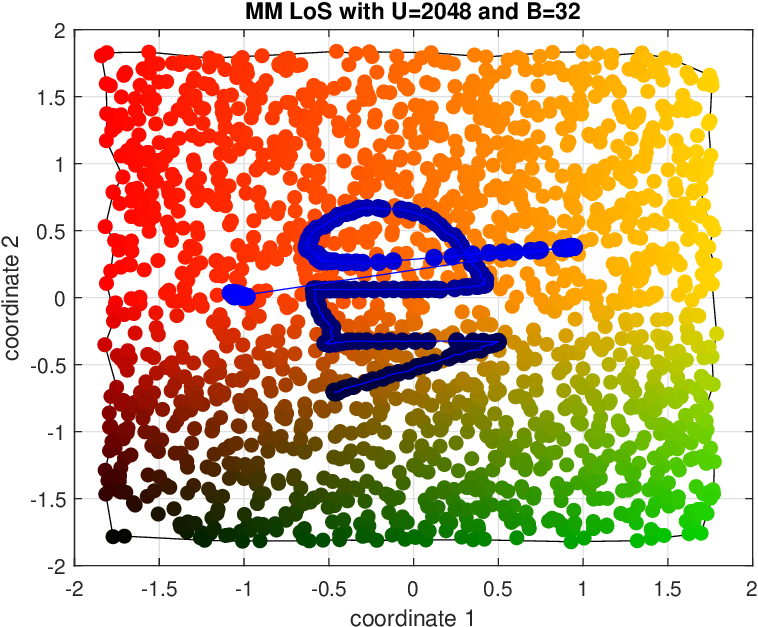} &
  \includegraphics[width=0.5in]{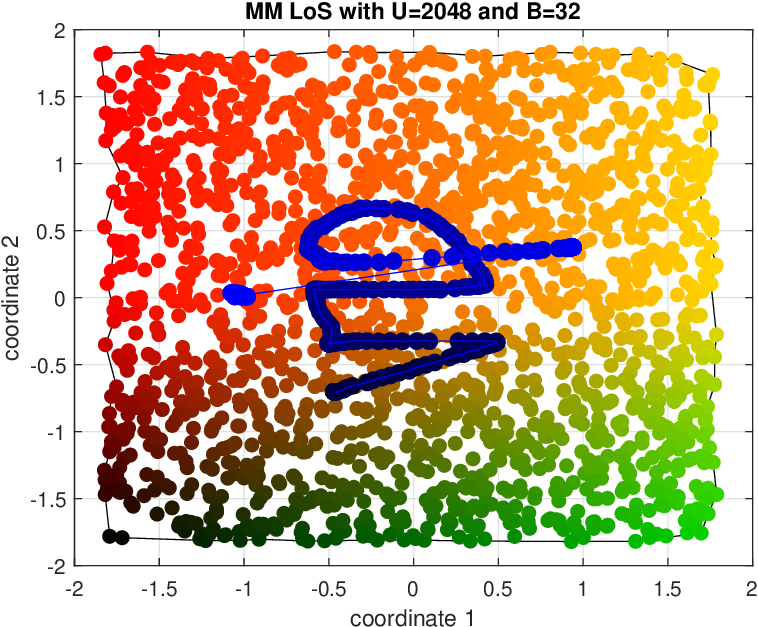} &
  \includegraphics[width=0.5in]{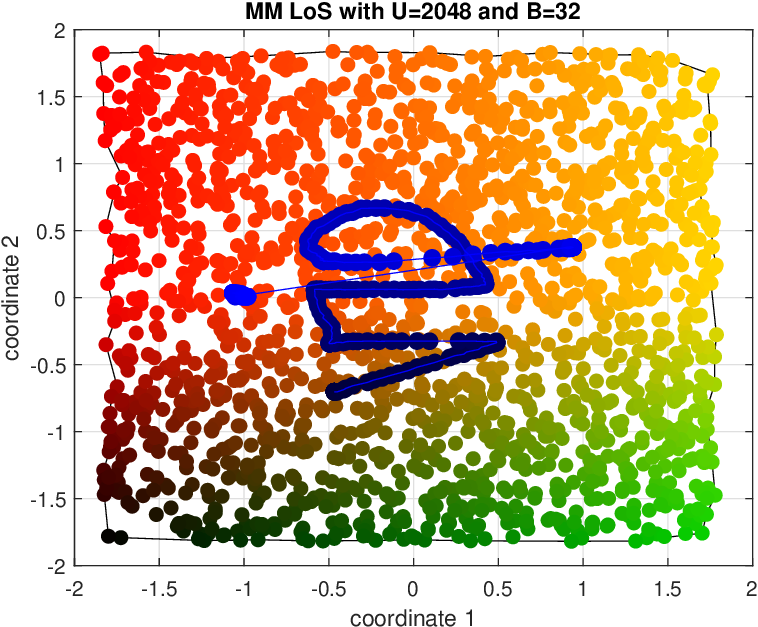} \\
  \addlinespace[2pt]
  \includegraphics[width=0.5in]{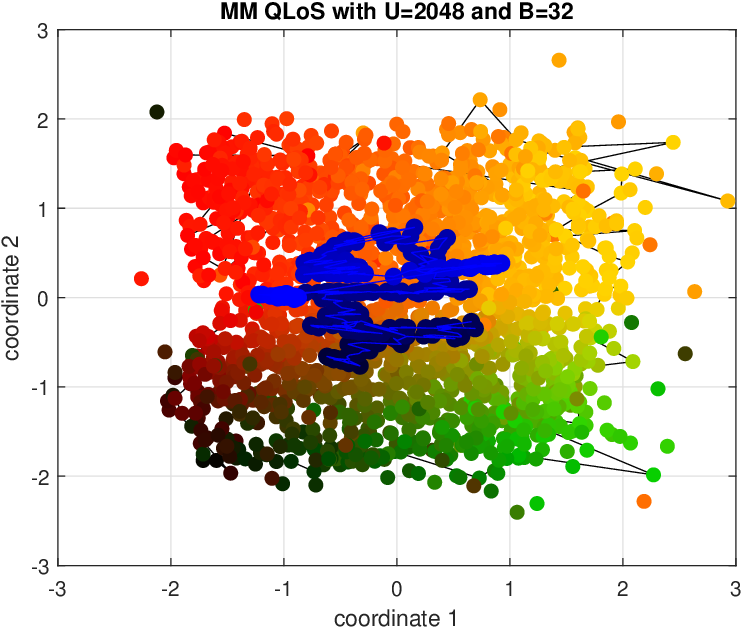} &
  \includegraphics[width=0.5in]{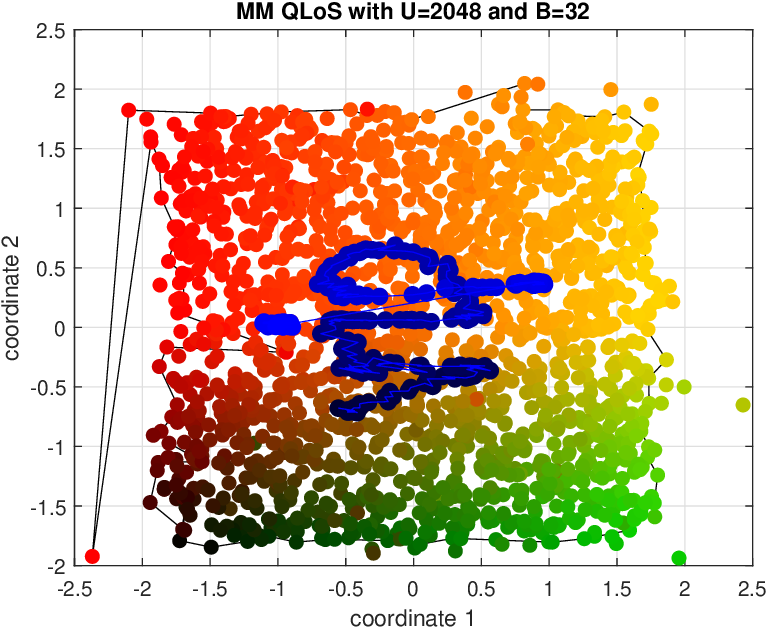} &
  \includegraphics[width=0.5in]{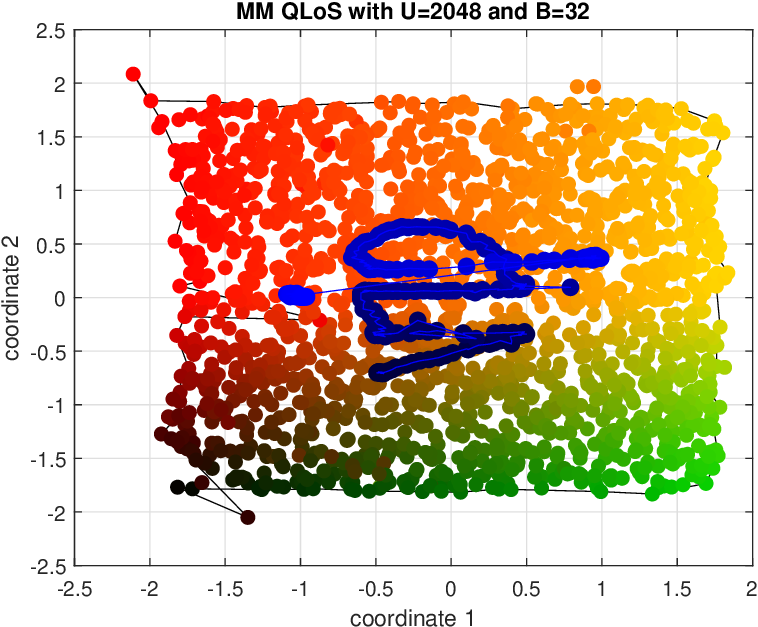} &
  \includegraphics[width=0.5in]{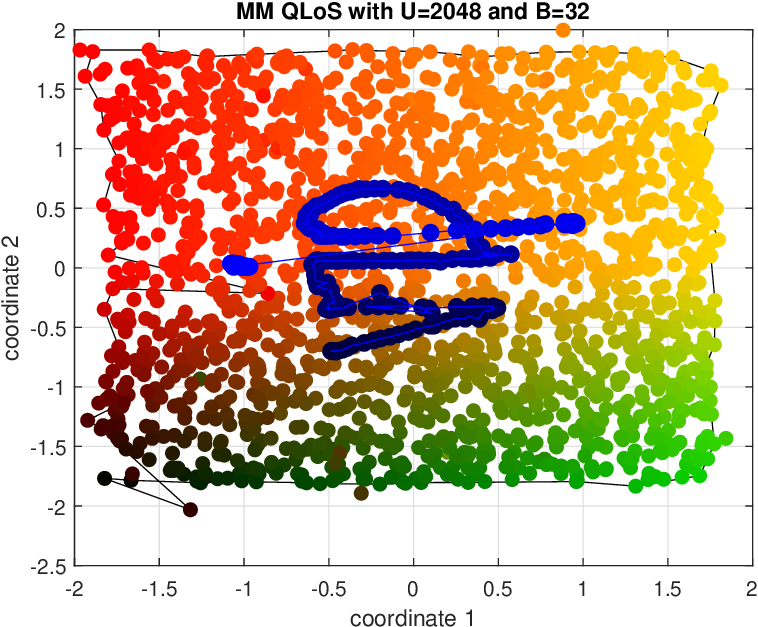} \\
  \addlinespace[2pt]
  \includegraphics[width=0.5in]{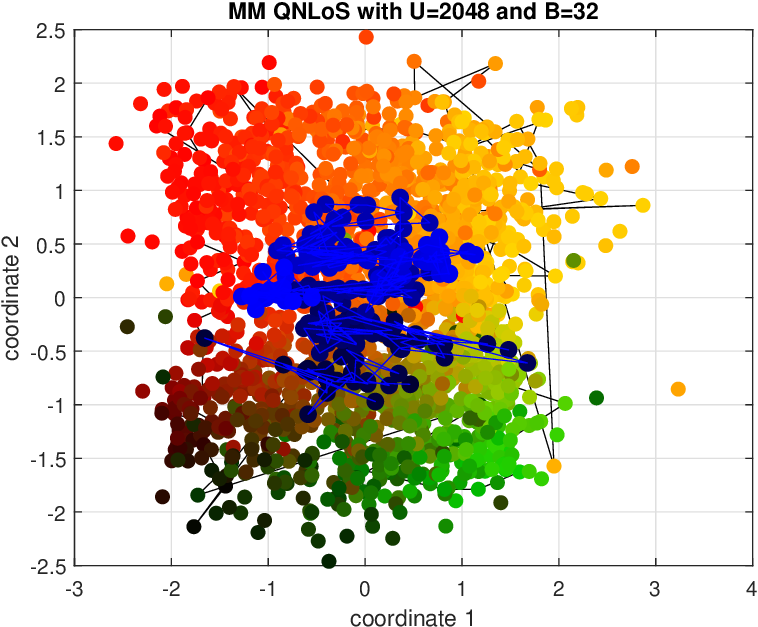} &
  \includegraphics[width=0.5in]{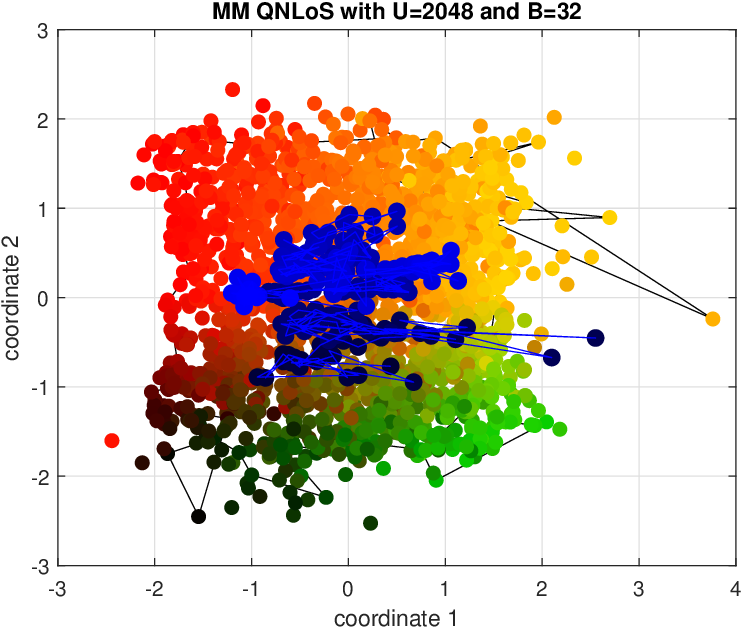} &
  \includegraphics[width=0.5in]{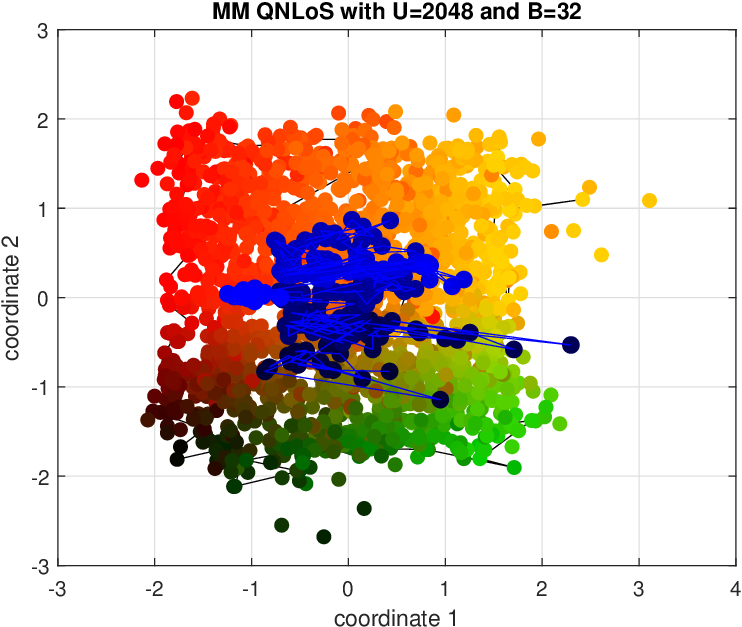} &
  \includegraphics[width=0.5in]{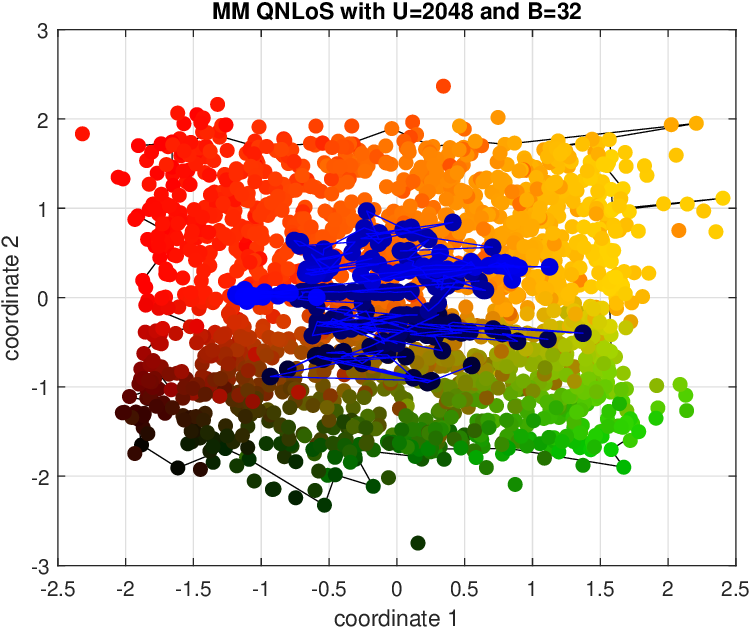} \\
  \end{tabular}
}
\caption{Channel charts with the MM algorithm for the 3D LOS, QLOS, and QNLOS channels at 2, 8, 20, and 32 subcarriers.}
  \label{fig46}
\end{figure*}

\begin{figure*}[htbp]
\centering 
\begin{adjustbox}{width=1.0\textwidth}
  \renewcommand{\arraystretch}{0}%
  \begin{tabular}{c}
     \addlinespace[1pt]
  \includegraphics[width=0.5in]{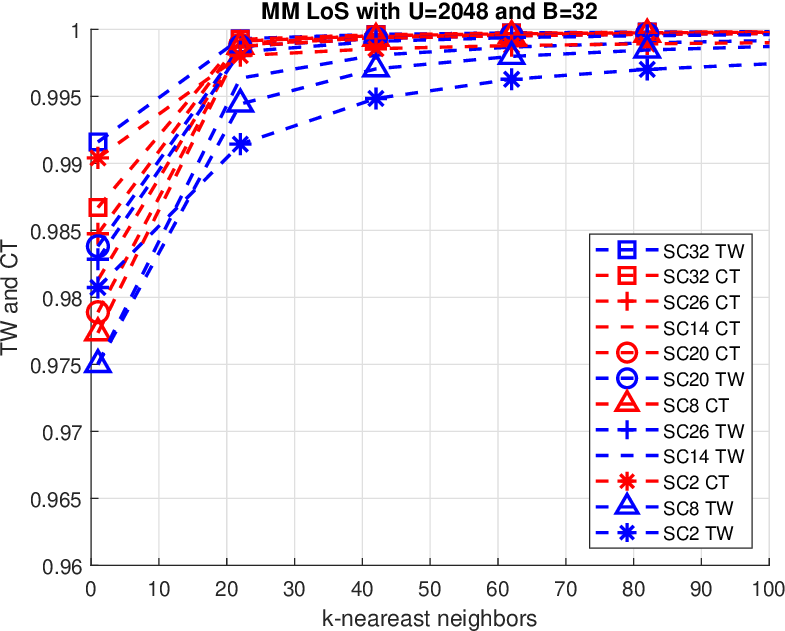}
  \includegraphics[width=0.5in]{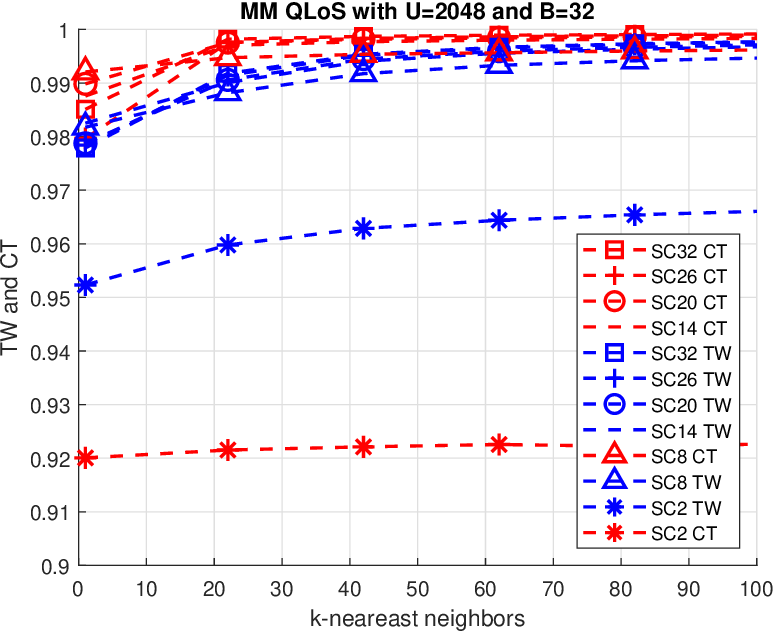}
  \includegraphics[width=0.5in]{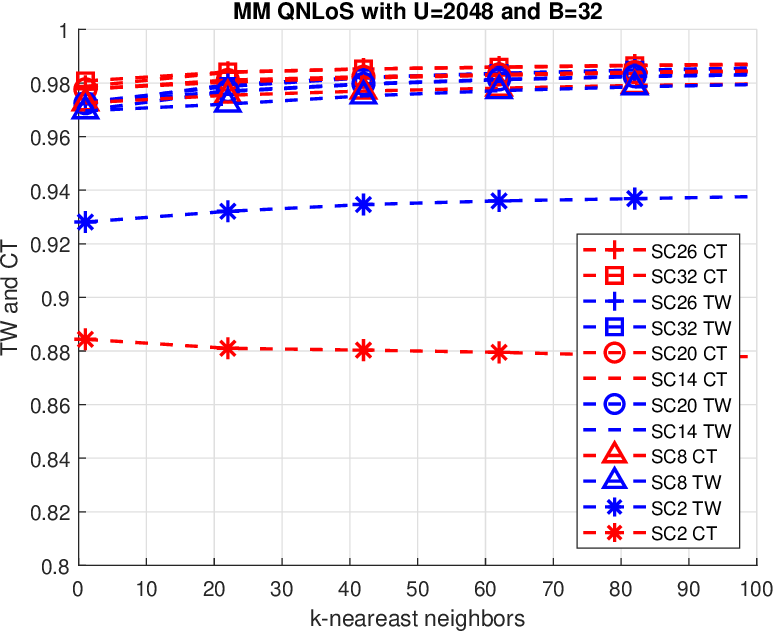} \\
  \end{tabular}
\end{adjustbox}
\caption{TW and CT performance against $k$-nearest neighbors for MM algorithm in 3D channel. Left: LOS, middle: QLOS, right: QNLOS.}
 \label{fig48}
 \end{figure*}
\fi

We will now discuss the performance of the MM algorithm. Table~\ref{tab44} presents the TW and CT results for $k$-nearest neighbors equal to 102 at 2, 8, 14, 20, 26, and 32 subcarriers. Then, Fig.~\ref{fig46} presents the channel charts at 2, 8, 20, and 32 subcarriers. Finally, Fig.~\ref{fig48}, presents TW and CT performance against $k$-nearest neighbors at 2, 8, 14, 20, 26, and 32 subcarriers. First, we note that, in most cases, increasing the number of subcarriers increases TW and CT performance as well as the visual quality of the channel charts, which is to be expected as more information is available with more subcarriers. One can also carry out a comparison for the same column in Table~\ref{tab44}, for all columns, to note that the performance of the MM algorithm with the same number of subcarriers is always the best for the LOS chanel, second best for the QLOS channel, and the worst for the QNLOS channel. Note that comparisons with algorithms PCA, SM, AE, LR, and AE are not present in Table~\ref{tab44}, Fig.~\ref{fig46}, and Fig.~\ref{fig48}. For that purpose, one needs to compare Table~\ref{tab44}, Fig.~\ref{fig46}, and Fig.~\ref{fig48} one-by-one with Table~\ref{tab42}, Fig.~\ref{fig12}, and Fig~\ref{fig9}, respectively.

Let us first consider TW and CT results for the MM algorithm in Table~\ref{tab44} at $k$-nearest = 102. We will compare these results against those in Table~\ref{tab42}. For the LOS channel, the TW results for the PCA, SM, and AE algorithms are in the range of 0.8272--0.8603, whereas for the LR and ISQ algorithms, they are in the range of 0.9885--0.9930. On the other hand, for the same channel, TW results with the MM algorithm are in the range of 0.9975--0.9998, a big difference. Looking at the TW performance for the QLOS and QNLOS channels, the performance of the PCA, SM, and AE algorithms are in the range of 0.8474--0.8574 while those of the LR and ISQ algorithms are in the 0.9029--0.9092 range. Yet, the TW results of the MM algorithm for these two channels are in the range of 0.9622--0.9976. 
We will now carry out the same analysis for the CT measure. In the LOS channel, the performance of PCA, SM, and AE are in the range of 0.8932--0.9288, whereas those of the LR and ISQ are in 0.9940--0.9968. On the other hand, for the same channel, the MM algorithm performance is in the range of 0.9992--1.000. For the QLOS and QNLOS channels, the CT performance of the PCA, SM, and AE algorithms are in the range 0.9055--0.9278, whereas those of the LR and ISQ algorithms are in 0.9220--0.9416. On the other hand, the performance of the MM algorithm is in the range 0.9626--0.9992, another big jump. As a result of all of this analysis, we can say that the MM algorithm significantly outperforms all of PCA, SM, AE, LR, and ISQ algorithms. The differences are very significant against those of the PCA, SM, and AE algorithms.

We will now study the channel charts created by the MM algorithm for the LOS, QLOS, and QNLOS channels at 2, 8, 20, and 32 subcarriers in Fig.~\ref{fig46}, comparing them with the channel charts by the PCA, SM, and AE algorithms on the same channels in Fig.~\ref{fig12}. For the LOS channel, clearly, the MM algorithm has the best visual outcome, preserving the word ``VIP'' very clearly. The quality of the channel chart generated by the MM algorithm is clearly much better than those by the PCA, SM, and AE algorithms, and better than those by LR and ISQ. It can be stated that this is the case even for the QLOS channel, the word ``VIP'' is being mostly visible for the MM algorithm, while that is not the case for PCA, SM, and AE, and not as clearly for LR and ISQ. Even for the QNLOS channel, the visual outcome by the MM algorithm is better than those by PCA, SM, and AE, albeit the clarity of the word ``VIP'' is no longer the case. It can be stated that, for this channel, the visual outcome by the MM algorithm is better than those by LR and ISQ, although the difference is not as clear as in the cases of the LOS and QLOS channels.

Finally, we will compare the TW and CT performance of the MM algorithm on LOS, QLOS, and QNLOS channels in Fig.~\ref{fig48} against those of the PCA, SM, AE, LR, and ISQ algorithms in Fig.~\ref{fig9}. The MM algorithm clearly outperforms the PCA, SM, and AE algorithms for all of LOS, QLOS, and QNLOS channels. It also outperforms the LR and ISQ algorithms on the QLOS and QNLOS channels. For the LOS channel, the performance of the LR and ISQ algorithms are close to that of the MM algorithm. In summary, comparison of Fig.~\ref{fig48} with Fig.~\ref{fig9} shows that the MM algorithm overperforms all of the PCA, SM, AE, LR, and ISQ algorithms, except for the LOS channel, where the MM, LR, and ISQ algorithms have similar TW and CT performance.
\subsection{JM Algorithm}
We will now discuss performance results of the JM algorithm introduced in Sec.~\ref{sec:JM}.
After some experimentation, we decided a subarray of dimensions $N_{Sa}= 4$ and $M_{Sa}=4$ worked best. We employed our thresholding and peak search algorithm. The complexity increase is very high.
We presented the results of the MM algorithm in Table~\ref{tab44}, Fig.~\ref{fig46}, and Fig~\ref{fig48}. We present similar results for the JM algorithm in Table~\ref{tab2022}, Fig.~\ref{fig446}, and Fig~\ref{fig448} except we do not present the case with 2 subcarriers because the smoothing subarray is $4 \times 4$ and the case of 2 subcarriers cannot run.

First, let us compare the TW and CT results at $k$-nearest = 102 for the MM algorithm in Table~\ref{tab44} and for the JM algorithm in Table~\ref{tab2022}. It can be observed that for the LOS and QLOS channels, the TW and CT performance of the MM algorithm and the JM algorithm are almost the same. For the QNLOS channel, there is a very slight advantage towards the JM algorithm, for both TW and CT. Because the MM and JM algorithm results are very close, we can say that the JM algorithm, like the MM algorithm, outperforms PCA, SM, AE, ISQ, and LR algorithms at $k$-nearest = 102 in terms of both TW and CT.

Second, we compare the channel charts for the MM algorithm in Fig.~\ref{fig46} and the JM algorithm in Fig.~\ref{fig446}. We deem that the visual outcome in both figures is very similar for the same channel and for the same number of subcarriers. And because of this similarity, we can say that the JM algorithm outperforms the PCA, SM, AE, ISQ, and LR algorithms in terms of channel charts.

Finally, we compare the TW and CT performance of the JM algorithm in Fig.~\ref{fig448} with that of the MM algorithm in Fig.~\ref{fig48} against $k$-nearest neighbors. For the LOS channel, it can be stated that the performance of both algorithms, in the left plots of Fig.~\ref{fig48} and Fig.~\ref{fig448}, are very close. On the other hand, a very slight advantage towards the JM algorithm can be observed in the case of QLOS and QNLOS channels in the middle and right subplots of Fig.~\ref{fig48} and Fig.~\ref{fig448}. Since the performance of the JM algorithm and the MM algorithm are very close, we can say that the JM algorithm outperforms the PCA, SM, AE, LR, and ISQ algorithms for TW and CT against $k$-nearest neighbors.

In summary, the performance of the MM algorithm and that of the JM algorithm are very close, with a very slight advantage towards the JM algorithm. But, in the face of the enormous complexity of the JM algorithm, we cannot say that it would be the preferred choice.
\subsection{RS Algorithm}
We have carried out extensive simulations on the performance of the RS algorithm as compared to the MM and JM algorithms. We will state that the performance of the three algorithms are very close, almost the same. In order not to be very repetitive, we only provide Table~\ref{tab5} for comparison with Table~\ref{tab44}. Clearly, the results are very close. Algorithms~\ref{alg:cap4} and \ref{alg:cap5} employ a large number of multiplications.
On the other hand, there may be a very important advantage of the RS algorithm against the MM and JM %
algorithms in terms of complexity. Both MM and JM algorithms employ the MUSIC algorithm which is based on an eigenvector and eigenvalue decomposition of an autocorrelation matrix. The computational complexity of this operation is very high. The RS algorithm avoids this operation.
For that reason, its implementation can be carried out in RTL whereas the implementation of MUSIC may require the use of an additional processor such as a digital signal processing (DSP) unit. DSP cores are available in embedded systems for wireless applications and there are implementations of MUSIC on them \cite{LC15}. Thus, although the operation of RS requires many multiplications, its implementation may be more straightforward than that of MUSIC.
%
\subsection{Runtime and Complexity Comparison}
When we compare the complexity of our algorithms ISQ and MM against the three algorithms used in \cite{b1}, i.e., PCA, SM, and AE, the most important advantage is that our algorithms do not require training or an abundant number of CSI to be able to reduce dimensionality efficiently. We can calculate the channel chart even for one UE data. This can make us calculate the channel chart sequentially in real time as the data are received. The alternative in \cite{b1} is to store the data of all UEs (2048 in our simulations as well as in \cite{b1}) and use it all at once as in the case of PCA, SM, or AE, which consumes a very large amount of memory and complexity. Please note that our algorithm LR requires a modest amount of training.

The other advantage is the latency. PCA, SM, and AE algorithms need to collect the data of all UEs, which can take some time. Furthermore, if the system is mobile, the geometry might have already changed by the time the channel chart is calculated. In our case, we can calculate each UE channel chart as we receive it, which makes our algorithms much more efficient.

As an indication of the complexity, we will compare the runtime (simulation time) for producing channel chart for 2048 UEs using different algorithms. This is provided in Table~\ref{tab222}.
\ifCLASSOPTIONonecolumn
\else
\clearpage
\begin{table*}[!ht]
\begin{center}
\caption{Performance comparison for TW and CT at $k$-nearest = 102 for JM algorithm in 3D channel at 8, 14, 20, 26, and 32 subcarriers.}
\begin{tabular}{|c|l|c|c|c|c|c|}
\hline
Measure & Channel & 8sc & 14sc & 20sc & 26sc & 32sc \\ \hline
   & LOS   &   0.9981 &   0.9984  &  0.9983  &  0.9982  &  0.9982 \\ \cline{2-7}
TW & QLOS  &   0.9936 &   0.9967  &  0.9966  &  0.9963  &  0.9958 \\ \cline{2-7}
   & QNLOS &   0.9798 &   0.9899  &  0.9888  &  0.9870  &  0.9894 \\
\hline
   & LOS   &   1.0000 &   1.0000  &  1.0000  &  1.0000  &  1.0000 \\ \cline{2-7}
CT & QLOS  &   0.9951 &   0.9985  &  0.9984  &  0.9981  &  0.9977 \\ \cline{2-7}
   & QNLOS &   0.9821 &   0.9916  &  0.9904  &  0.9886  &  0.9910 \\
\hline
\end{tabular}
\label{tab2022}
\end{center}
\end{table*}
\begin{figure*}
\centering 
\resizebox{0.8\textwidth}{!}{%
  \renewcommand{\arraystretch}{0}%
  \begin{tabular}{@{}c@{\hspace{1pt}}c@{\hspace{1pt}}c@{\hspace{1pt}}c@{}}
  \includegraphics[height=0.4in]{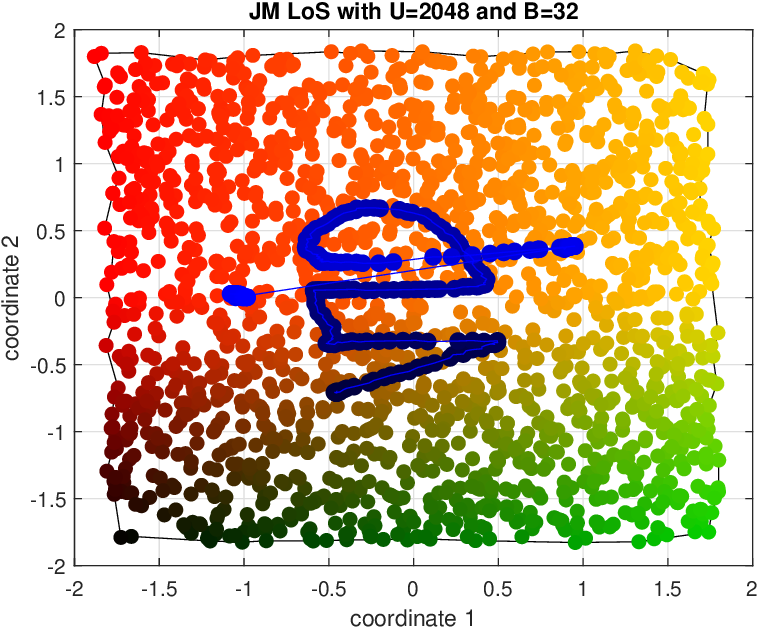} &
  \includegraphics[height=0.4in]{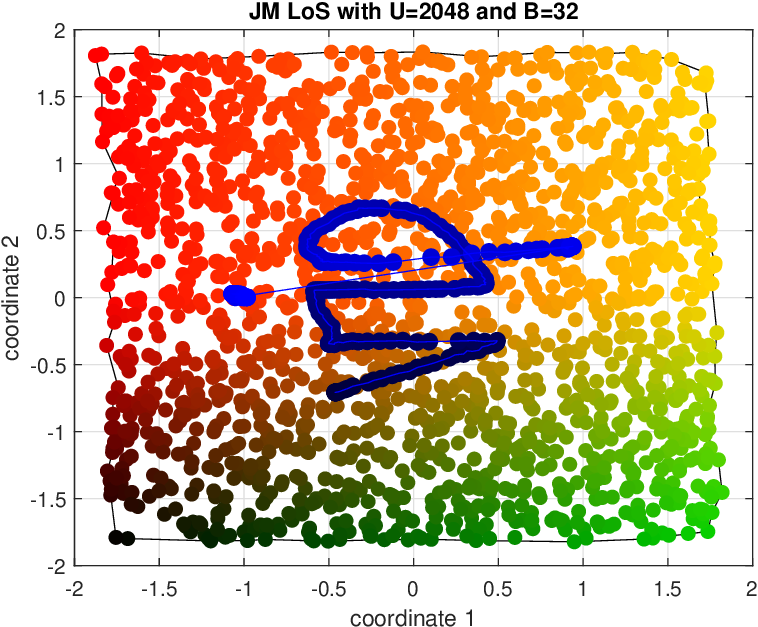} &
  \includegraphics[height=0.4in]{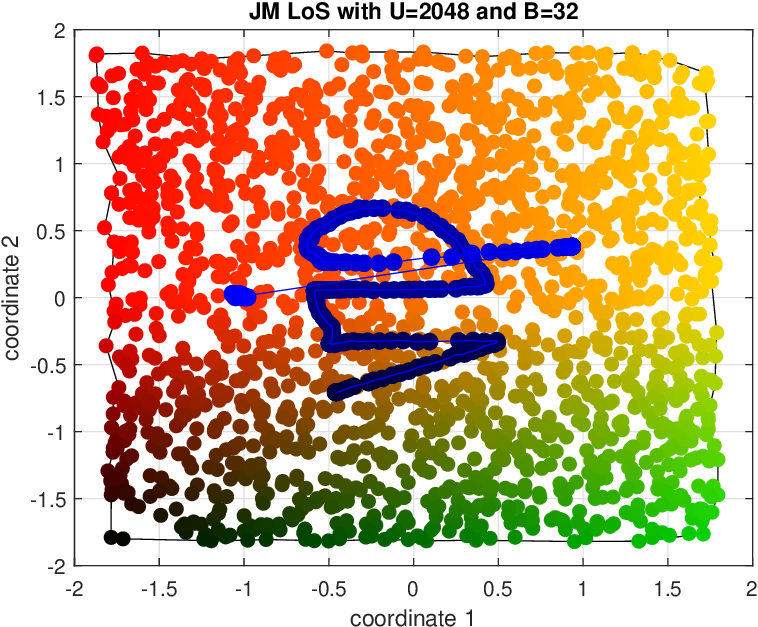} \\
  \addlinespace[2pt]
  \includegraphics[height=0.4in]{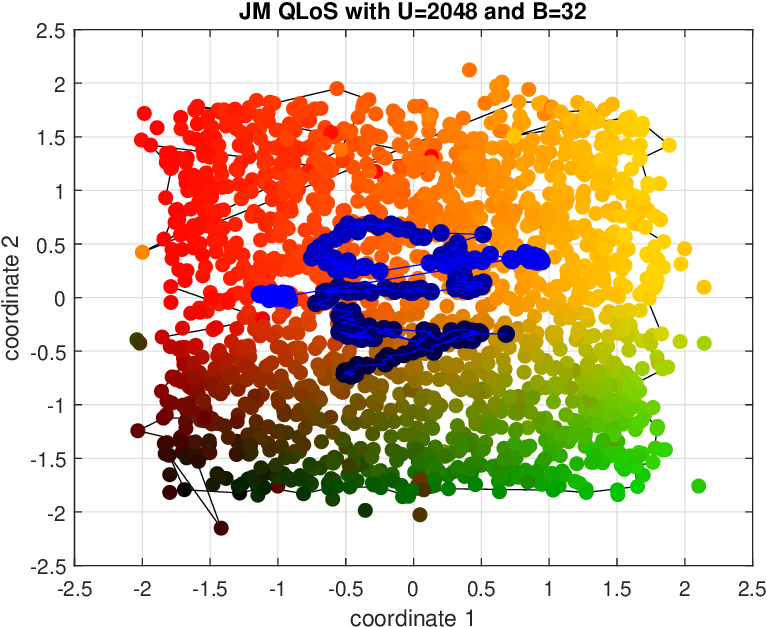} &
  \includegraphics[height=0.4in]{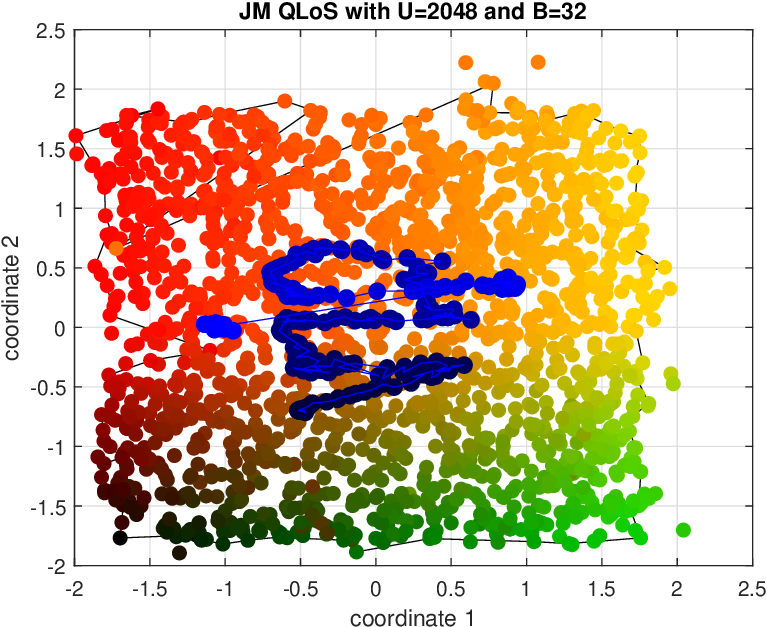} &
  \includegraphics[height=0.4in]{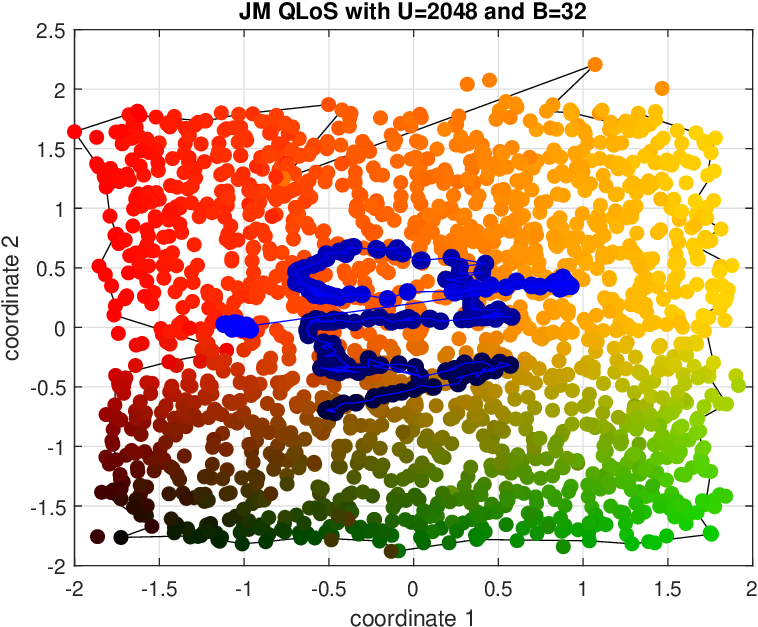}\\
  \addlinespace[2pt]
  \includegraphics[height=0.4in]{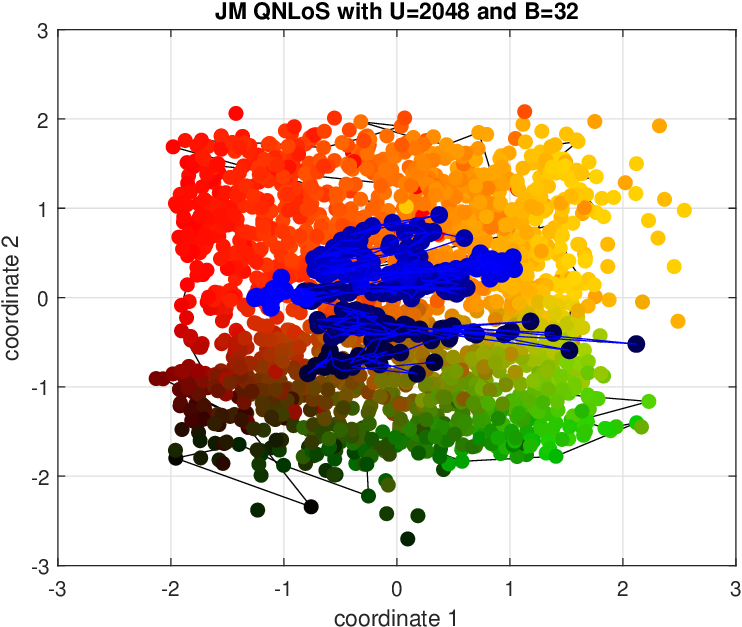} &
  \includegraphics[height=0.4in]{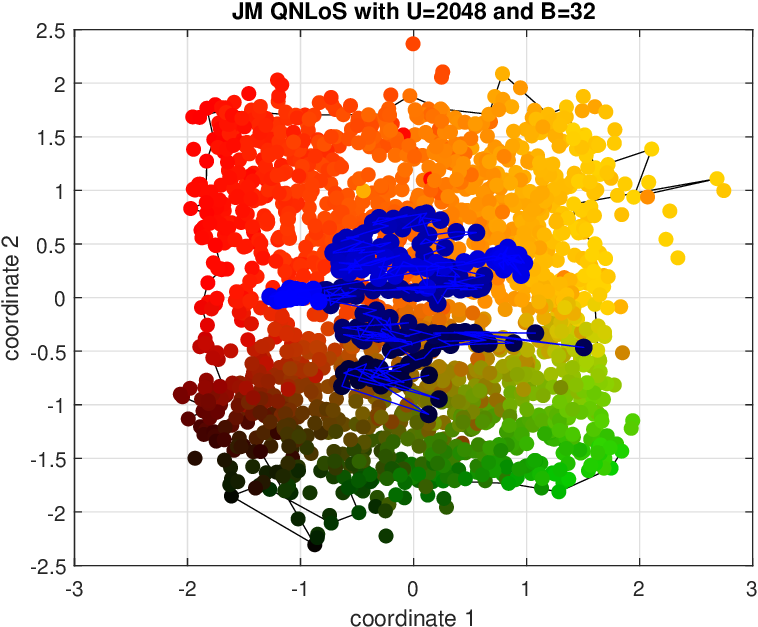} &
  \includegraphics[height=0.4in]{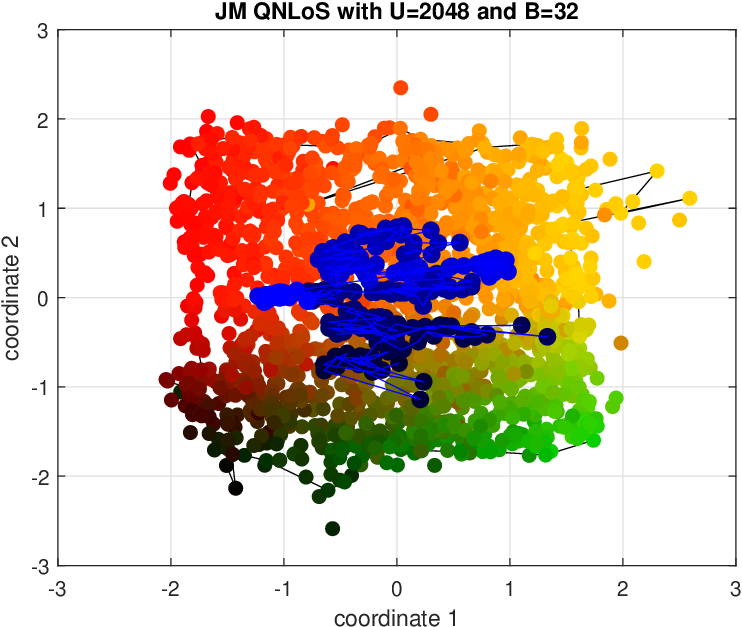}\\
\end{tabular}
}
\caption{Channel charts with the JM algorithm for the 3D LOS, QLOS, and QNLOS channels at 8, 20, and 32 subcarriers.}
  \label{fig446}
\end{figure*}

\begin{figure*}
\centering 
\begin{adjustbox}{width=1.\textwidth}
  \renewcommand{\arraystretch}{0}%
  \begin{tabular}{c}
     \addlinespace[1pt]
  \includegraphics[width=0.5in]{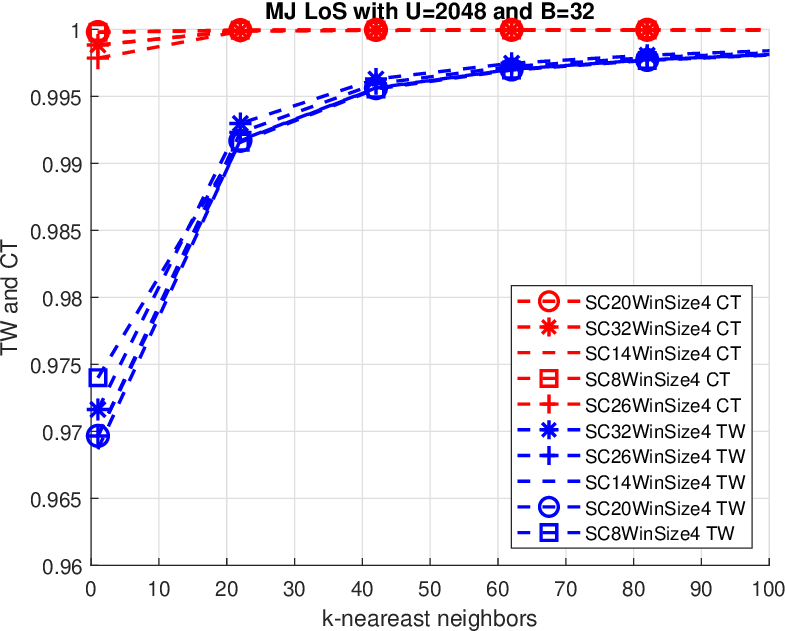}
  \includegraphics[width=0.5in]{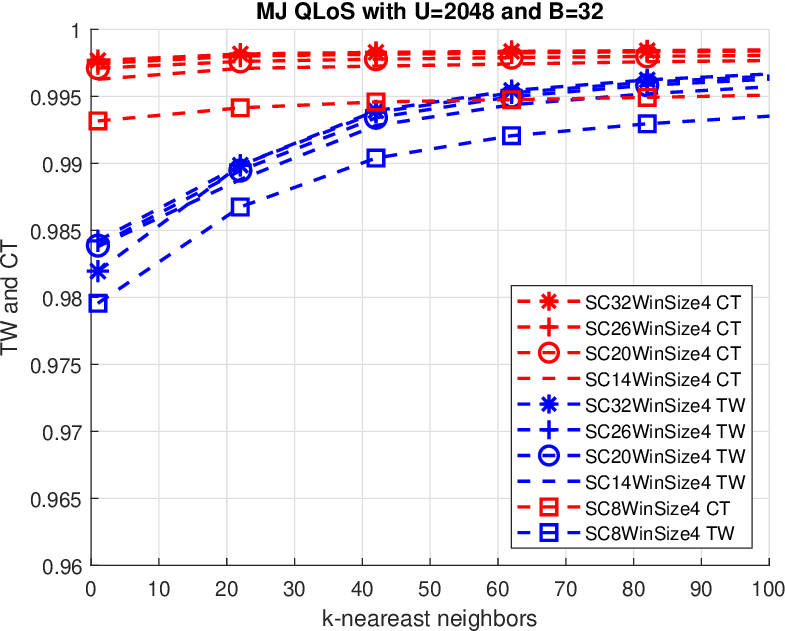}
  \includegraphics[width=0.5in]{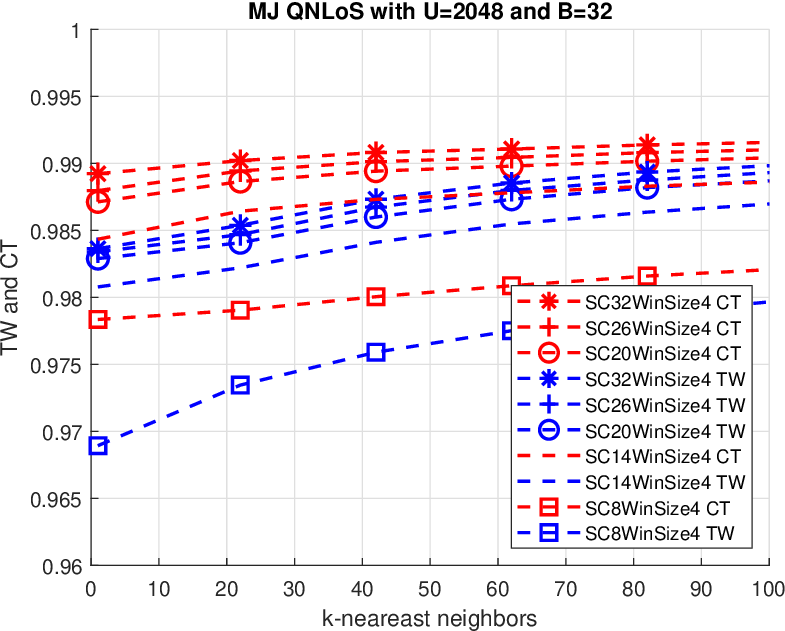} \\
  \end{tabular}
\end{adjustbox}
\caption{TW and CT performance against $k$-nearest neighbors for JM algorithm in 3D channel. Left: LOS, middle: QLOS, right: QNLOS.}
 \label{fig448}
 \end{figure*}
 \clearpage\newpage
\fi
\ifCLASSOPTIONonecolumn
\else
\begin{table*}[ht!]
\begin{center}
\caption{Performance comparison for TW and CT at $k$-nearest = 102 for RS algorithm in 3D channel at 2, 8, 14, 20, 26, and 32 subcarriers.}
\begin{tabular}{|c|l|c|c|c|c|c|c|}
\hline
Measure & Channel & 2sc & 8sc & 14sc & 20sc & 26sc & 32sc \\ \hline
   & LOS   &   0.9972  &  0.9986  &  0.9991  &  0.9996  &  0.9997  &  0.9998 \\ \cline{2-8}
TW & QLOS  &   0.9648  &  0.9903  &  0.9933  &  0.9943  &  0.9946  &  0.9950 \\ \cline{2-8}
   & QNLOS &   0.9421  &  0.9664  &  0.9753  &  0.9779  &  0.9804  &  0.9821 \\ \hline
   & LOS   &   0.9982  &  0.9997  &  0.9998  &  0.9998  &  0.9998  &  0.9998 \\ \cline{2-8}
CT & QLOS  &   0.9572  &  0.9920  &  0.9951  &  0.9959  &  0.9965  &  0.9968 \\ \cline{2-8}
   & QNLOS &   0.9337  &  0.9680  &  0.9784  &  0.9813  &  0.9837  &  0.9854 \\ \hline
\end{tabular}
\label{tab5}
\end{center}
\end{table*}
\fi
\ifCLASSOPTIONonecolumn
\else
\begin{table}[!ht]
\begin{center}
\caption{Simulation times.}
\begin{tabular}{||c | c||}
 \hline
 Algorithm & Simulation time (seconds) \\ 
 \hline\hline
 PCA & 0.817 \\
 \hline
 SM & 12.2 \\
  \hline
 AE & 53.9  \\
 \hline
 LR & 7.15  \\
 \hline
 ISQ & 7.09 \\
 \hline
MM & 20.4 \\
 \hline
JM & 442.0 \\
\hline
\end{tabular}
\end{center}
\label{tab222}
\end{table}
\fi
\ifCLASSOPTIONonecolumn
\else
\begin{table}[!ht]
\begin{center}
\caption{Number of multiplications employed by the algorithms in the paper. $N:$ Number of vectors of feature size $D$, to be reduced to size $d$. $N_R:$ Number of antennas at the BS, $N_S:$ Number of subcarriers. $P$: Number of potential AOAs to determine $\theta$, $Q:$ Number of potential distances to determine $\rho$.}
\begin{tabular}{||c|c||}
\hline
Algorithm & Number of Multiplications \\
\hline\hline
PCA & $O(ND^2+D^3)$\\
\hline
SM & $\gg \textrm{PCA}$\\
\hline
AE & $\gg \textrm{LR, ISQ}$\\
\hline
LR & $O(N_R^2 P)$\\
\hline
ISQ & $O(N_R^2 P)$\\
\hline
MM & $O(N_R^2 P + N_R^2 N_S + N_S^2 Q+N_S^2 N_R)$\\
\hline
RS & $O(P (2N_R^2+1)+N_SN_R^2+Q (2N_S^2+1) +N_RN_S^2)$\\
\hline
\end{tabular}
\end{center}
\label{tbl:Complexity}
\end{table}
\fi
The simulation time is not dependent on the channel, the scale,  or the geometry of the environment. The simulation time only relies on the number of UEs, the number of antennas, and the number of subcarriers. In Table~\ref{tab222}, the number of UEs is 2048, the number of BS antennas is 32, and the number of subcarriers is 32.

We note that PCA has very small simulation time, much lower than all of the other five algorithms. However, we know from earlier sections that LR, ISQ, and especially MM beats it in terms of performance. SM and AE not only are beaten by LR and ISQ in terms of performance, but also, in terms of simulation time. MM has the best performance by far but its simulation time is about 2.5 times those of LR and ISQ and about 1.5 times that of SM. It has less simulation time than AE. Although its simulation time is at a disadvantage as compared to PCA, SM, LR, and ISQ, the performance gains with it are substantial. We note that the simulation time for PCA is very small but this is due to the fact that PCA is an optimized routine in Matlab. Assume we have $N$ vectors of feature size $D$, to be reduced to a size $d$. The computationally intensive operations of PCA consist of two major steps: {\em i)\/} computing the covariance matrix, and {\em ii)\/} computing the eigenvalue decomposition of the covariance matrix. The first step involves the multiplication of two matrices of size $D\times N$ and $N\times D$, resulting in $O(ND^2)$. The second step involves the eigenvalue decomposition of a $D\times D$ matrix, contributing $O(D^3)$, with a total complexity of $O(ND^2 + D^3)$, see e.g., \cite{Banerjee20}.

Although it has advantages over PCA, SM is stated to be substantially more complex than PCA \cite{b1}. We note that SM is nonlinear and needs multiple runs until convergence, whose number is not known prior to running the algorithm \cite{LGAD00}. The high complexity of the SM method is further emphasized in the literature, e.g., \cite{BJD81,Yin02,Yin03}.

The complexity of the MUSIC algorithm is given as $O(M^2 P + M^2 N)$ where $M$ is the number of antennas, $N$ is the number of snapshots or multiple measurement vectors, and $P$ is the number of potential AOAs searched \cite{SMMN15}. In the case of ISQ, we have $M=N_R$, $N=1$, and $P$ is the number of potential AOAs for the MUSIC algorithm to determine $\theta$, with $P\gg 1$. As a result, we have $O(N_R^2 P)$ to determine $\theta$, where we ignore a second term $N_R^2 \cdot 1$ since $P\gg 1$. With ISQ, to determine $\rho$, we need $N_R$ instantiations of two squares, a sum, and a square root. These need to be summed, a square root and a reciprocal operation needs to take place. As a result, there are $2N_R$ multiplications, $N_R+1$ square roots, $2N_R-1$ adds, and a reciprocal needed to determine $\rho$. We note that it is stated that a square root operation can be performed with the same complexity as a multiplication \cite{Alt79}. In any case, the complexity of the ISQ algorithm will be dominated by the MUSIC algorithm for $\theta$, which has $O(N_R^2 P)$, as discussed above. The LR algorithm employs MUSIC to determine $\theta$, thus it has the same complexity as ISQ for $\theta$. After LR parameters $a$ and $b$ are determined, it needs $2N_R+1$ multiplications, $N_R$ square roots, $N_R$ adds, and a $\log$ operation, similar to ISQ. In addition, there is the computational complexity of calculating the linear regression, based on real $\rho$ values of a number of UEs (in our case 256), shared by all UEs (in our case, 2048). As a result, LR is more complicated than ISQ, but with only a small gain over it. We consider its complexity to be dominated by the MUSIC algorithm for estimating $\theta$, or $O(N_R^2 P)$, as in the ISQ algorithm. For the MM algorithm, we use MUSIC to estimate both $\theta$ and $\rho$. For $\theta$, the number of snapshots is equal to $N_S$, unlike 1 in the case of ISQ and LR.
Therefore, the complexity of MUSIC for $\theta$ in the MM algorithm is $O(N_R^2 P + N_R^2 N_S)$. For $\rho$, the roles of $N_R$ and $N_S$ are reversed and $P$ is replaced by $Q$ where $Q$ is the number of potential distances for the MUSIC algorithm to determine $\rho$. Thus the complexity of the MUSIC algorithm for $\rho$ is $O(N_S^2 Q+N_S^2 N_R)$ .

As a result, we can state that the ISQ and LR algorithms are dominated by the MUSIC algorithm for $\theta$, at $O(N_R^2 P)$. On the other hand, MM employs two MUSIC algorithms, for $\theta$ and $\rho$, and has complexity $O(N_R^2 P + N_R^2 N_S + N_S^2 Q + N_S^2 N_R)$.

We will now compare two training-based algorithms AE and LR in terms of complexity. For LR, estimating AOA is carried out by using the MUSIC algorithm for $\theta$. For $\rho$, linear regression is performed. Let the training set be $({\bf x}_1, y_1),$ $({\bf x}_2, y_2), \ldots,$ $({\bf x}_N, y_N),$ where ${\bf x}_i\in {\mathbb R}^d$, $y_i\in {\mathbb R}$, $i=1,2,\ldots , N$. Note that, in our case, $d=2$, $N=256$. Then, the training computational complexity is given as $O(d^2 N + d^3)$, see e.g., \cite{ChoHsieh19}. On the other hand, it is difficult to come up with a simple formulation for the training complexity of the AE algorithm as it depends on many factors such as the number of layers, the number of neurons at each layer, the number of training symbols, the number of epochs, and the algorithm used for training. However, the AE in \cite{b1} consists of ten layers with 500, 100, 50, 20, $D'$, $D'$, 20, 50, 100, and 500 neurons in its layers where $D'=2$ or 3. It uses the Tensorflow algorithm for learning. It is very clear that the computational complexity of training or running this AE is substantially higher than the LR algorithm, including the MUSIC algorithm employed to calculate the AOA. The prediction complexity of the LR algorithm has $O(N_R^2P)$ due to the MUSIC algorithm for $\theta$. In addition, $2N_R+1$ multiplications, $N_R$ square roots, $N_R$ adds, and a $\log$ operation are needed. Whereas, the prediction complexity of the AE algorithm in \cite{b1} is extremely high due to the large number of neurons employed, i.e., $2\times 500$, $2\times 100$, $2\times 50$, $2\times 20$, $2\times D'$.

To address the complexity of the RS algorithm, consider Algorithm~\ref{alg:cap4}. Calculation of the ${\bf S}$ matrix takes $N_S N_R^2$ multiplications. As before, let $P$ be the number of $\phi$ values to be tested (given as 180 in Algorithm~\ref{alg:cap4}). Then, first, $P$ times $N_R^2 + 1$ multiplications are needed. Then, to calculate $C(\phi )$, $P$ times $N_R^2$ multiplications are needed. As a result, the total number of multiplications for the RS algorithm to estimate $\theta$ is $P (2N_R^2 + 1) + N_S N_R^2$.

Algorithm~\ref{alg:cap5} is based on the same technique as Algorithm~\ref{alg:cap4}, but using column vectors of the CSI matrix, instead of row vectors. It then provides an estimate of $\rho$. We will now calculate its complexity in a fashion similar to Algorithm~\ref{alg:cap4}. Calculation of the ${\bf S}$ matrix takes $N_R N_S^2$ multiplications. As before, let $Q$ be the number of $\sigma$ values to be tested (given as 1000 in Algorithm~\ref{alg:cap5}). Then, $Q$ times $N_S^2 + 1$ multiplications are needed for the ${\bf B}$ matrices. Ignoring division of ${\bf S}$ by $N_R$, to calculate $C(\sigma)$, $Q$ times $N_S^2$ multiplications are needed. As a result, the total number of multiplications for the RS algorithm to estimate $\rho$ is $N_\sigma (2N_S^2 + 1) + N_R N_S^2$.

We would like to note that, both Algorithm~\ref{alg:cap4} and \ref{alg:cap5}, the numbers 180 and 1000 are just examples. Instead of uniformly spacing the search numbers $\phi$ and $\sigma$, nonuniform search techniques can be developed. Also, the closeness of the matrices ${\bf S}$ and ${\bf B}$ can be measured by other techniques, such as a norm of the difference of ${\bf S}$ and ${\bf B}$.

We provide a summary of the required number of multiplications in the algorithms we discussed in Table~\ref{tbl:Complexity}.
\subsection{Performance at Different SNR}
\ifCLASSOPTIONonecolumn
\else
\begin{table*}[!t]
\begin{center}
\caption{Performance comparison for TW and CT at $k$-nearest = 102 in 3D channel, SNR = -10 dB.}
\begin{tabular}{|c|l|c|c|c|c|c|c|c|c|}
\hline
Measure & Channel & PCA & SM & AE & LR & ISQ & MM & JM & RS\\ \hline
   & LOS   &     0.7494 &   0.8132  &  0.8155  &  0.9275 &   0.9146 & 0.9991 & 0.9990 & 0.9958\\ \cline{2-10}
TW & QLOS  &     0.6665 &   0.6968  &  0.7260  &  0.7467 &   0.7470 & 0.9774 & 0.9362 & 0.9612\\ \cline{2-10}
   & QNLOS &     0.6860 &   0.6813  &  0.6981  &  0.7223 &   0.7192 & 0.9361 & 0.9652 & 0.9133\\
\hline
   & LOS   &     0.7919 &   0.8594  &  0.8467  &  0.9403 &   0.9364 & 0.9998 & 0.9998 & 0.9942\\ \cline{2-10}
CT & QLOS  &     0.7570 &   0.7852  &  0.8035  &  0.7670 &   0.7686 & 0.9809 & 0.9456 & 0.9725\\ \cline{2-10}
   & QNLOS &     0.7649 &   0.7679  &  0.7773  &  0.7404 &   0.7356 & 0.9465 & 0.9687 & 0.9372\\
\hline
\end{tabular}
\end{center}
\label{tab4233}
\end{table*}
\fi
\ifCLASSOPTIONonecolumn
\else
\begin{table*}[!ht]
\begin{center}
\caption{Performance comparison for TW and CT at $k$-nearest = 102 in 3D channel, phase noise -100 dBc/Hz @  1 MHz offset, SNR = 0 dB.
Numbers in the second rows for each entry are the percentage differences with the entries in Tables~\ref{tab42}--\ref{tab5}.}
\begin{tabular}{|c|l|c|c|c|c|c|c|c|c|}
\hline
Measure & Channel & PCA & SM & AE & LR & ISQ & MM & JM & RS\\ \hline
\multirow{6}{*}{TW}   &\multirow{2}{*}{LOS}&     0.8566 &   0.7741 &  0.8402  &  0.9932 &   0.9888 & 0.9994 & 0.9984 & 0.9960\\
   &       &   0.43\% & 6.4\%  & -1.4\% &-0.020\%&-0.030\%&0.040\%&-0.020\%&0.38\% \\ \cline{2-10}
   &\multirow{2}{*}{QLOS}&     0.8450 &   0.8307 &  0.8536  &  0.9078 &   0.9085 & 0.9977 & 0.9969 & 0.9939\\
   &       &   0.28\% & 2.4\%  &  4.4\% & 0.12\%& 0.077\%&-0.010\%&-0.11\%&0.11\% \\\cline{2-10}
   &\multirow{2}{*}{QNLOS}&     0.8487 &   0.8451 &  0.8528  &  0.9059 &   0.9068 & 0.9859 & 0.9910 & 0.9820\\
   &       &   0.18\% & 0.059\% & -0.38\% & -0.33\% & -0.30\% & -0.030\% & -0.16\% & 0.010\%\\
\hline
\multirow{6}{*}{CT}&\multirow{2}{*}{LOS}&   0.9276 &   0.8868 &  0.8534  &  0.9968 &   0.9941 & 0.9999 & 1.0000 & 0.9943\\
   &       &0.13\% & 2.0\% & 4.5\% & 0 & -0.010\% & 0.010\% & 0 & 0.55\% \\\cline{2-10}
   &\multirow{2}{*}{QLOS}&     0.9215 &   0.8881 &  0.9268  &  0.9402 &   0.9292 & 0.9991 & 0.9987 & 0.9933\\
   &       & 0.087\% & 4.3\% & -2.4\% & 0.15\% & 0.13\% & 0.010\% & -0.10\% & 0.35\% \\\cline{2-10}
   &\multirow{2}{*}{QNLOS}&     0.9239 &   0.9219 &  0.9108  &  0.9283 &   0.9254 & 0.9874 & 0.9929 & 0.9829\\
   &       & -0.22\% & -0.022\% & -0.56\% & -0.40\% & -0.37\% & -0.091\%  & -0.19\% & 0.25\% \\
\hline
\end{tabular}
\end{center}
\label{tab4284}
\end{table*}
\fi
We would like to compare the performance of the eight algorithms discussed in this paper at SNR values different than 0 dB. To that end, first, we provide results for the eight algorithms at SNR = -10 dB in Table~\ref{tab4233}. These results are obtained employing the number of antennas equal to 32 and the number of subcarriers equal to 32. Thus, they should be compared with Table~\ref{tab42} for PCA, SM, AE, LR, and ISQ, and the last columns of Tables~\ref{tab44}--\ref{tab5}. First, we observe that that there are significant drops in TW and CT performance for the first five algorithms PCA, SM, AE, LR, and ISQ. As can be expected, the drops are more significant for the QLOS and QNLOS channels. On the other hand, although there are some drops in the TW and CT performance of the MM, JM, and RS algorithms, they are much less as compared to the first five algorithms PCA, SM, AE, LR, and ISQ. We ran simulations at SNR = 10 dB as well, but found out that the improvements in TW and CT remained modest.

We conclude that although PCA, SM, AE, LR, and ISQ algorithms present TW and CT performance degradation in the case of SNR drop from 0 dB to -10 dB, the algorithms MM, JM, and RS, albeit showing some drop, remain quite robust, much more so than PCA, SM, AE, LR, and ISQ.
\subsection{Effect of Phase Noise}
As the techniques LR, ISQ, MM, and JM in this paper depend on the MUSIC algorithm for determination of the AOA, the effect of phase noise on them can be questioned. To that end, we refer to \cite{Ericsson20}, which specifies that,
at 1 MHz offset, the phase noise of the phase-locked loop should be at or below -102 dBc/Hz (about -100 dBc/Hz).
Incorporating phase noise to the matrix ${\bf R}$ via a Matlab function, we tabulated the TW and CT performance of the eight algorithms in this paper with phase noise equal to -100 dBc/Hz at 1 MHz offset in Table~\ref{tab4284} with an SNR = 0 dB. 
In its second row for each entry in Table~\ref{tab4284}, we have made a comparison of the percentage changes in the first row of the entry in Table~\ref{tab4284} with the corresponding value in Tables~\ref{tab42}--\ref{tab5}.
These changes can be positive, indicating we have {\em worse\/} performance with phase noise, or negative, indicating {\em better\/} performance with phase noise, which may be considered counterintuitive. We note that the randomness in the process does give rise to both positive and negative entries. First,
the TW and CT performance of all the eight algorithms remains within close proximity (positive or negative) of the performance without phase noise. Second, we note that in Table~\ref{tab4284}, there are 21 negative entries out of 48 total. When we increase the phase noise to -70 dBc/Hz at 1 MHz with an SNR = 0 dB, the number of negative entries reduces to 6 out of 48, indicating worsening performance with increased phase noise, as expected. Nevertheless, as we stated above, we conclude from Table~\ref{tab4284} that the performance with phase noise when it is within acceptable limits indicated in \cite{Ericsson20}, employing the algorithms studied in this work, remains within close proximity of the performance without phase noise on TW and CT performance. Note that TW and CT are neighborhood measures and although on the average they become worse with increasing phase noise, this behavior happens not to be uniform as discussed above. On the other hand, we measured the Error Vector Magnitude (EVM) performance with the simulated system, which illustrated the uniformly degrading EVM performance with increased phase noise, as expected.

%% file: conclusion.tex
\section{Conclusion}\label{ch:5}
The LR, ISQ, and MM algorithms presented in this paper significantly outperform the three algorithms in the seminal paper \cite{b1}, PCA, SM, and AE, in terms of performance. As in \cite{b1}, we measure the performance in terms of connectivity (CT) and trustworthiness (TW). An important advantage of ISQ and MM over the three algorithms from \cite{b1} is that we can calculate each UE data independently as it comes, so it is much faster and simpler. In the case of LR, a similar advantage exists, however a number of UE data is first needed in order to perform the regression. The MM algorithm has more complexity than LR and ISQ but the advantage it provides in terms of TW and CT measures is significantly better than those of ISQ and LR. In addition, the MM algorithm results in channel charts with significantly better visual outcome. We studied an algorithm from the literature, intended for a different application, we call the JM algorithm. This algorithm has only very slight improvement over the MM algorithm at a very large increase in computational complexity. Finally, we introduced the RS algorithm whose performance is about the same as the MM and JM algorithms. The advantage of this algorithm is that it does not need the eigenvector and eigenvalue decomposition needed by the MM and JM algorithms, and furthermore, its RTL implementation may be more straightforward as it does not require eigenvalue and eigenvector analysis.

Note that the MUSIC algorithm for $\theta$ is model based. In addition, the ISQ algorithm for $\rho$ is also model based. Finally, the MM, JM, and RS algorithms are completely model based for both $\theta$ and $\rho$. The performance of ISQ, MM, JM, and RS are better than those of PCA, SM, and AE of \cite{b1}. PCA and SM employ standard techniques of dimensionality reduction whereas AE is training based.

Reference \cite{b1} discusses an extension of the SM algorithm, called the SM+ algorithm. This algorithm employs a penalty in the objective function that keeps temporally adjacent points in the local geometry nearby in the channel chart. In \cite[Fig.~6 and Fig.~7]{b1}, it is stated that the TW and CT performance of SM and SM+ are close while SM+ provides more visually satisfying results than PCA, SM, and AE. Yet, comparing \cite[Fig.~6]{b1} with our Fig.~\ref{fig12} and Fig~\ref{fig46}, one can observe that LR, ISQ, and MM provide better visual outcomes than SM+. We have observed the same conclusion for the RS algorithm, which we are not presenting due to avoiding repetitive figures. For that reason, we did not attempt to simulate the SM+ algorithm in this paper. One can conclude based on the comparisons discussed above that MM, JM, and RS will outperform SM+ in TW and CT as well as the visual quality of the channel charts.

We have carried out extensive simulations for the bias and variability of our estimation algorithms. The results show that their bias and variability vanish asymptotically. A theoretical analysis of these facts is beyond the scope of this paper and is left as potential future work.

\section{Acknowledgment}
The authors would like to thank the editor and the anonymous reviewers whose comments improved the presentation in the paper.
\ifCLASSOPTIONonecolumn
\clearpage\newpage
\begin{table}
\vspace{10mm}
\begin{center}
\begin{tabular}{|c|l|c|c|c|c|c|}
\hline
Measure & Channel & PCA & SM & AE & LR & ISQ \\ \hline
   & LOS   &     0.8603 &   0.8272  &  0.8286  &  0.9930 &   0.9885 \\ \cline{2-7}
TW & QLOS  &     0.8474 &   0.8512  &  0.8574  &  0.9089 &   0.9092 \\ \cline{2-7}
   & QNLOS &     0.8502 &   0.8456  &  0.8496  &  0.9029 &   0.9041 \\
\hline
   & LOS   &     0.9288 &   0.9051  &  0.8932  &  0.9968 &   0.9940 \\ \cline{2-7}
CT & QLOS  &     0.9223 &   0.9278  &  0.9055  &  0.9416 &   0.9304 \\ \cline{2-7}
   & QNLOS &     0.9237 &   0.9217  &  0.9057  &  0.9246 &   0.9220 \\
\hline
\end{tabular}
\end{center}
\caption{Performance comparison for TW and CT at $k$-nearest = 102 for LR and ISQ algorithms in 3D channel.}
\label{tab42}
\end{table}
\begin{figure*}
\centering 
\resizebox{\textwidth}{!}{%
  \renewcommand{\arraystretch}{0}%
\begin{tabular}{@{}c@{\hspace{1pt}}c@{\hspace{1pt}}c@{\hspace{1pt}}c@{\hspace{1pt}}c@{}}
  \includegraphics[width=0.5in]{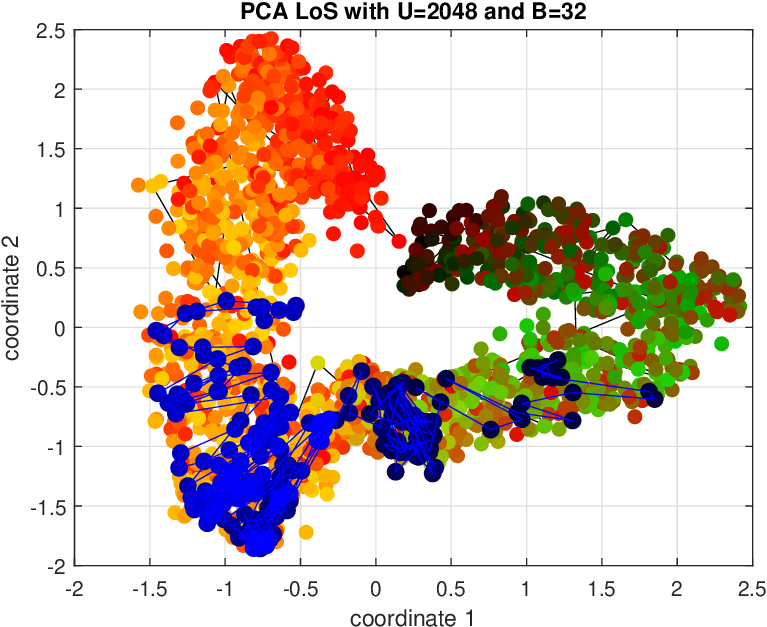} &
  \includegraphics[width=0.5in]{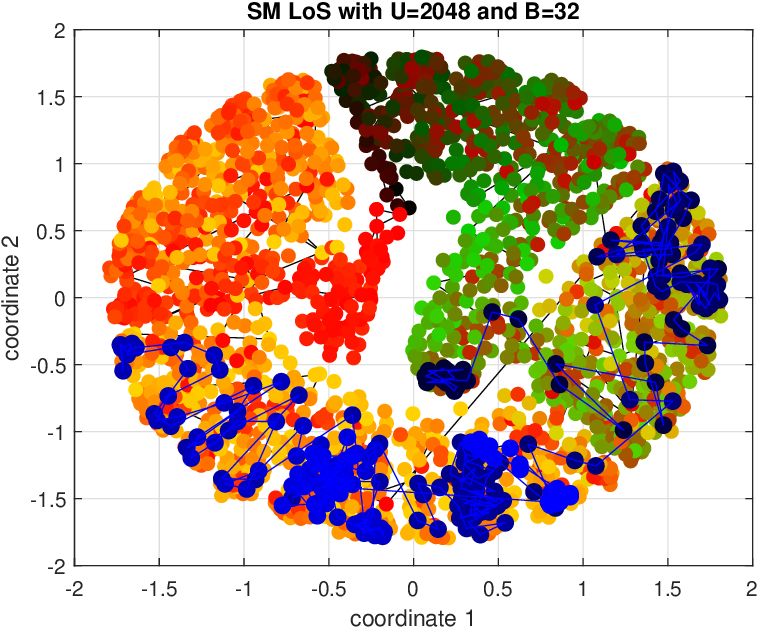} &
  \includegraphics[width=0.5in]{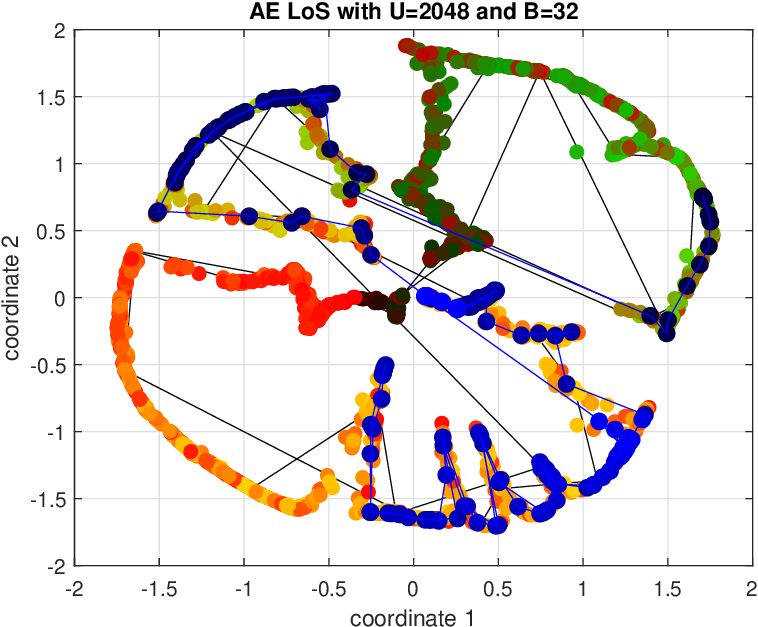} &
  \includegraphics[width=0.5in]{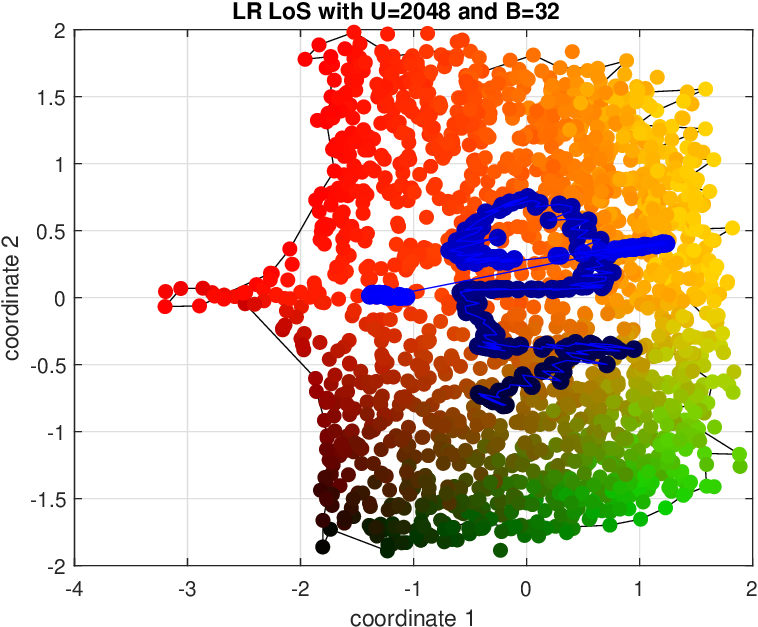} &
  \includegraphics[width=0.5in]{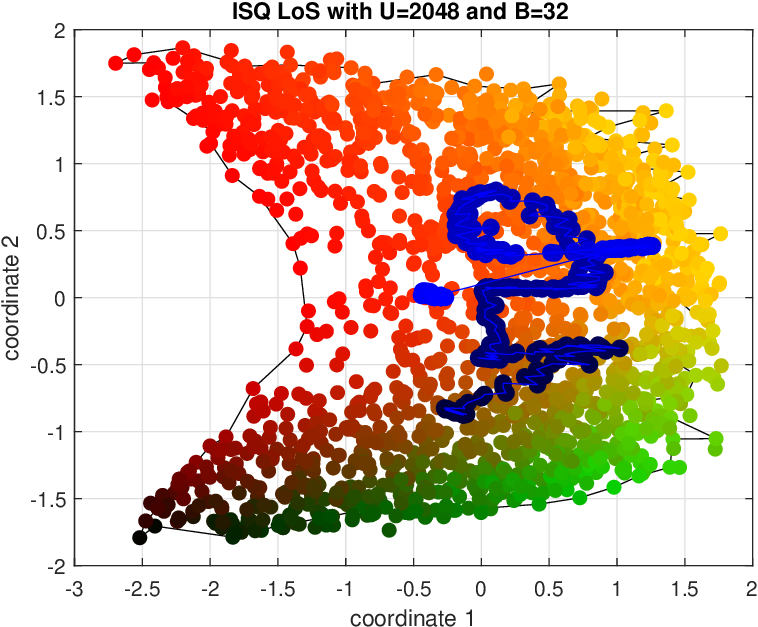} \\
  \addlinespace[2pt]
  \includegraphics[width=0.5in]{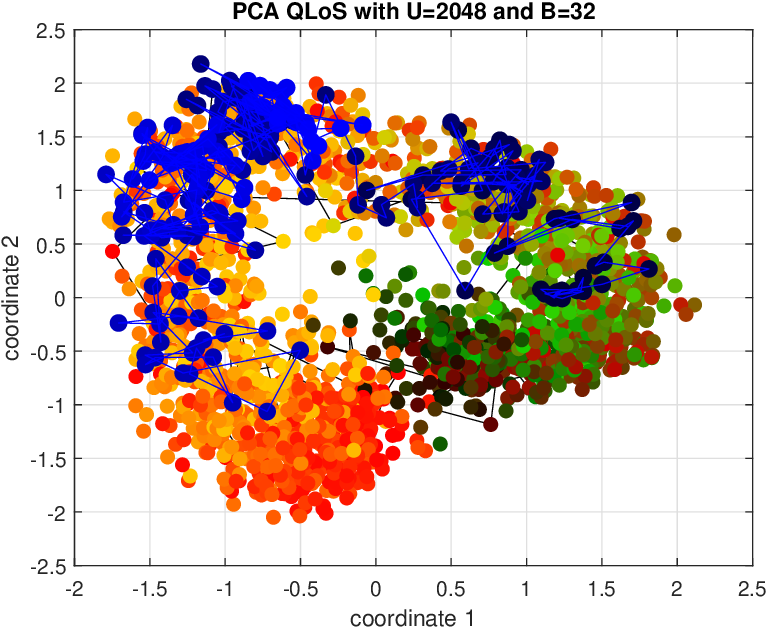} &
  \includegraphics[width=0.5in]{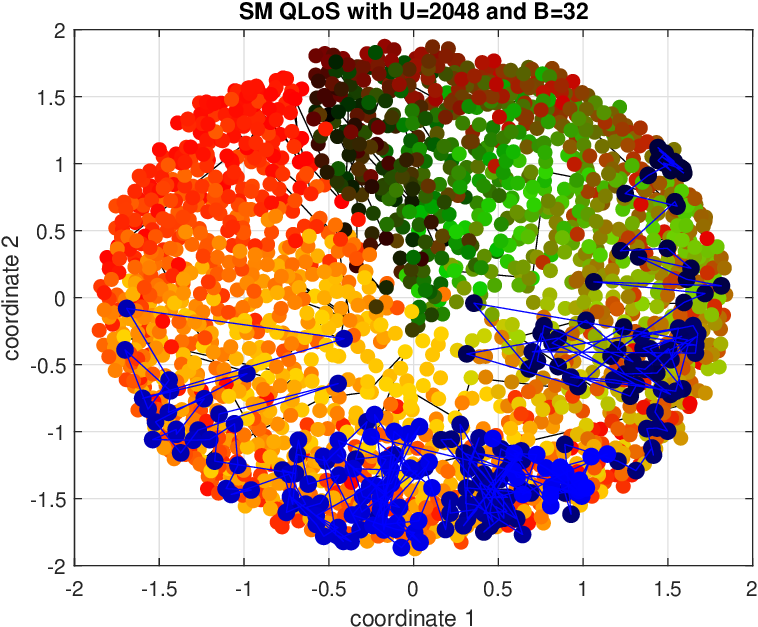} &
  \includegraphics[width=0.5in]{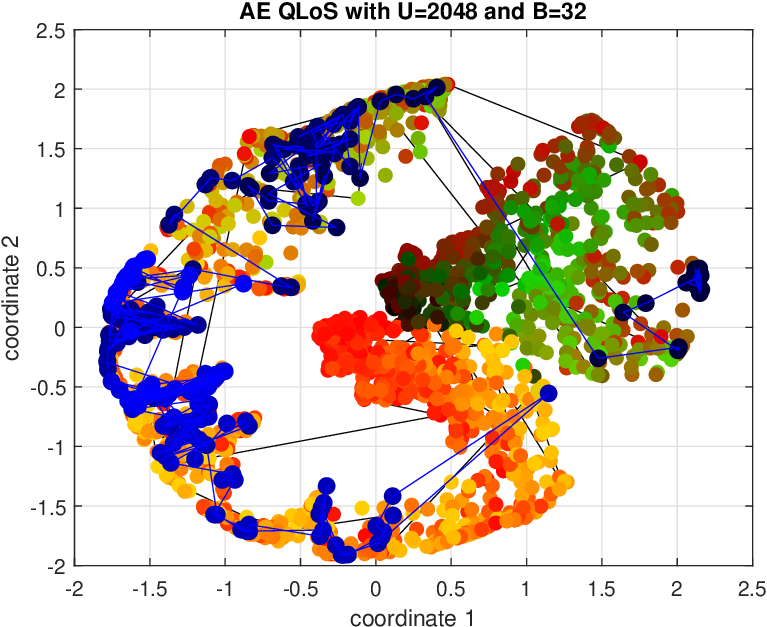} &
  \includegraphics[width=0.5in]{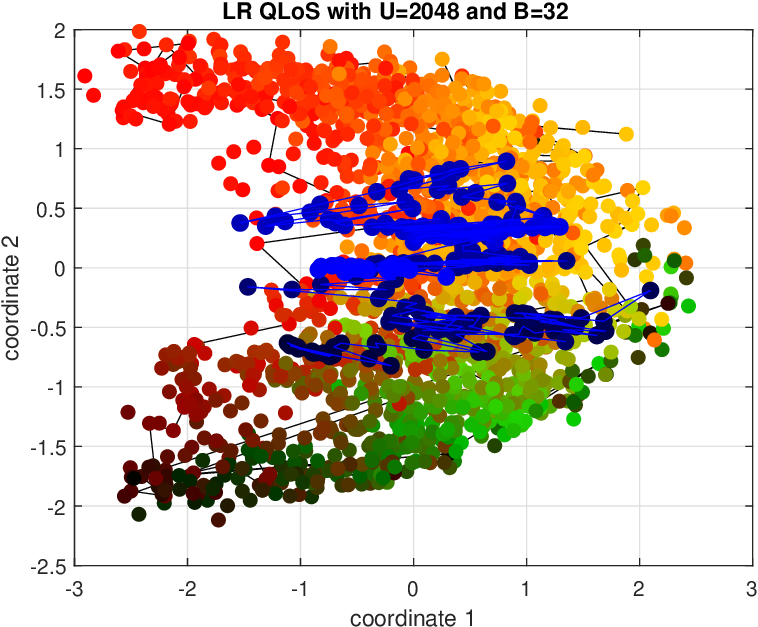} &
  \includegraphics[width=0.5in]{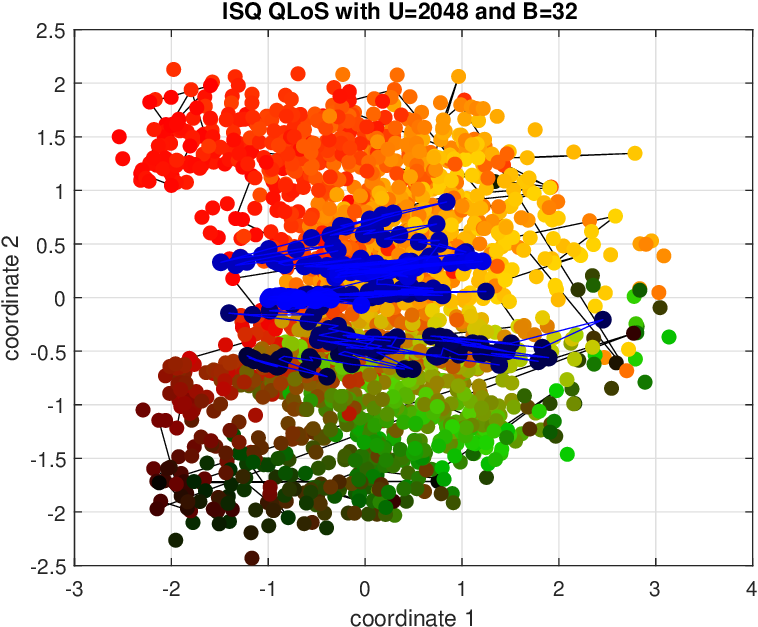} \\
  \addlinespace[2pt]
  \includegraphics[width=0.5in]{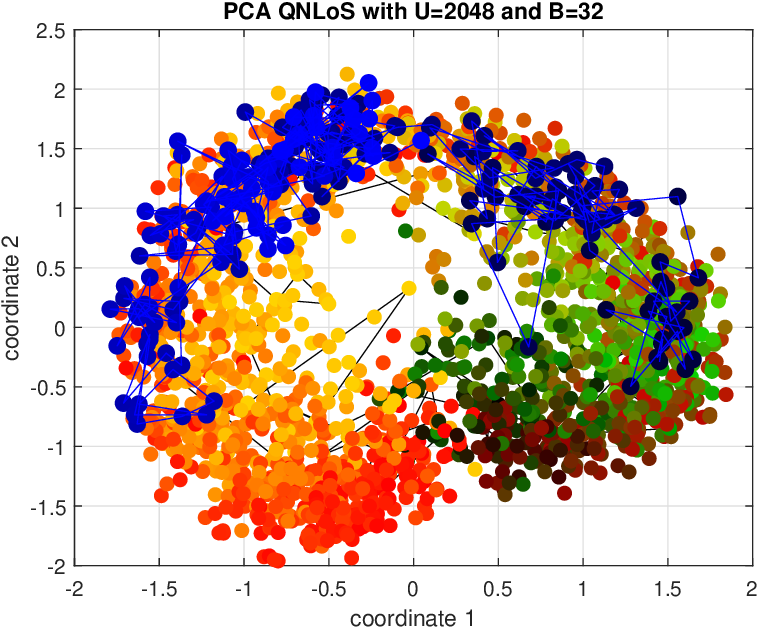} &
  \includegraphics[width=0.5in]{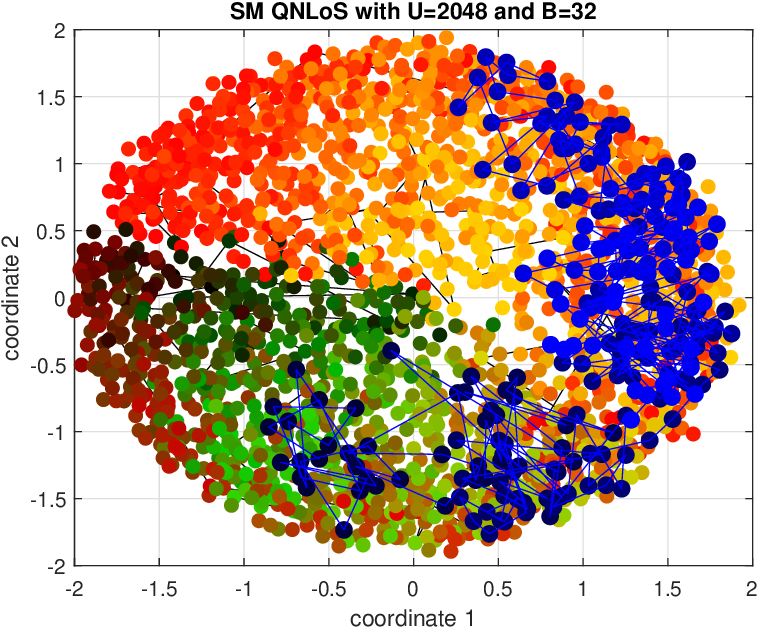} &
  \includegraphics[width=0.5in]{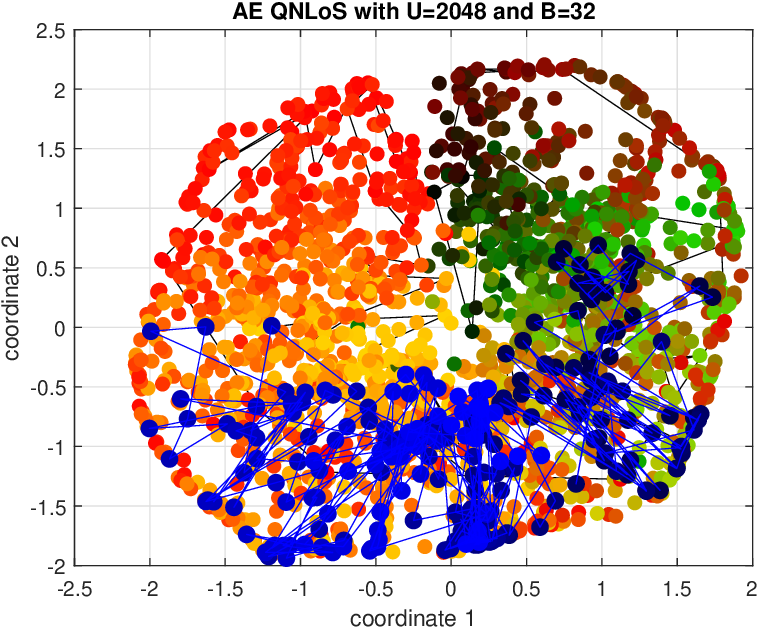} &
  \includegraphics[width=0.5in]{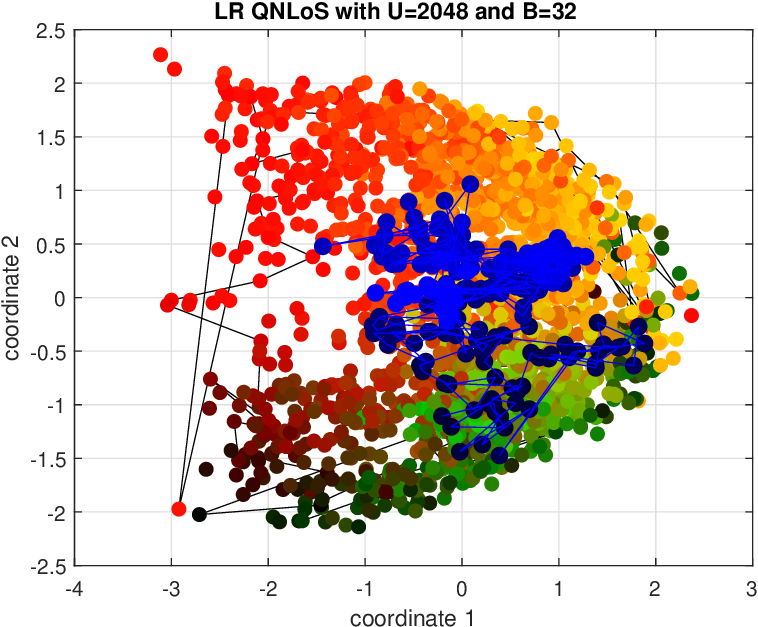} &
  \includegraphics[width=0.5in]{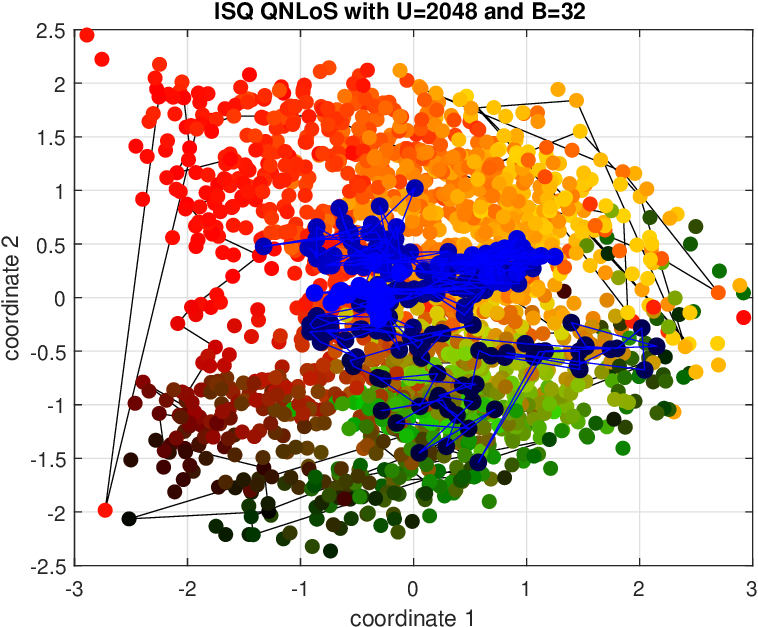} \\
  \end{tabular}
}
 \caption{Channel charts with PCA, SM, AE, LR, and ISQ algorithms for the 3D LOS, QLOS, and QNLOS channels.}
  \label{fig12}
\end{figure*}
\begin{figure*}[htbp]
\centering 
\begin{adjustbox}{width=1.0\textwidth}
  \renewcommand{\arraystretch}{0}%
  \begin{tabular}{c}
     \addlinespace[1pt]
  \includegraphics[width=0.5in]{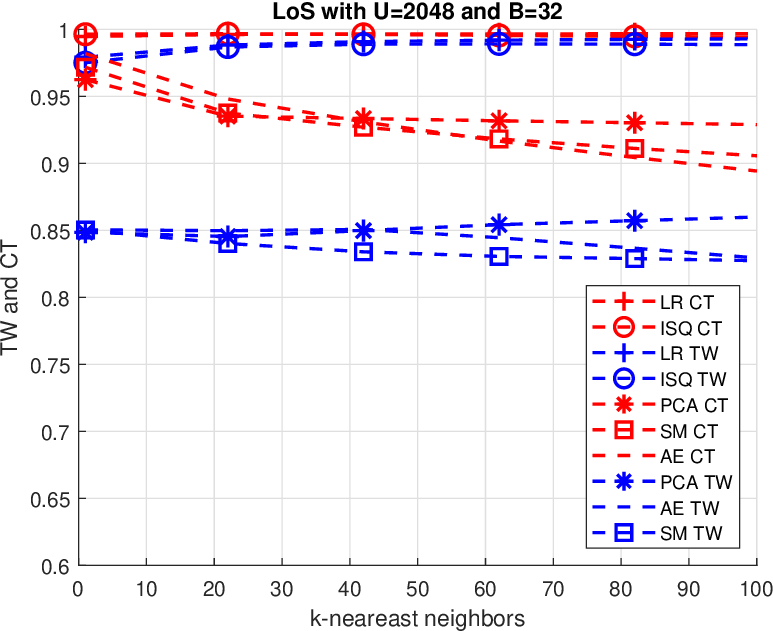}
  \includegraphics[width=0.5in]{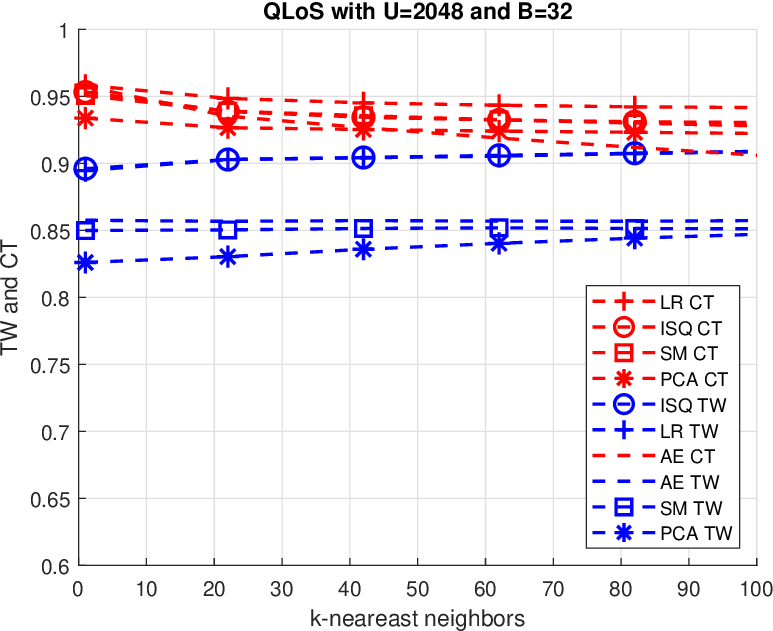}
   \includegraphics[width=0.5in]{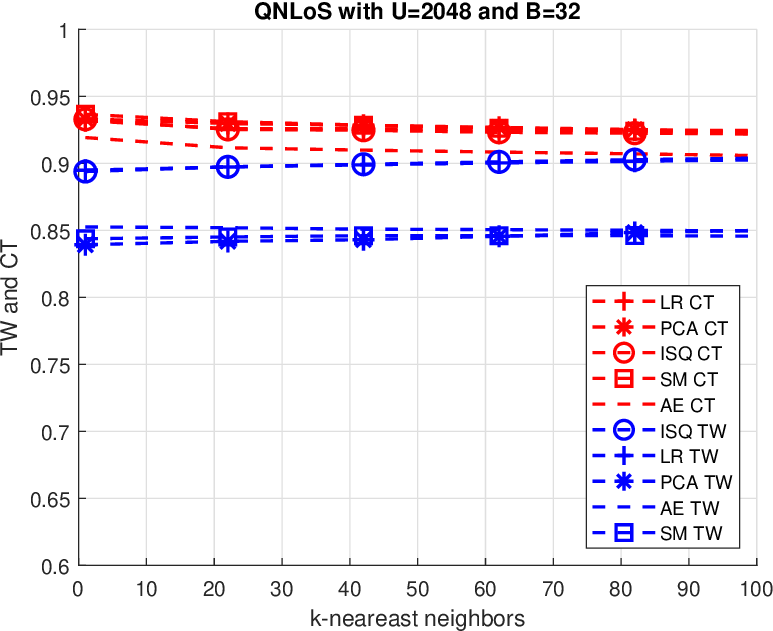} \\
  \end{tabular}
\end{adjustbox}
\caption{TW and CT performance against $k$-nearest neighbors for LR and ISQ algorithms in 3D. Left: LOS, middle: QLOS, right: QNLOS.}
 \label{fig9}
 \end{figure*}

\FloatBarrier

\fi

\ifCLASSOPTIONonecolumn
\begin{table}
\begin{center}
\begin{tabular}{|c|l|c|c|c|c|c|c|}
\hline
Measure & Channel & 2sc & 8sc & 14sc & 20sc & 26sc & 32sc \\ \hline
   & LOS   & 0.9975 &   0.9986 &   0.9992 &   0.9997  &  0.9996  &  0.9998 \\ \cline{2-8}
TW & QLOS  & 0.9817 &   0.9958 &   0.9970 &   0.9973  &  0.9976  &  0.9976 \\ \cline{2-8}
   & QNLOS & 0.9622 &   0.9802 &   0.9818 &   0.9825  &  0.9837  &  0.9856 \\
\hline
   & LOS   & 0.9992 &   0.9999 &   1.0000 &   0.9999  &  0.9999  &  1.0000 \\ \cline{2-8}
CT & QLOS  & 0.9816 &   0.9972 &   0.9985 &   0.9990  &  0.9993  &  0.9992 \\ \cline{2-8}
   & QNLOS & 0.9626 &   0.9803 &   0.9821 &   0.9833  &  0.9845  &  0.9865 \\
\hline
\end{tabular}
\end{center}
\caption{Performance comparison for TW and CT at $k$-nearest = 102 for MM algorithm in 3D channel at 2, 8, 14, 20, 26, and 32 subcarriers.}
\label{tab44}
\end{table}

\begin{figure*}
\centering 
\resizebox{\textwidth}{!}{%
  \renewcommand{\arraystretch}{0}%
  \begin{tabular}{@{}c@{\hspace{1pt}}c@{\hspace{1pt}}c@{\hspace{1pt}}c@{}}
  \includegraphics[width=0.5in]{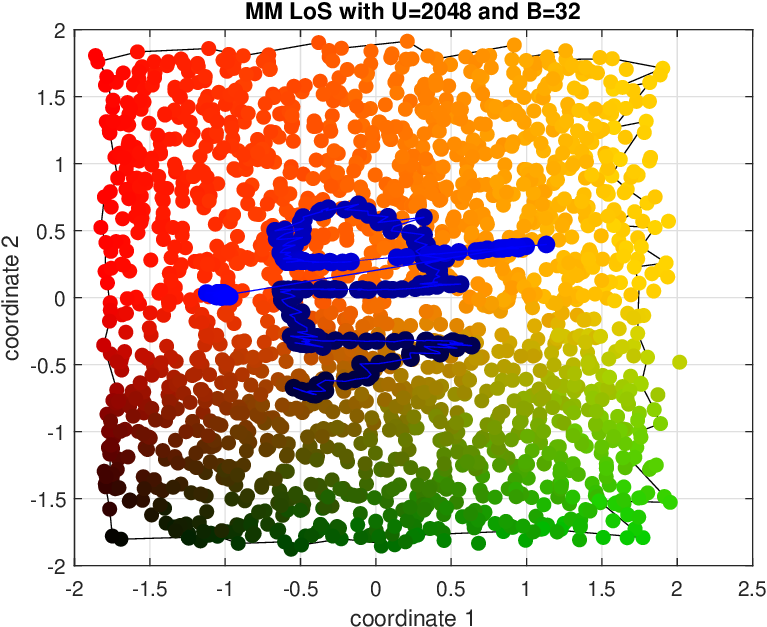} &
  \includegraphics[width=0.5in]{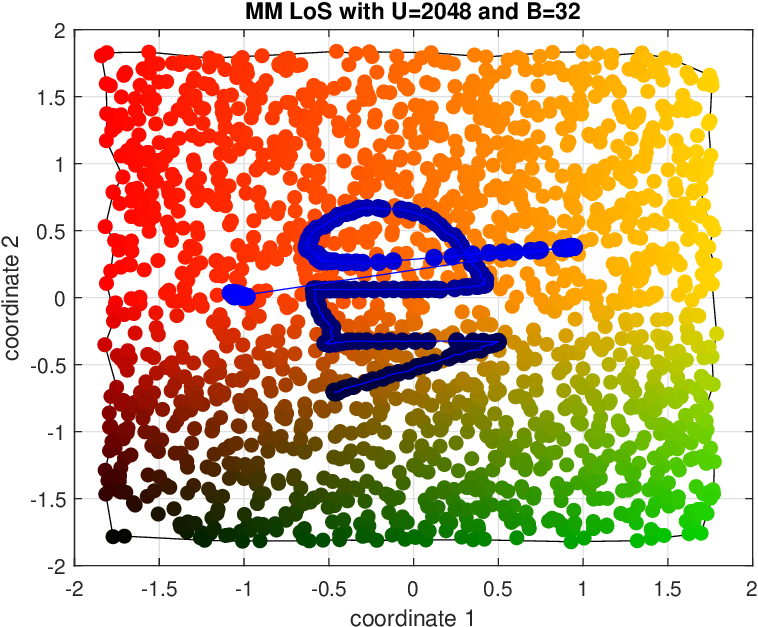} &
  \includegraphics[width=0.5in]{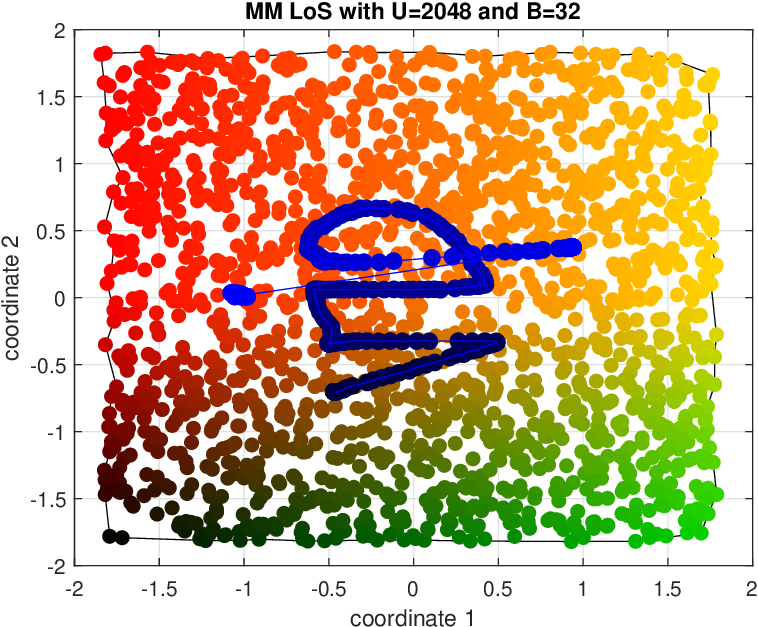} &
  \includegraphics[width=0.5in]{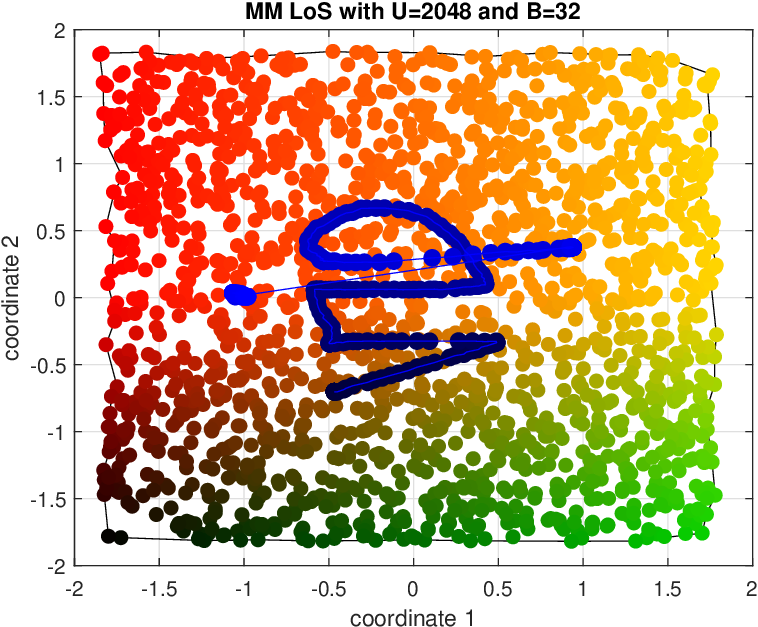} \\
  \addlinespace[2pt]
  \includegraphics[width=0.5in]{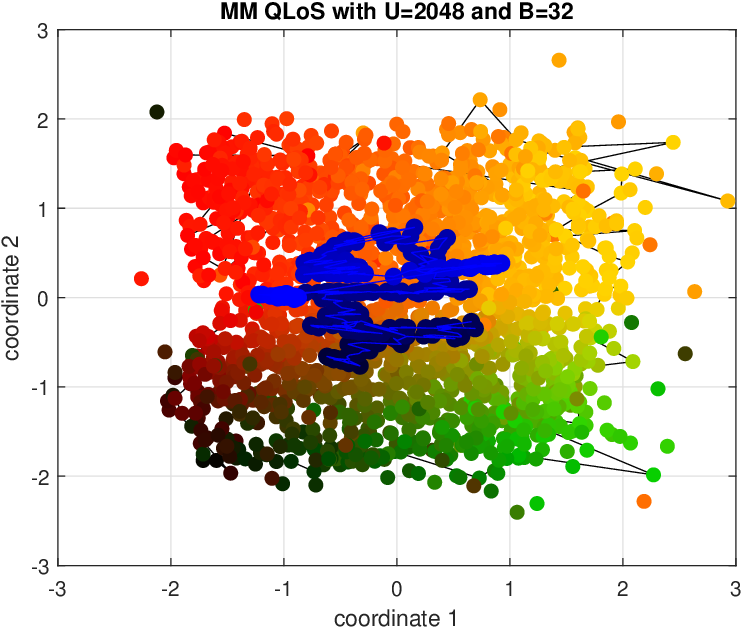} &
  \includegraphics[width=0.5in]{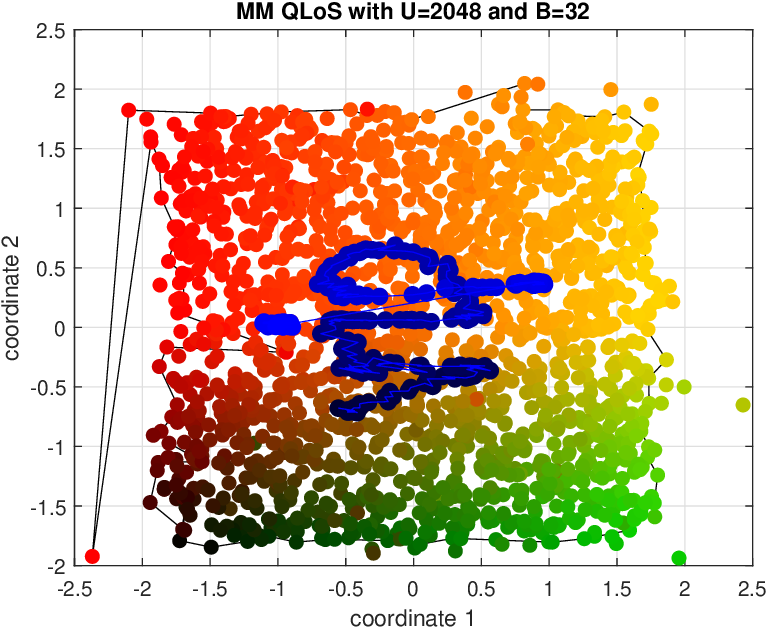} &
  \includegraphics[width=0.5in]{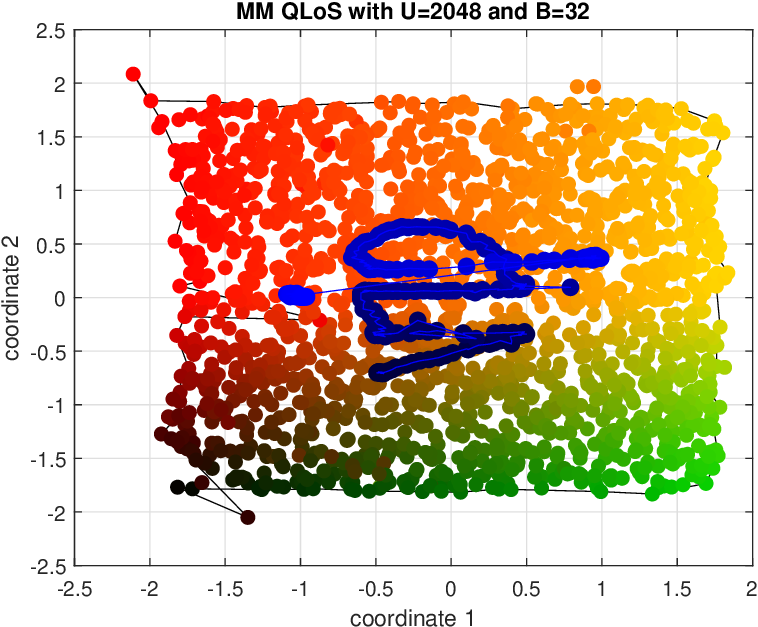} &
  \includegraphics[width=0.5in]{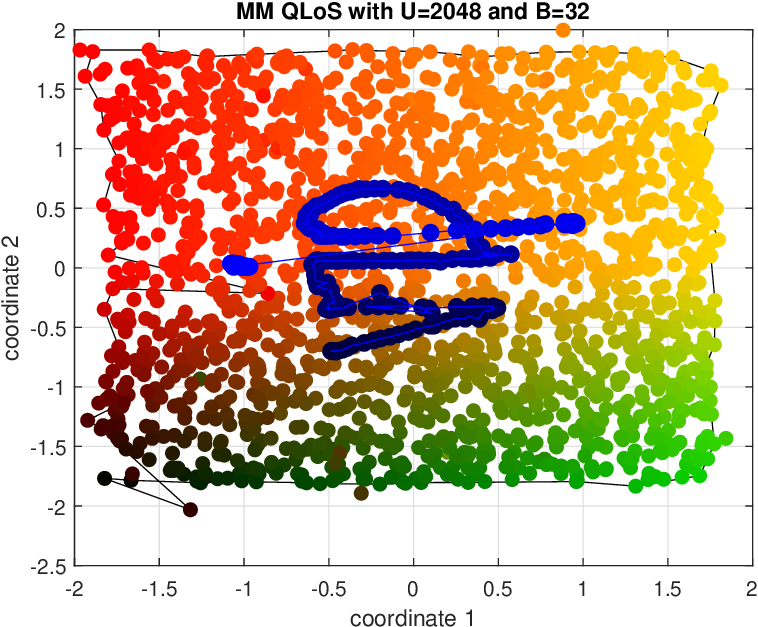} \\
  \addlinespace[2pt]
  \includegraphics[width=0.5in]{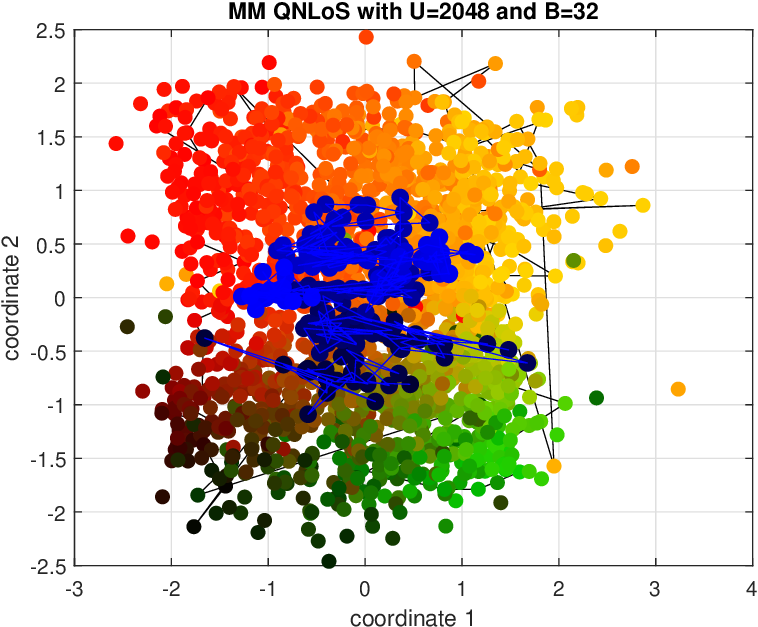} &
  \includegraphics[width=0.5in]{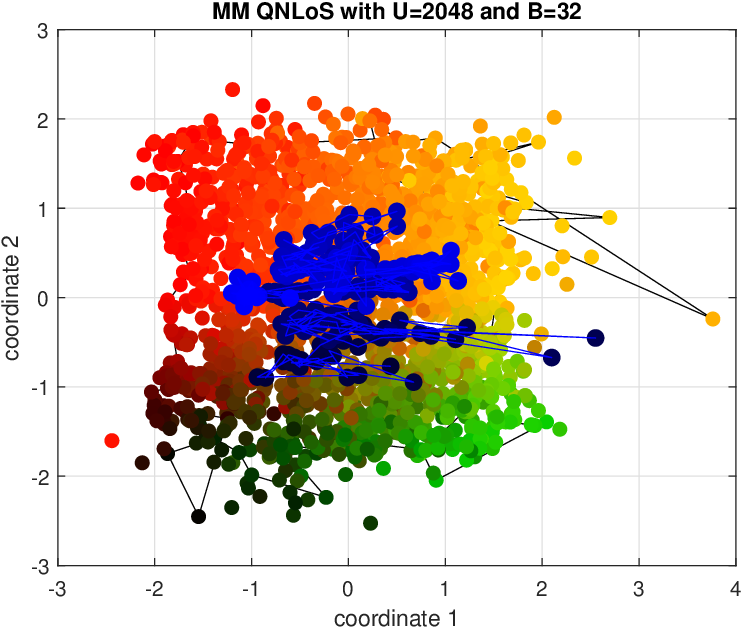} &
  \includegraphics[width=0.5in]{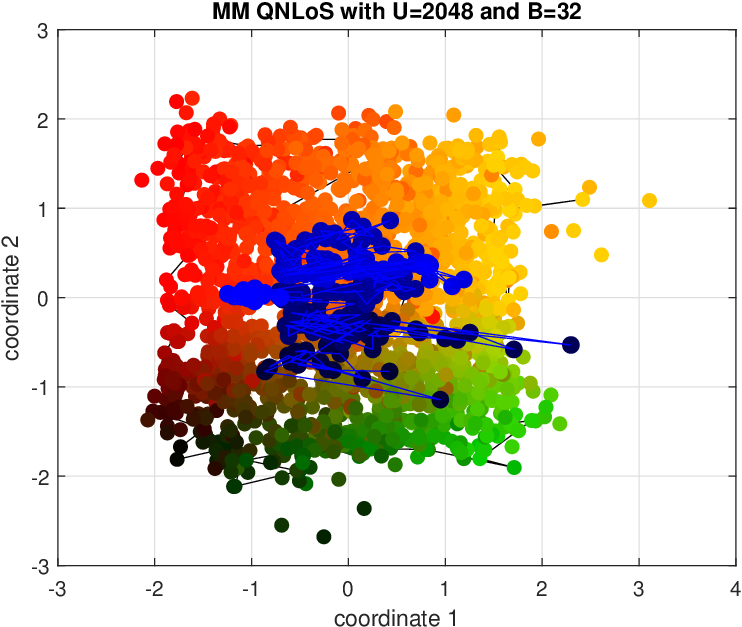} &
  \includegraphics[width=0.5in]{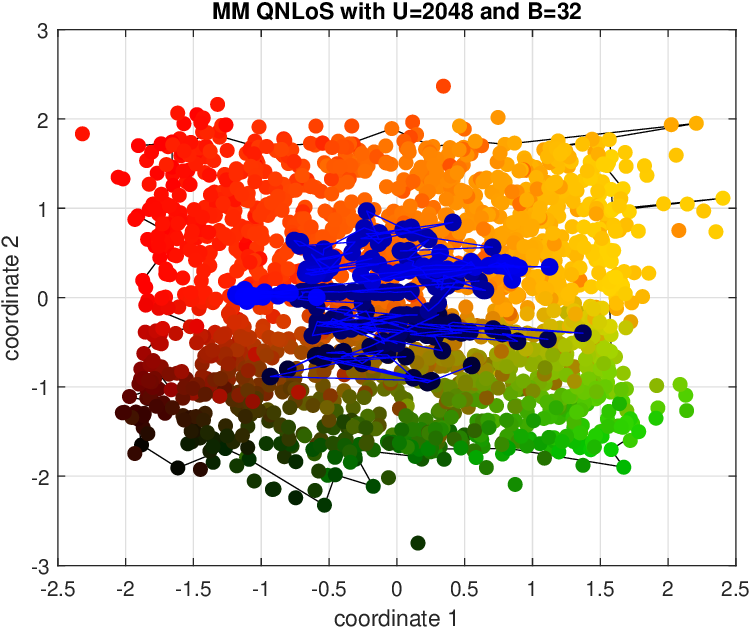} \\
  \end{tabular}
}
\caption{Channel charts with the MM algorithm for the 3D LOS, QLOS, and QNLOS channels at 2, 8, 20, and 32 subcarriers.}
  \label{fig46}
\end{figure*}

\begin{figure*}[htbp]
\centering 
\begin{adjustbox}{width=1.0\textwidth}
  \renewcommand{\arraystretch}{0}%
  \begin{tabular}{c}
     \addlinespace[1pt]
  \includegraphics[width=0.5in]{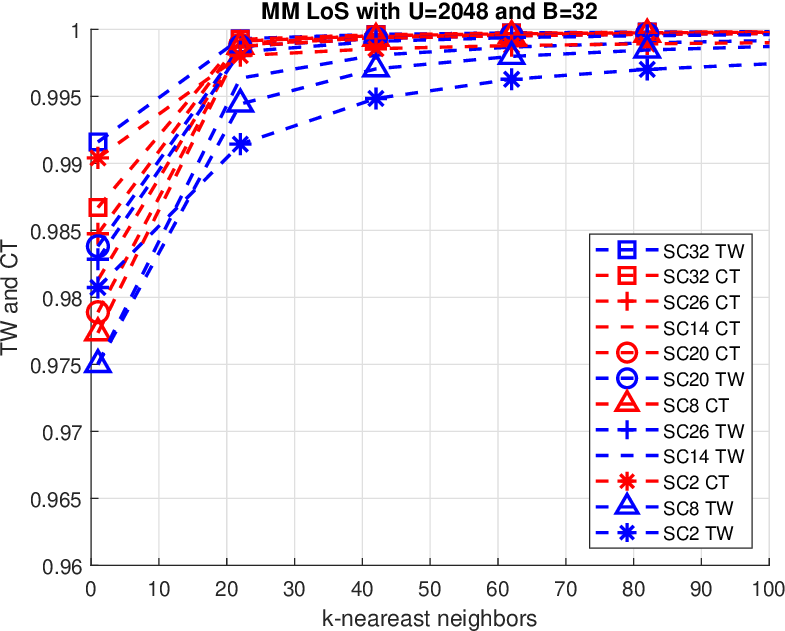}
  \includegraphics[width=0.5in]{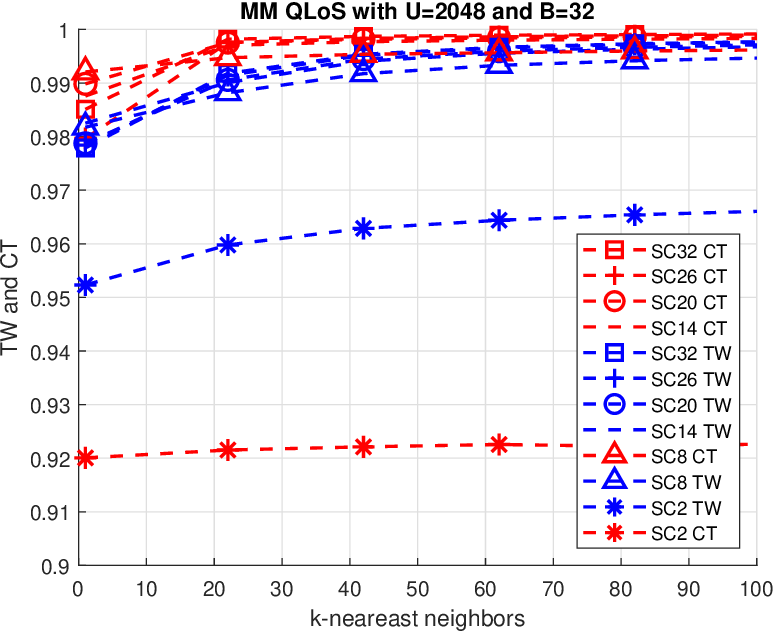}
  \includegraphics[width=0.5in]{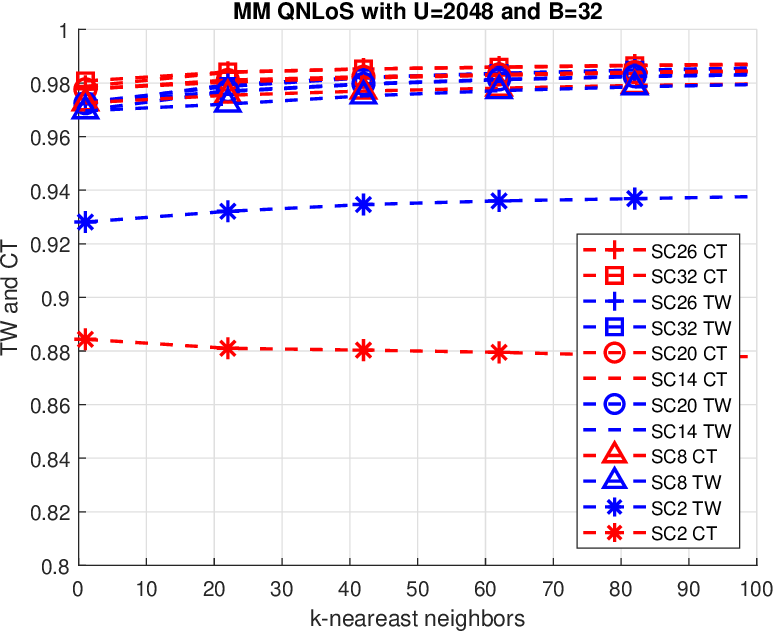} \\
  \end{tabular}
\end{adjustbox}
\caption{TW and CT performance against $k$-nearest neighbors for MM algorithm in 3D channel. Left: LOS, middle: QLOS, right: QNLOS.}
 \label{fig48}
 \end{figure*}
\fi

\ifCLASSOPTIONonecolumn
\clearpage\newpage

\begin{table}[!t]
\begin{center}
\begin{tabular}{|c|l|c|c|c|c|c|}
\hline
Measure & Channel & 8sc & 14sc & 20sc & 26sc & 32sc \\ \hline
   & LOS   &   0.9981 &   0.9984  &  0.9983  &  0.9982  &  0.9982 \\ \cline{2-7}
TW & QLOS  &   0.9936 &   0.9967  &  0.9966  &  0.9963  &  0.9958 \\ \cline{2-7}
   & QNLOS &   0.9798 &   0.9899  &  0.9888  &  0.9870  &  0.9894 \\
\hline
   & LOS   &   1.0000 &   1.0000  &  1.0000  &  1.0000  &  1.0000 \\ \cline{2-7}
CT & QLOS  &   0.9951 &   0.9985  &  0.9984  &  0.9981  &  0.9977 \\ \cline{2-7}
   & QNLOS &   0.9821 &   0.9916  &  0.9904  &  0.9886  &  0.9910 \\
\hline
\end{tabular}
\caption{Performance comparison for TW and CT at $k$-nearest = 102 for JM algorithm in 3D channel at 8, 14, 20, 26, and 32 subcarriers.}
\label{tab2022}
\end{center}
\end{table}
\begin{figure*}
\centering 
\resizebox{0.8\textwidth}{!}{%
  \renewcommand{\arraystretch}{0}%
  \begin{tabular}{@{}c@{\hspace{1pt}}c@{\hspace{1pt}}c@{\hspace{1pt}}c@{}}
  \includegraphics[height=0.4in]{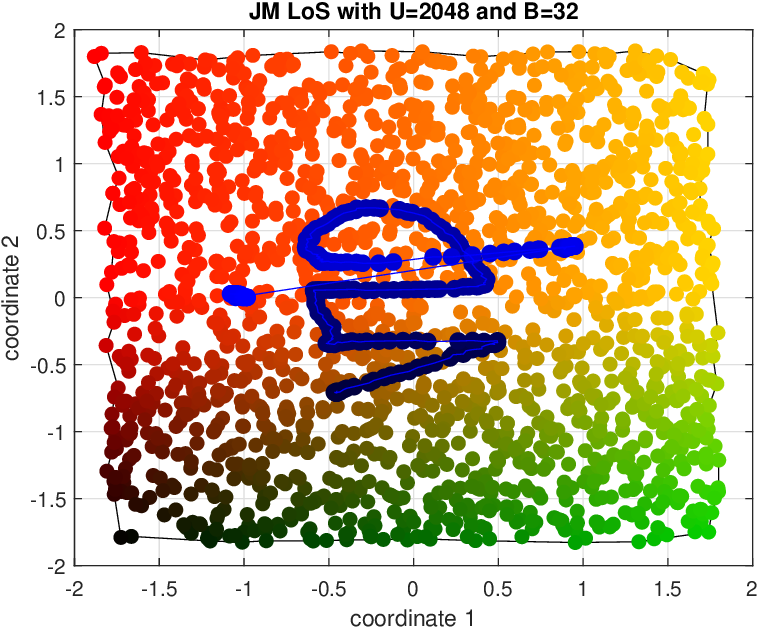} &
  \includegraphics[height=0.4in]{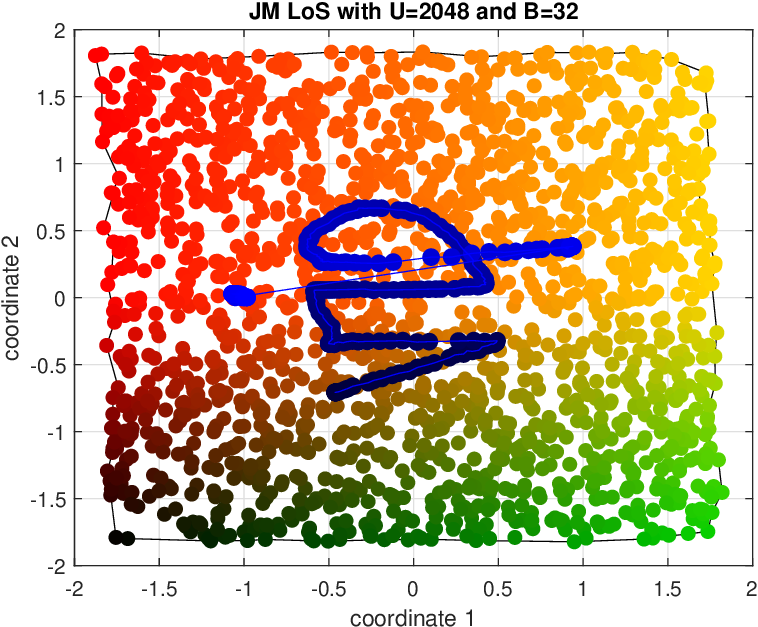} &
  \includegraphics[height=0.4in]{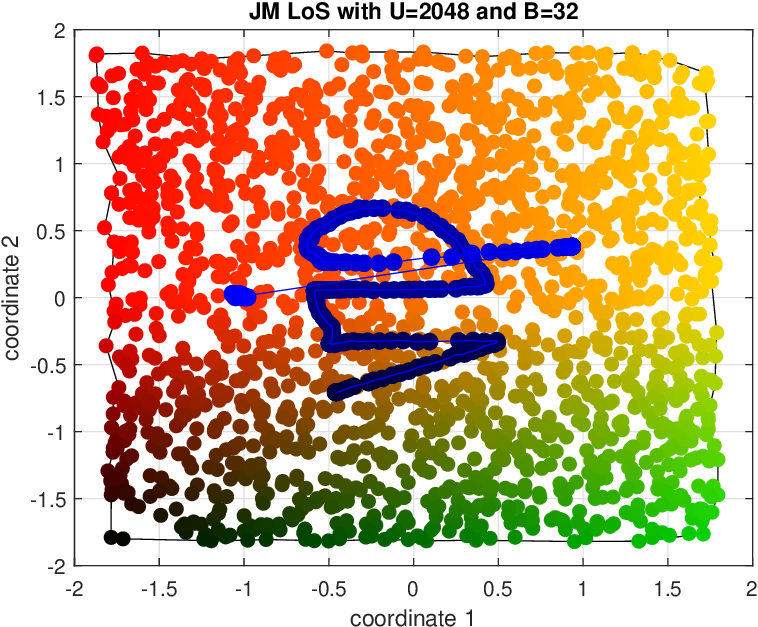} \\
  \addlinespace[2pt]
  \includegraphics[height=0.4in]{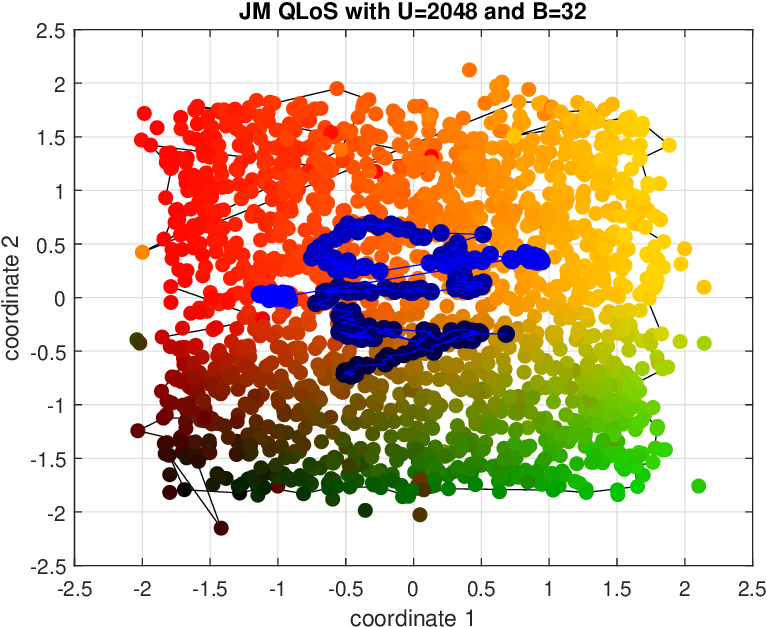} &
  \includegraphics[height=0.4in]{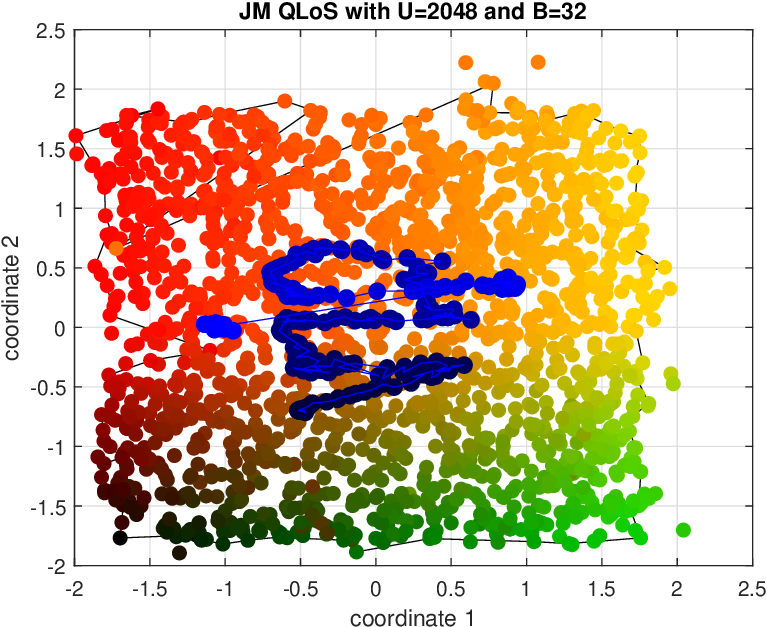} &
  \includegraphics[height=0.4in]{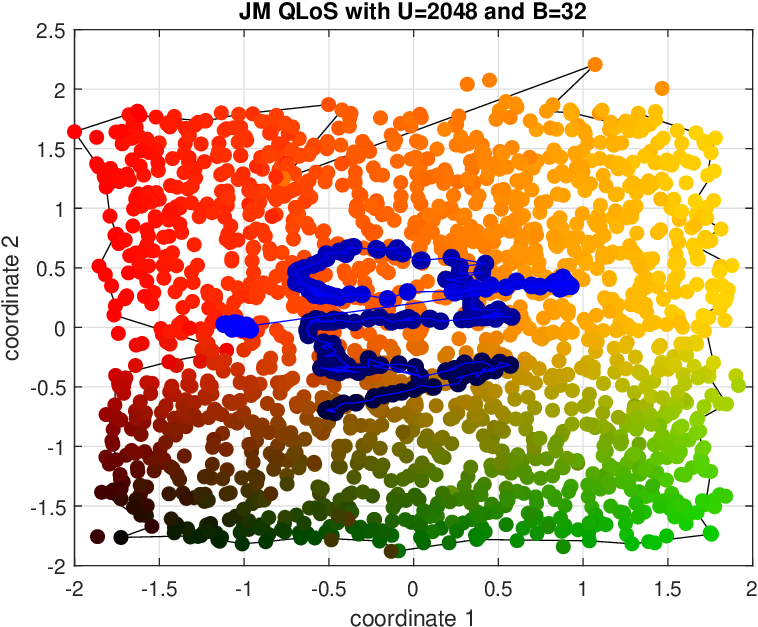}\\
  \addlinespace[2pt]
  \includegraphics[height=0.4in]{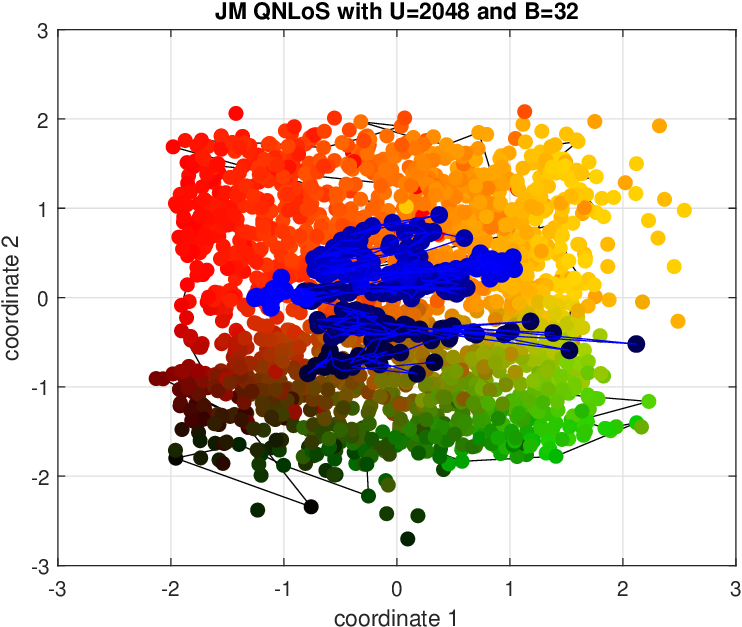} &
  \includegraphics[height=0.4in]{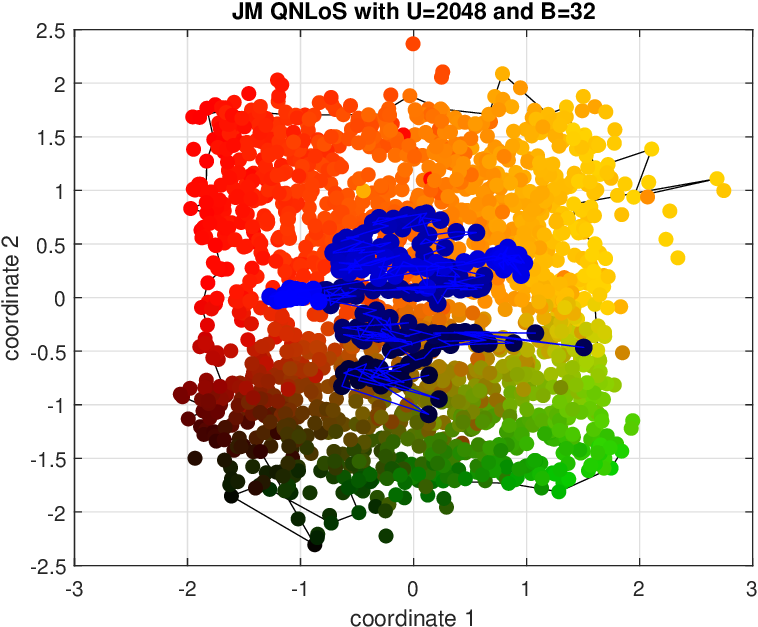} &
  \includegraphics[height=0.4in]{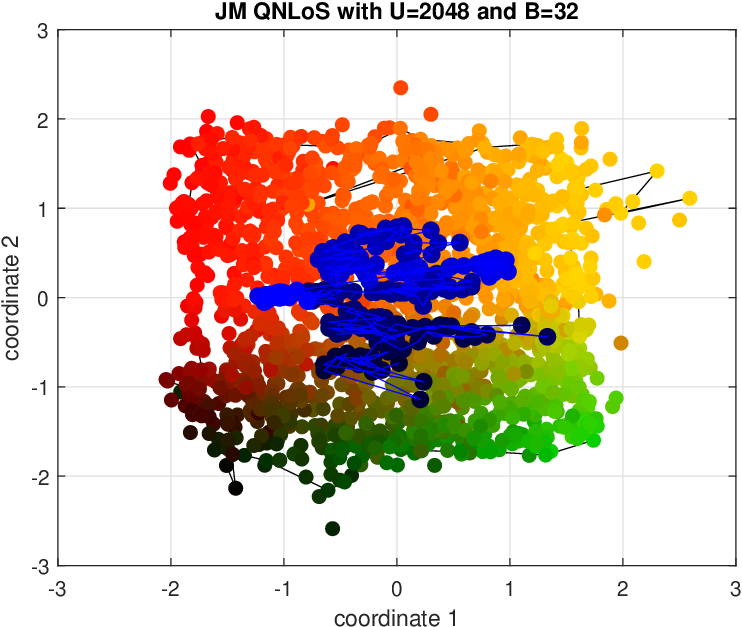}\\
\end{tabular}
}
\caption{Channel charts with the JM algorithm for the 3D LOS, QLOS, and QNLOS channels at 8, 20, and 32 subcarriers.}
  \label{fig446}
\end{figure*}

\begin{figure*}
\centering 
\begin{adjustbox}{width=1.\textwidth}
  \renewcommand{\arraystretch}{0}%
  \begin{tabular}{c}
     \addlinespace[1pt]
  \includegraphics[width=0.5in]{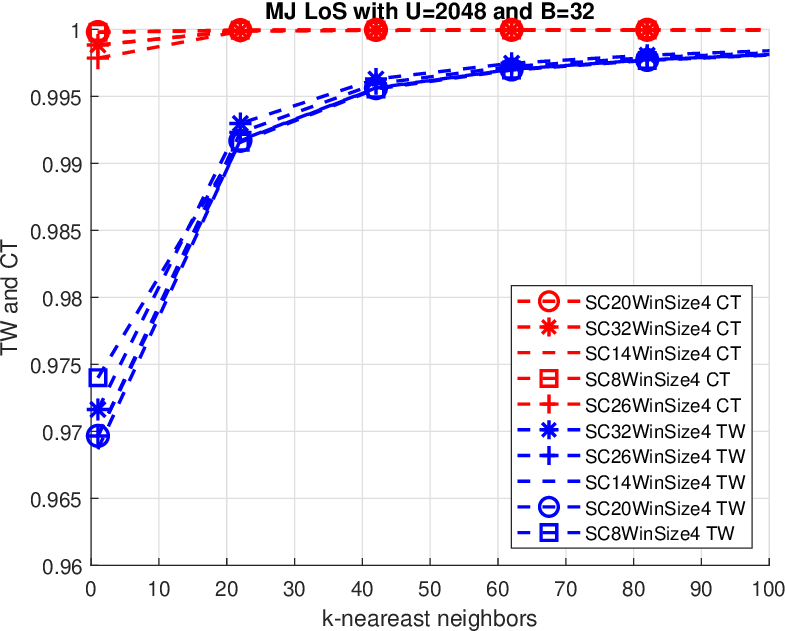}
  \includegraphics[width=0.5in]{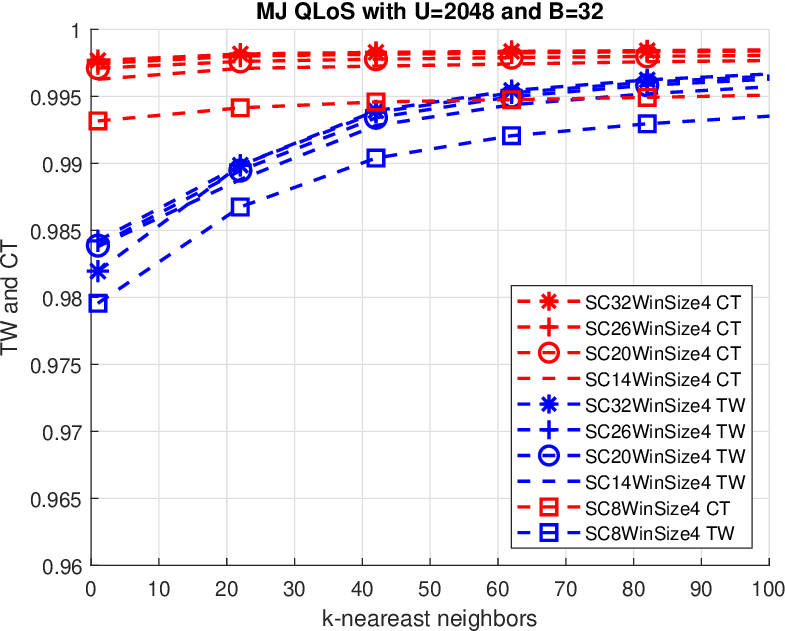}
  \includegraphics[width=0.5in]{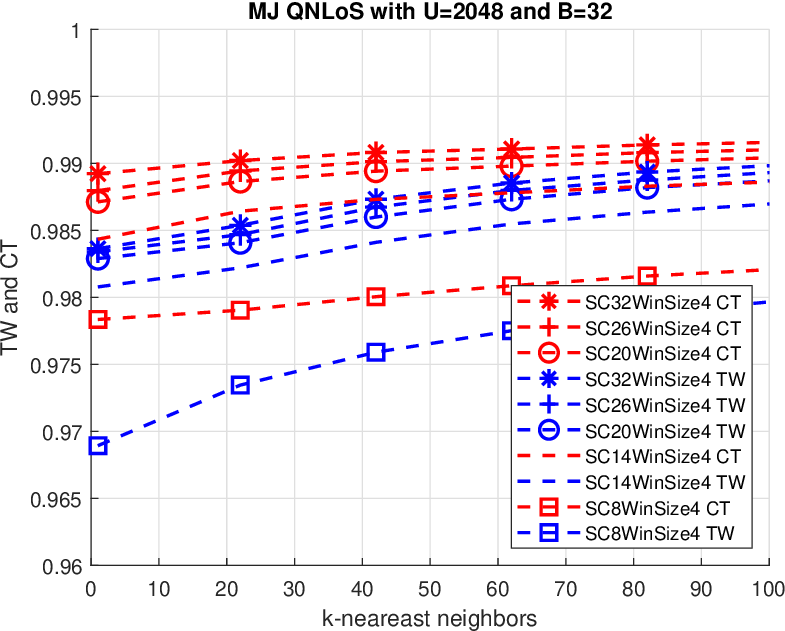} \\
  \end{tabular}
\end{adjustbox}
\caption{TW and CT performance against $k$-nearest neighbors for JM algorithm in 3D channel. Left: LOS, middle: QLOS, right: QNLOS.}
 \label{fig448}
 \end{figure*}
\FloatBarrier
\fi

\ifCLASSOPTIONonecolumn
\begin{table}[!t]
\begin{center}
\begin{tabular}{|c|l|c|c|c|c|c|c|}
\hline
Measure & Channel & 2sc & 8sc & 14sc & 20sc & 26sc & 32sc \\ \hline
   & LOS   &   0.9972  &  0.9986  &  0.9991  &  0.9996  &  0.9997  &  0.9998 \\ \cline{2-8}
TW & QLOS  &   0.9648  &  0.9903  &  0.9933  &  0.9943  &  0.9946  &  0.9950 \\ \cline{2-8}
   & QNLOS &   0.9421  &  0.9664  &  0.9753  &  0.9779  &  0.9804  &  0.9821 \\ \hline
   & LOS   &   0.9982  &  0.9997  &  0.9998  &  0.9998  &  0.9998  &  0.9998 \\ \cline{2-8}
CT & QLOS  &   0.9572  &  0.9920  &  0.9951  &  0.9959  &  0.9965  &  0.9968 \\ \cline{2-8}
   & QNLOS &   0.9337  &  0.9680  &  0.9784  &  0.9813  &  0.9837  &  0.9854 \\ \hline
\end{tabular}
\caption{Performance comparison for TW and CT at $k$-nearest = 102 for RS algorithm in 3D channel at 2, 8, 14, 20, 26, and 32 subcarriers.}
\label{tab5}
\end{center}
\end{table}
\fi

\ifCLASSOPTIONonecolumn
\begin{table}[!t]
\vspace{3mm}
\begin{center}
\begin{tabular}{||c | c||}
 \hline
 Algorithm & Simulation time (seconds) \\ 
 \hline\hline
 PCA & 0.817 \\
 \hline
 SM & 12.2 \\
  \hline
 AE & 53.9  \\
 \hline
 LR & 7.15  \\
 \hline
 ISQ & 7.09 \\
 \hline
MM & 20.4 \\
 \hline
JM & 442.0 \\
\hline
\end{tabular}
\end{center}
\caption{Simulation times.}
\label{tab222}
\end{table}
\fi

\ifCLASSOPTIONonecolumn
\begin{table}[!t]
\begin{center}
\begin{tabular}{||c|c||}
\hline
Algorithm & Number of Multiplications \\
\hline\hline
PCA & $O(ND^2+D^3)$\\
\hline
SM & $\gg \textrm{PCA}$\\
\hline
AE & $\gg \textrm{LR, ISQ}$\\
\hline
LR & $O(N_R^2 P)$\\
\hline
ISQ & $O(N_R^2 P)$\\
\hline
MM & $O(N_R^2 P + N_R^2 N_S + N_S^2 Q+N_S^2 N_R)$\\
\hline
RS & $O(P (2N_R^2+1)+N_SN_R^2+Q (2N_S^2+1) +N_RN_S^2)$\\
\hline
\end{tabular}
\end{center}
\caption{Number of multiplications employed by the algorithms in the paper. $N:$ number of vectors of feature size $D$, to be reduced to size $d$. $N_R:$ Number of antennas at the BS, $N_S:$ number of subcarriers. $P$: Number of potential AOAs to determine $\theta$, $Q:$ number of potential distances to determine $\rho$.}
\label{tbl:Complexity}
\end{table}
\fi

\ifCLASSOPTIONonecolumn
\begin{table}
\begin{center}
\begin{tabular}{|c|l|c|c|c|c|c|c|c|c|}
\hline
Measure & Channel & PCA & SM & AE & LR & ISQ & MM & JM & RS\\ \hline
   & LOS   &     0.7494 &   0.8132  &  0.8155  &  0.9275 &   0.9146 & 0.9991 & 0.9990 & 0.9958\\ \cline{2-10}
TW & QLOS  &     0.6665 &   0.6968  &  0.7260  &  0.7467 &   0.7470 & 0.9774 & 0.9362 & 0.9612\\ \cline{2-10}
   & QNLOS &     0.6860 &   0.6813  &  0.6981  &  0.7223 &   0.7192 & 0.9361 & 0.9652 & 0.9133\\
\hline
   & LOS   &     0.7919 &   0.8594  &  0.8467  &  0.9403 &   0.9364 & 0.9998 & 0.9998 & 0.9942\\ \cline{2-10}
CT & QLOS  &     0.7570 &   0.7852  &  0.8035  &  0.7670 &   0.7686 & 0.9809 & 0.9456 & 0.9725\\ \cline{2-10}
   & QNLOS &     0.7649 &   0.7679  &  0.7773  &  0.7404 &   0.7356 & 0.9465 & 0.9687 & 0.9372\\
\hline
\end{tabular}
\end{center}
\caption{Performance comparison for TW and CT at $k$-nearest = 102 in 3D channel, SNR = -10 dB.}
\label{tab4233}
\end{table}
\fi

\ifCLASSOPTIONonecolumn
\begin{table}[!ht]
\begin{center}
\begin{tabular}{|c|l|c|c|c|c|c|c|c|c|}
\hline
Measure & Channel & PCA & SM & AE & LR & ISQ & MM & JM & RS\\ \hline
\multirow{6}{*}{TW}   &\multirow{2}{*}{LOS}&     0.8566 &   0.7741 &  0.8402  &  0.9932 &   0.9888 & 0.9994 & 0.9984 & 0.9960\\
   &       &   0.43\% & 6.4\%  & -1.4\% &-0.020\%&-0.030\%&0.040\%&-0.020\%&0.38\% \\ \cline{2-10}
   &\multirow{2}{*}{QLOS}&     0.8450 &   0.8307 &  0.8536  &  0.9078 &   0.9085 & 0.9977 & 0.9969 & 0.9939\\
   &       &   0.28\% & 2.4\%  &  4.4\% & 0.12\%& 0.077\%&-0.010\%&-0.11\%&0.11\% \\\cline{2-10}
   &\multirow{2}{*}{QNLOS}&     0.8487 &   0.8451 &  0.8528  &  0.9059 &   0.9068 & 0.9859 & 0.9910 & 0.9820\\
   &       &   0.18\% & 0.059\% & -0.38\% & -0.33\% & -0.30\% & -0.030\% & -0.16\% & 0.010\%\\
\hline
\multirow{6}{*}{CT}&\multirow{2}{*}{LOS}&   0.9276 &   0.8868 &  0.8534  &  0.9968 &   0.9941 & 0.9999 & 1.0000 & 0.9943\\
   &       &0.13\% & 2.0\% & 4.5\% & 0 & -0.010\% & 0.010\% & 0 & 0.55\% \\\cline{2-10}
   &\multirow{2}{*}{QLOS}&     0.9215 &   0.8881 &  0.9268  &  0.9402 &   0.9292 & 0.9991 & 0.9987 & 0.9933\\
   &       & 0.087\% & 4.3\% & -2.4\% & 0.15\% & 0.13\% & 0.010\% & -0.10\% & 0.35\% \\\cline{2-10}
   &\multirow{2}{*}{QNLOS}&     0.9239 &   0.9219 &  0.9108  &  0.9283 &   0.9254 & 0.9874 & 0.9929 & 0.9829\\
   &       & -0.22\% & -0.022\% & -0.56\% & -0.40\% & -0.37\% & -0.091\%  & -0.19\% & 0.25\% \\
\hline
\end{tabular}
\end{center}
\caption{Performance comparison for TW and CT at $k$-nearest = 102 in 3D channel, phase noise -100 dBc/Hz @  1 MHz offset, SNR = 0 dB.
Numbers in the second rows for each entry are the percentage differences with the entries in Tables~\ref{tab42}--\ref{tab5}.}
\label{tab4284}
\end{table} 
\fi